\documentstyle[11pt,aaspp4]{article}

\def\distrbn{distribution }
\def\diff{different }
\def\rec{reconstruction }
\def\gau{Gaussianization }
 
\def\gal{galaxy }
\def\nnbr{nearest neighbor }
\def\z{redshift }
\def\vel{velocity }

\def\be{\begin{equation}}
\def\ee{\end{equation}}
\def\hmpc{{h^{-1}\;{\rm Mpc}}}
\def\hubunits{km \hskip -2pt s$^{-1}$ \hskip -2pt Mpc$^{-1}$}
\begin{document}

{\title{RECONSTRUCTION ANALYSIS OF GALAXY REDSHIFT SURVEYS: 
A HYBRID RECONSTRUCTION METHOD}
\author{\bf
Vijay K. Narayanan and
David H. Weinberg
}
\affil{Department of Astronomy, The Ohio State University, Columbus, OH 43210;}
\affil{ E-mail: vijay,dhw@astronomy.ohio-state.edu}

\bigskip
\bigskip
\bigskip
\centerline{\bf ABSTRACT}
\medskip

In reconstruction analysis of a galaxy redshift survey, one works
backwards from the observed galaxy distribution to the primordial
density field in the same region of space, then evolves the
primordial fluctuations forward in time with an N-body code.
A reconstruction incorporates assumptions about the values of
cosmological parameters, the properties of primordial fluctuations,
and the ``biasing'' relation between galaxies and mass. These assumptions
can be tested by comparing the reconstructed galaxy distribution
to the observed distribution, and to peculiar velocity data when available.
This paper presents a hybrid reconstruction method that combines
the ``Gaussianization'' technique of Weinberg (1992)
with the dynamical schemes of Nusser \& Dekel (1992) and Gramann (1993a).
We test the method on N-body simulations and on N-body mock catalogs
designed to mimic the depth and geometry of the Point Source Catalog
Redshift Survey and the Optical Redshift Survey.
The hybrid method is more accurate than
Gaussianization or dynamical reconstruction alone.
Matching the observed morphology of clustering can set limits on the
bias factor $b$ independently of $\Omega$.
Matching cluster velocity dispersions and the redshift space distortions
of the correlation function $\xi(s,\mu)$ constrains the parameter
combination $\beta \approx \Omega^{0.6}/b$.
Relative to linear or quasi-linear approximations, a fully
non-linear reconstruction makes more accurate predictions of 
$\xi(s,\mu)$ for a given $\beta$, reducing the systematic biases
of $\beta$ measurements and offering further possibilities for
breaking the degeneracy between $\Omega$ and $b$.  Reconstruction
also circumvents the cosmic variance noise that limits conventional
analyses of $\xi(s,\mu)$, since the orientations of large,
coherent structures in the observed galaxy distribution are reproduced
in the reconstruction.  Finally, reconstruction can improve the
determination of $\Omega$ and $b$ from joint analyses of redshift 
and peculiar velocity surveys because it provides a fully non-linear
prediction of the peculiar velocity distribution at each point
in redshift space.

\keywords{cosmology: theory, galaxies: clustering, large scale structure of
the Universe}
\slugcomment{Submitted to ApJ, \date}

\section{INTRODUCTION}

The standard approach to testing theories for the formation of
large scale structure uses analytic approximations or
numerical simulations to predict volume-averaged statistical
properties of galaxy clustering.
A complete theoretical model specifies the properties of primordial
fluctuations, the values of cosmological parameters like $H_0$
and $\Omega$, and the ``biasing'' relation between the galaxy distribution
and the underlying mass distribution.  If the model is correct in all
of its details, then the statistical properties of the 
predicted clustering should match those of the observed
clustering to within the measurement uncertainties, which are usually
dominated by the finite volume of the data sample.  However, one could 
not expect a simulation started from random initial conditions to
reproduce the detailed arrangement of observed structures --- the
Local Supercluster and the Perseus-Pisces filament, for example ---
even if the statistical properties of these initial conditions were correct.

In this paper we focus on reconstruction analysis of galaxy redshift
surveys, a complementary approach to the study of large scale structure.
Here one works backwards from the observed galaxy distribution
to the initial fluctuations in the same region of space, then evolves
these model initial conditions forward in time to the present day.
A reconstruction of this sort incorporates assumptions --- about cosmological
parameters, about bias, and perhaps about the statistical properties
of the initial conditions --- and these assumptions are tested by
comparing the evolved reconstruction to the original galaxy redshift data.
The strength of this approach is that a reconstruction with correct
assumptions should reproduce the specific structure in the region
probed by the survey, eliminating finite volume statistical 
fluctuations (a.k.a.\ ``cosmic variance'') as a source of uncertainty
in the comparison between theory and data.
Even the properties of individual clusters, superclusters, and voids
can serve as diagnostics for the success of a reconstruction.
Reconstruction analysis can therefore be a valuable supplement
to traditional statistical studies of the galaxy distribution,
by more fully exploiting the information present in redshift surveys.
Reconstruction can also be a powerful tool in the comparison between
galaxy density and peculiar velocity fields, since a reconstruction
of a redshift survey provides a fully non-linear prediction of the 
peculiar velocity distribution throughout the survey volume.

The limitation of reconstruction analysis is that no method can recover
the initial fluctuations with perfect accuracy, so even a
reconstruction with correct assumptions will not produce an exact 
match to the input data.  The magnitude of expected errors can be
calibrated on numerical simulations, but the discriminatory power
of reconstruction analysis is clearly greater if the reconstruction
method is more accurate.  Proposed methods for recovering initial
fluctuations from redshift survey data fall into three general categories:
the ``Gaussianization'' technique of Weinberg (1992, hereafter W92),
which monotonically maps the smoothed galaxy density field to smoothed
initial conditions with a Gaussian probability distribution;
dynamical methods based on the Zel'dovich (1970) approximation
(\cite{nd92}; \cite{gr93a}), which integrate the gravitational potential
or velocity potential backward in time; and dynamical methods
based on the least action principle (\cite{lap89}; \cite{giav93}; 
\cite{shaya95}; \cite{piza97}),
which attempt to construct dynamically self-consistent galaxy orbits
with appropriate boundary conditions.
In this paper we describe a hybrid reconstruction method that combines
many of the best features of the first two approaches.  In the case
where galaxies are assumed to be unbiased tracers of the underlying
mass distribution, our hybrid method is broadly similar to the 
technique used in Kolatt et al.'s (1996) reconstruction of the
1.2 Jy IRAS redshift survey, though the two methods differ in numerous 
details.  Our method for incorporating the possibility of biased
galaxy formation is novel; it allows us to recover the initial
{\it mass} density field without assuming a detailed model of the
relation between galaxies and mass today.  The hybrid method
is more accurate than Gaussianization alone, and it is more flexible
than the dynamical methods because it works further into the non-linear
regime and can be applied to biased galaxy distributions.

Much of the power of reconstruction analysis derives from the fact
that non-linear gravitational evolution transfers power from large scales
to small scales.  Structure on $\sim 1$ Mpc scales of the evolved mass
distribution is largely determined by the collapse of initial fluctuations
on a scale of several Mpc.  One consequence is that reconstruction
cannot recover details of the initial fluctuations on scales much smaller 
than the present day scale of non-linearity (Fourier wavenumbers
$k > k_{\rm nl}$); information about these fluctuations is effectively
erased by non-linear evolution (Little, Weinberg, \& Park 1991, hereafter
\cite{lwp91}).  The encouraging converse is that a reconstruction 
that recovers the initial fluctuations up to $k=k_{\rm nl}$ can
reproduce the evolved structure with reasonable accuracy 
even on smaller ($k>k_{\rm nl}$) scales (see \cite{lwp91}).
Taking advantage of this transfer-of-power effect requires that 
the reconstruction method work for smoothing lengths where the rms 
fluctuation of the smoothed galaxy density field is $\sigma_s \sim 1$.
For the observed galaxy distribution this scale is $\sim 8\hmpc$ for
a tophat smoothing window (\cite{dp83}) or $\sim 3-4\hmpc$ for a 
Gaussian smoothing window (where $h \equiv H_0 / 100\;$\hubunits).
Reconstructions that start from more
heavily smoothed fields can still be useful, but they cannot
reproduce collapsed structure nearly as well, and they therefore have
less power to test the validity of different assumptions, especially
with regard to biased galaxy formation.  In this paper we will therefore
compare reconstruction methods using a Gaussian smoothing length
of $3\hmpc$, and the hybrid method is specifically designed to
function at this scale.

The plan of the paper is as follows.  In \S 2.1, we briefly review
the Gaussianization and dynamical reconstruction methods, and in \S 2.2
we describe the hybrid reconstruction scheme, a combination of these
two approaches.  In \S 3.1 we test the hybrid scheme on cosmological
N-body simulations, comparing its accuracy to that of Gaussianization
or dynamical reconstruction alone.  In \S 3.2 we apply the hybrid
scheme to simulations with biased galaxy formation, focusing on the
ability of reconstruction analysis with the hybrid method to discriminate
between models with different degrees of bias.
All of the N-body data sets used in \S 3 are periodic, real space cubes.
In \S 4 we apply the hybrid method to mock redshift catalogs with the
depth and geometry of the Point Source Catalog Redshift Survey
(PSCZ, \cite{pscz1}) and the Optical Redshift Survey (ORS, \cite{ors1}).
This section describes how we account for peculiar velocity distortions
in redshift space, and the mock catalog tests focus on the ability of 
reconstruction analysis to constrain values of $\Omega$ and the bias
factor and on its accuracy in predicting the galaxy peculiar velocity field.
In \S 5 we summarize our results and discuss the potential applications
of reconstruction analysis.

\section{ A HYBRID RECONSTRUCTION SCHEME}
\subsection{\it Gaussianization and Dynamical Reconstruction Schemes}

All density fluctuations grow at the same rate when they are in the linear 
regime of gravitational instability (characterized by 
$\vert \delta \vert \ll 1$).
This universal behavior is destroyed once the density fluctuations
become non-linear ($\vert \delta \vert \geq 1$).
The Gaussianization \rec method (W92) is based 
on the approximation that the rank order of the mass density field,
smoothed over scales of a few Mpc, is
preserved even under non-linear gravitational evolution.
The method employs a monotonic mapping of the smoothed
final  density field to a smoothed initial mass
density field that has a Gaussian one-point probability 
distribution function (PDF).
By construction, this procedure imposes a Gaussian PDF
on the initial mass density field.
The high overdensities in extreme non-linear regions are mapped
to the positive tail of the Gaussian distribution, while the 
voids are assigned density values in the negative tail (see W92,
figure 3, for a graphical illustration of this procedure).
This method works 
satisfactorily even on moderately non-linear scales, and it can be used to 
recover the initial density fields with smoothing
lengths as small as $R_{s}=3h^{-1}$Mpc (Gaussian filter radius).
To the extent that the recovered initial density field is
accurate, an N-body simulation started from these initial conditions
should reproduce the true properties of the final mass distribution,
including the locations and masses of individual structures.

If the monotonic relation between the smoothed initial
and final density fields were exact, then Gaussianization 
would recover the smoothed initial density field perfectly.
However, as shown by W92, non-linear effects tend 
to suppress small scale power in the reconstructed initial
density field, beyond the suppression due to the smoothing filter.
We correct for this effect using the ``power restoration'' procedure
of W92.
Using an ensemble of N-body simulations, we compute (ensemble averaged)
correction factors $C(k)$ defined by 
\be
C(k) = \left[\frac{P_{r}(k)}{P_{i}(k)}\right]^{1/2}~,
\label{eqn:ckdef}
\ee
where $P_i(k)$ is the power spectrum of a simulation's smoothed
initial conditions and $P_r(k)$ is the power spectrum of the density
field recovered by Gaussianizing the simulation's smoothed final density
field.  When applying the reconstruction procedure, we multiply
each Fourier mode of the Gaussianized final density field by $C(k)$
and also multiply by $\exp(k^2 R_s^2/2)$ in order to remove
the effect of the original Gaussian filtering.
Above some wavenumber $k_{\rm corr} \sim \pi/R_{\rm nl}$, 
where $R_{\rm nl}$ is the scale on which rms fluctuations are $\approx 1$,
non-linear evolution erases the phase information in the
initial density field (\cite{lwp91}; \cite{rg91}) to the
point that Gaussianization cannot recover it.
For $k>k_{\rm corr}$, therefore, we simply add random phase
Fourier modes with an assumed power spectrum.
More specifically, we assume a {\it shape} for the primordial
power spectrum and normalize it by fitting the power spectrum
of the recovered density field up to the wavenumber $k_{\rm corr}$.
We then add random phase small scale waves in the range 
$k_{\rm corr} < k \leq k_{\rm Nyq}$, 
where $k_{\rm Nyq}$ is the Nyquist frequency of the grid
on which the initial density field is recovered.
Thus, the long wavelength modes of the Gaussianized,
power-restored density field preserve 
the phase information of the true initial density field, while the 
small scale modes have random phases by construction.

We determine the overall amplitude of the initial fluctuations
by evolving them forward with an N-body code until they
reproduce the amplitude of fluctuations in the
input (non-linear) density field.  Specifically, we require the
reconstruction to reproduce $\sigma_8$, the rms fluctuation in $8\hmpc$ 
spheres, which is related to the power spectrum of the input density field by
\be
 {\sigma}^{2}_{8} = \int_{0}^{\infty} 4\pi k^{2}P(k)\tilde W^{2}(kR)dk,
\label{eqn:s8def}
\ee
where $\tilde W(kR)$ is the Fourier transform of a spherical tophat
with radius $R=8\hmpc$.
The values of the correction factors $C(k)$ themselves depend (mildly)
on the amplitude of the initial fluctuations, so we require that
the correction factors and the recovered initial power spectrum
be self-consistent (see W92 for further discussion).

Any reconstruction of the observed galaxy distribution should also account
for the possibility that the galaxy distribution is a biased tracer
of the underlying mass distribution.  
As long as the bias between the mass and \gal distributions preserves 
the rank order of the smoothed mass density field,
the effects of biased galaxy formation can be easily reversed by 
Gaussianization: the procedure does not assume any specific biasing
model, only that regions of higher galaxy density are also regions
of higher mass density.
However, a detailed knowledge of the bias mechanism is
necessary in the amplitude normalization step, as biasing can change the 
shape and the amplitude of the mass power spectrum, and hence the
value of $\sigma_{8}$.  
Thus, when the power restored mass density field is evolved forward in time,
we require an explicit biasing prescription
to convert the evolved mass \distrbn to the galaxy distribution, 
before we can compare it  with the true final galaxy distribution.

The procedure for reconstructing a galaxy distribution by the 
Gaussianization method can be summarized as follows:
\begin{description}
\item [{(G1)}:] Smooth the final galaxy density 
field with a Gaussian filter of radius $R_{s}$.
\item[\ \ \ \ \ ] The smoothing length should be large enough
to suppress shot noise caused by the discreteness of the galaxy
distribution and to suppress very strong non-linearities.
In all of our tests below, we use a mean galaxy density
$n_{g}=  0.01 h^{3}$Mpc$^{-3}$ and a Gaussian smoothing length 
$R_{s} = 3\hmpc$, yielding an rms fluctuation of the smoothed
galaxy density field $\sigma_s \sim 1.3$.
\item [{(G2)}:] Monotonically map this smoothed final galaxy density 
field to field with a Gaussian PDF.
\item [{(G3)}:] Restore power.
\item[\ \ \ \ \ ] We multiply all modes of the Gaussianized density
field with $k\leq k_{\rm corr}$ by the empirically
determined correction factors $C(k)$ and by $\exp(k^2 R_s^2/2)$.
In the small wavelength regime, $k_{\rm corr} < k \leq k_{\rm Nyq}$,
we add random phase waves that are 
drawn from an assumed power spectrum normalized to match the large scale modes.
\item[{(G4)}:] Evolve this power-restored density field forward in time,
assuming a value for $\Omega$.
Select galaxies from this evolved mass distribution
either in an unbiased manner or with an assumed biasing prescription.
Fix the normalization of the reconstructed initial conditions by 
requiring that the reconstructed galaxy distribution have the same 
$\sigma_{8}$ as the original galaxy distribution.
\item[{(G5)}:] Compare the local and global properties of this 
reconstructed galaxy distribution with those of the original
galaxy distribution.
\item[\ \ \ \ \ ] We can constrain the value of $\Omega$ and the bias 
parameter (or parameters) by requiring that we accurately recover 
the observed properties of the galaxy distribution.
\end{description}

There is one obvious source of inaccuracy in the Gaussianization method.
Since it maps the final galaxy density field to a Gaussian initial 
mass density field at the same Eulerian position, it ignores any 
bulk displacements of galaxies during gravitational evolution.
In regions where a large concentration of galaxies has
moved significantly during
gravitational evolution, the recovered initial density value
at an Eulerian position will correspond to the true initial density value
at a different position.
These displacements are typically small ($\sim$ a few Mpc),
and they are therefore not fatal to the Gaussianization procedure.
However, we can improve the accuracy of the Gaussianization method
if we can account for these displacements.

Alternatives to Gaussianization that naturally correct for the 
displacements during gravitational evolution include the two related methods
that we refer to as ``dynamical'' \rec schemes.
These methods attempt to reverse the effects of gravitational evolution
by treating the mass density field as a self gravitating fluid.
Under this assumption, the second-order differential equation that governs 
the growth of density fluctuations in an expanding universe has
both growing and decaying mode solutions (\cite{lss80}).
Direct attempts to run gravity backwards will be stymied 
by the decaying mode, which, when evolved back in time, blows up any noise 
present in the final density field.
The dynamical schemes overcome this problem by approximating the evolution
of velocity or gravitational potentials using {\it first}-order differential 
equations that have only growing mode solutions.
The first such scheme was proposed by Nusser \& Dekel (1992) and is 
based on the Euler 
momentum conservation equation and the approximation that 
the comoving trajectories of mass particles are straight lines
(the Zel'dovich [1970] approximation).
The Zel'dovich-Bernoulli equation, as derived by Nusser \& Dekel (1992),
combines 
the Zel'dovich approximation, the assumption of an irrotational velocity 
field, and the Euler momentum 
conservation equation, yielding a first-order differential equation for the
evolution of the velocity potential $\phi_{v}$,
\be
\frac{\partial \phi_{v}}{\partial D} = \frac{1}{2} \vert \nabla \phi_{v}\vert
^{2}, 
\label{eqn:zelber}
\ee
where $D(t)$ is the growth rate of  density fluctuations in linear theory.
In the linear regime, this velocity potential is related to the perturbed 
gravitational potential $\phi_{g}$ by
\be
\phi_{v}({\bf x},t) = \frac{2f(\Omega)}{3H\Omega}\phi_{g}({\bf x},t),
\label{eqn:phivg}
\ee
where $\Omega$ is the density parameter, $H$ is the Hubble constant,
and $f(\Omega) = \dot{D}/HD$.

Gramann (1993a), showed that the initial gravitational potential can be
recovered more accurately using  the Zel'dovich-continuity 
equation of Nusser et al. (1991), which combines the Zel'dovich displacements 
with the mass continuity equation.
Under the assumption of an irrotational velocity field, 
the evolution of the gravitational potential is then described by the 
equation 
\be
\frac{\partial \phi_{g}}{\partial D} = \frac{1}{2} \vert \nabla \phi_{g}\vert
^{2} + C_{g},
\label{eqn:zelcon}
\ee
where $C_{g}$ is the solution of the Poisson type equation
\be
\nabla ^{2}C_{g} = \sum_{i=1}^{i=3}\sum_{j=i+1}^{j=3}\left[
  {\partial^2 \phi_g \over \partial x_i^2} 
  {\partial^2 \phi_g \over \partial x_j^2} - 
  {\left(\partial^2 \phi_g \over \partial x_i \partial x_j \right)^2} 
  \right] .
\label{eqn:cgdef}
\ee
Once we recover  the initial gravitational potential by integrating
backwards in time to $D = 0$, we can derive the initial density field 
from it using the Poisson equation
\be
\nabla ^{2}\phi_{g} = -\delta.
\label{eqn:poi}
\ee

These dynamical reconstruction schemes have so far been used mainly
to recover the initial density fluctuations from the present day galaxy 
density or peculiar velocity field (\cite{nd92}; \cite{kolatt96}).
The properties of these reconstructed initial fluctuations can then be 
compared directly to the theoretical expectations of any model for the
origin of these fluctuations.
However, these methods could also be used in a full fledged \rec
of a galaxy redshift catalog in much the same way as the Gaussianization method
described above.
The steps in such a scheme can be summarized as follows:
\begin{description}
\item[{(D1):}] Smooth the final density field with a filter large enough 
to remove any gross non-linearities, so that $\sigma_s \la 1$.
Compute the smoothed
final velocity potential $\phi_{v}$ or gravitational potential 
$\phi_{g}$, depending on whether the \rec will be
based on the Zel'dovich-Bernoulli equation
or the Zel'dovich-continuity equation.
\item[{(D2):}] Calculate the smoothed
initial velocity potential or gravitational
potential by integrating equation ~(\ref{eqn:zelber}) 
or~(\ref{eqn:zelcon}) backwards in time to $D = 0$.
If the Zel'dovich-Bernoulli equation~(\ref{eqn:zelber}) is used,
compute the initial gravitational potential from the velocity potential
using equation~(\ref{eqn:phivg}).
Derive the initial density field from the initial gravitational 
potential by solving the Poisson equation (\ref{eqn:poi}).
\item[{(D3):}] Restore power.  Same as (G3).
\item[{(D4):}] Evolve forward and normalize.  Same as (G4).
\item[{(D5):}] Compare the reconstruction to the input data.  Same as (G5).
\end{description}

If we were to evolve the dynamically  reconstructed initial density field 
forward in time using the Zel'dovich approximation, we would be guaranteed 
to reproduce the smoothed final mass density field, as the dynamical \rec
schemes apply the Zel'dovich approximation in reverse.
However, if we evolve this density field forward by an N-body code that 
follows fully non-linear evolution, we can 
get more information about the evolved mass distribution on small scales
because of the transfer of power from large scales to small scales.
This means that we can recover small scale non-linearities
in the final mass distribution that  cannot be reproduced using
linear or quasi-linear approximations.

The dynamical schemes naturally correct for bulk displacements 
during gravitational evolution, unlike the Gaussianization 
method, which performs an  Eulerian mapping of the final density  
to the initial density at the same position.
Thus, dynamical schemes lead to more accurate locations 
of density structures when the reconstructed initial fields are evolved
forward in time.
In addition, since there is no {\it a priori} constraint 
on the PDF of the initial fluctuations, the dynamical schemes can 
be used to check if the initial density fluctuations derived
from redshift or peculiar velocity surveys are indeed Gaussian distributed 
(\cite{ndy95}).

The dynamical schemes can recover the 
initial density fields from peculiar velocity data in a straightforward 
manner, as  the velocity potential constructed from the peculiar velocity 
catalogs can be easily integrated back in time using 
equation~(\ref{eqn:zelber}).
There are, however, two major disadvantages in applying the dynamical
schemes to reconstruct galaxy {\it redshift surveys}.
The first drawback is the need to smooth the density fields over fairly
large scales, since the perturbation theory expansions break down in regions of
high density contrast.
As a result, dynamical reconstruction
cannot accurately recover the initial density field 
down to the non-linear scale $k_{\rm nl}$,
and when evolving forward in time, it cannot get the full benefit
of non-linear transfer of power from large to small scales.
As we shall see below, this transfer of power helps to 
break the degeneracy between bias and dynamical evolution.
The other drawback of the dynamical schemes is that they need
the final {\it mass} density field as the input field.
Thus, before reconstructing the initial mass density field,
one must either assume that galaxies trace mass or adopt an 
explicit biasing model to 
convert the galaxy number density fluctuations to mass density fluctuations.
Gaussianization, by contrast, recovers the initial density field using
only the very general assumption that regions of higher galaxy density
are regions of higher mass density; it substitutes a strong
assumption about the PDF of primordial fluctuations in place of
a strong assumption about the relation between galaxies and mass.
Regardless of how the initial fluctuations are recovered, an explicit
biasing scheme is required in the 
forward evolution and normalization step
(G4 or D4), in order to 
derive the final reconstructed \gal \distrbn from the evolved mass
\distrbn before comparison to the input galaxy distribution.

An entirely different approach to reconstructing the initial density 
field uses the least action principle to compute particle
orbits in an expanding universe (\cite{lap89}).
This principle was used by Shaya, Peebles \& Tully (1995)
to reconstruct the orbits of galaxies in the Local Group assuming 
that they had vanishingly small initial peculiar velocities.
Giavalisco et al. (1993) combined the generalized Zel'dovich approximation 
with the least action principle to derive a parametrization for the 
particle orbits.
Croft \& Gazta\~{n}aga (1997) demonstrated that the Zel'dovich approximation 
{\it is} the least action solution when the particle trajectories are 
approximated by rectilinear paths, and they 
used this result to derive the Path Interchange 
Zel'dovich Approximation (PIZA) reconstruction method.
PIZA recovers the initial 
density field quite accurately from unbiased galaxy distributions,
although its applicability to biased galaxy density fields needs further study.
In this paper, we will restrict our attention to the Gaussianization and 
dynamical \rec schemes alone, leaving the analysis of the PIZA
method to a future study (Narayanan \& Croft, in preparation).

\subsection{\it Hybrid Scheme}

The Gaussianization and dynamical \rec schemes that we described above 
have complementary desirable features.
This motivates us to derive a hybrid \rec method, which retains the 
large scale accuracy present in the dynamical methods, gives robust 
reconstructions 
in the non-linear regime ($ \vert \delta \vert > 1$), and does not require 
strong assumptions about biasing in order to recover the initial fluctuations.
We will first describe a hybrid \rec method that can be applied when
galaxies trace mass, then consider modifications of this procedure to
allow for the possibility of biased galaxy formation.
We will demonstrate the superiority of this hybrid method
using N-body simulations in \S3, and we will test it on mock redshift
catalogs drawn from N-body simulations in \S4.

In developing the hybrid method, we 
began by testing the performance of the two dynamical \rec schemes
on a final density field obtained by gravitationally evolving a known 
initial density field using an N-body simulation.
We derived the gravitational potential from the $3h^{-1}$ Mpc Gaussian smoothed 
final density field by solving the Poisson equation, then evolved it backwards 
in time to $D = 0$ using the Zel'dovich-continuity equation~(\ref{eqn:zelcon}).
In general, we found that the Zel'dovich-continuity equation tends to over 
correct for the dynamical displacements of the mass particles.
This effect is quite prominent in the high density regions, with the 
result that the peaks in the reconstructed initial density field are flatter
than the corresponding peaks in the true initial density field.
To reduce this effect, 
we modified the implementation of the Zel'dovich-continuity method in the
following manner.
When we integrate the gravitational potential backwards in time,
we use a smoother potential 
for the source term in the right hand side of equation~(\ref{eqn:zelcon}).
We derive this smoother potential from a more heavily
smoothed final density field
and integrate this smoother potential backwards simultaneously.
We tested with different values of the smoothing length used in 
deriving the smoother potential and found that a Gaussian smoothing of 
$R_{s} = 4 h^{-1}$Mpc led to the best recovery of the initial density 
field, when the final density field is smoothed with a Gaussian filter 
of radius $R_{s} = 3 h^{-1}$Mpc.
We also found that the Zel'dovich-Bernoulli scheme yields a comparable 
recovery of the initial density field if we use the empirical 
relationship derived by Nusser et al.\ (1991) for the relation
between the density and velocity fields in the quasi-linear regime,
\be
\nabla\cdot {\bf v} = -\left(\frac{\delta}{1+0.18\delta}\right).
\label{eqn:vdndbb}
\ee
In what follows, we will always use our modified implementation of the 
 Zel'dovich-continuity equation as the canonical dynamical \rec scheme because
it evolves the final gravitational potential, which can be directly computed 
from the final mass density field without using any empirical approximations.
The Zel'dovich-Bernoulli scheme would be our method of choice if
we started from a peculiar velocity catalog instead of a galaxy redshift 
catalog.

Although the use of a smoother potential improves the initial field recovery 
of the Zel'dovich-continuity method, it is still quite inaccurate in the
non-linear regions where the underlying perturbation theory expansions 
break down.
Therefore, we use a hybrid method, in which we Gaussianize the dynamically 
reconstructed {\it initial} density field to robustly recover the initial 
density field in the non-linear regions.
Relative to the W92 method of Gaussianizing the {\it final} galaxy
density field, the hybrid method recovers more accurate locations
of features in the initial conditions, as we will
demonstrate in \S 3.1 below.  Note that we could not reverse the
order of the Gaussianization and dynamical reconstruction steps
of the hybrid method because we would then over-correct for non-linear 
evolution, producing a non-Gaussian initial density field.
Our hybrid method for reconstructing unbiased galaxy distributions
can be summarized as follows:
\begin{description}
\item[{(H1):}] Smooth the galaxy density field with a Gaussian filter of 
radius $R_{s}$. 
\item[\ \ \ \ \ ] Since we would like to accurately recover the structures 
even on small scales, we use a smoothing length of $R_{s} = 3h^{-1}$Mpc.
\item[{(H2):}] For an unbiased reconstruction, do nothing.
\item[\ \ \ \ \ ] This ``null step'' will be replaced by a critical
procedure in the case of a biased reconstruction, as we will explain below.
\item[{(H3):}] Derive the gravitational potential from the smoothed final
density field using the Poisson equation.
Evolve this gravitational potential backwards in time using the 
 modified implementation of the Zel'dovich-continuity dynamical scheme. 
Compute the dynamically reconstructed initial density field as the 
negative Laplacian of this initial gravitational potential.
\item[{(H4):}] Gaussianize this dynamically reconstructed {\it initial} mass 
density field to recover an initial density field that is accurate even in the 
non-linear regions.
\item[{(H5):}] Restore power to the recovered initial density field in the same 
manner as described in step (G3) for the Gaussianization \rec procedure.
\item[{(H6):}] Evolve this power-restored density field forward in time 
using an N-body simulation and choose galaxies in an unbiased manner from the 
evolved mass distribution.
Fix the amplitude of the initial fluctuations so that the 
$\sigma_{8}$ of the evolved density field matches that of the input 
density field.
\item[{(H7):}] Compare the properties of this reconstructed galaxy 
distribution to those of the input galaxy distribution.
\end{description}
If the dynamical step (H3) recovers a field with a Gaussian PDF,
then step (H4) has no effect.  Step (H4) can be viewed as a
``regularization'' that improves the robustness of the 
Zel'dovich-continuity method (and thereby allows it to be applied
on smaller smoothing scales) by introducing a prior assumption
that the initial fluctuations are Gaussian.

The dynamical \rec in step (H3) requires the smoothed {\it mass}
density field as its input.
If we want to allow for the possibility of biased galaxy formation,
we first need to compute the smoothed final mass density field from the
input \gal data.
We begin by assuming that there is a monotonic biasing relation between the 
smoothed galaxy density field and the smoothed mass density field.
We also assume that the initial mass density fluctuations have a 
Gaussian PDF.
We quantify the bias by the bias factor $b$, defined as
\be
 b = \frac{\sigma_{8g}}{\sigma_{8m}},
\label{eqn:bdef}
\ee
where $\sigma_{8g}$ and $\sigma_{8m}$ are the rms fluctuations in 
$8h^{-1}$Mpc spheres in the non-linear galaxy 
density field and the linear mass density field, respectively.
Note that, with this definition of the bias factor, $ b = 1 $ 
does not necessarily mean that the galaxy distribution is an unbiased 
tracer of the mass distribution, only that it has the same rms
fluctuation amplitude at $8\hmpc$.

The step (H2), which is a null step in the unbiased case, is modified 
in the biased case to the following:
\begin{description}
\item[{(H2B):}] Monotonically map the galaxy density field onto an empirically 
determined PDF of the underlying mass distribution.
\item[\ \ \ \ \ ] We first estimate the $\sigma_{8m}$ of the 
linear mass fluctuations using equation~(\ref{eqn:bdef}), 
assuming a bias factor $b$.
We evolve an ensemble of initial mass density fields
forward in time using N-body simulations,
all of them drawn from the same assumed power spectrum,
and all normalized to this value of $\sigma_{8m}$.
We then derive an ensemble-averaged PDF of the smoothed
final mass fluctuations from the final mass density fields 
of these simulations.
While reconstructing an input final galaxy distribution, we derive a 
smoothed final mass density field  by  monotonically mapping 
the smoothed final galaxy density field to this average PDF.
The resulting smoothed mass density fluctuation field should therefore have 
the same amplitude and PDF as the true mass density field
underlying the input galaxy distribution. 
This smoothed mass density field can be evolved backwards in time 
using the dynamical scheme as in step (H3).
\end{description}

We also replace the final step (H6) in the unbiased case by the following 
step in the biased reconstruction:
\begin{description}
\item[{(H6B):}] Fix the linear theory
amplitude of fluctuations in the power-restored initial density
field to the value $\sigma_{8m} = \sigma_{8g}/b$, before evolving it
forward using an N-body simulation.
Use an explicit biasing scheme to convert the evolved 
mass distribution to a galaxy distribution, choosing the
free parameter (or parameters) of the biasing scheme so that the reconstructed
galaxy distribution has the observed value of $\sigma_{8g}$.
\item[\ \ \ \ \ ] This normalization guarantees that the final mass 
density field has the degree of dynamical evolution that is 
consistent with the adopted bias factor and the 
amplitude of the input galaxy density fluctuations.
\end{description}

The other steps in the hybrid reconstruction of biased \gal distributions are 
the same as in the unbiased case.
Note that step (H2B) is much like the key step (G2) of the
Gaussianization method, except that it attempts to recover the final
mass density field instead of jumping directly to the initial conditions.
In effect, step (H2B) implicitly derives and corrects for 
the only monotonic biasing relation that is simultaneously consistent 
with the smoothed input data, the adopted bias factor $b$, 
and the assumption of  Gaussian initial density fluctuations.
Because the forward evolution should recover structure on scales
smaller than $R_s$, again thanks to the transfer of power from 
large scales to small scales, we do not expect the reconstruction
to reproduce the non-linear properties of the input data unless
the Gaussian assumption and the adopted $b$ are approximately correct.

As an aside, we note that the bias factor as defined in 
equation~(\ref{eqn:bdef}) can
be less than one, corresponding to an anti-bias, in which case 
the mass is more {\it strongly} clustered than the galaxies.
A value of $b$ less than one is also consistent with the assumption of a
monotonic relation between the mass and galaxy densities, since
the galaxy density
$\rho_{g}$ is an increasing function of the mass density $\rho_{m}$ as long 
as the efficiency of the galaxy formation process does not fall faster than 
$\rho_{m}^{-1}$.
Finally, we note that a hybrid \rec assuming that the galaxy \distrbn is 
biased with a bias factor $b = 1$ may be different from a \rec assuming an 
unbiased galaxy \distrbn because we can have a non-linear relation
between the mass and galaxy
density fields that does not change the rms fluctuation amplitude at a 
particular scale.
The two reconstructions will be similar only when the PDF of the \gal \distrbn 
is identical to that of the underlying mass distribution.

We will now test this hybrid \rec method on final galaxy density fields 
that are derived 
from simulations in which the input assumptions are known {\it a priori}.
The hybrid \rec analysis of a real galaxy redshift catalog should also 
take account 
of the distortions in redshift space that are caused by the peculiar velocities
of galaxies.
We will describe our method for correcting these distortions in \S4,
where we test the \rec method on artificial redshift catalogs.
Before that, however, we will test the \rec method 
in a more controlled setting, where the density fields are 
constructed from the periodic, real space, final \gal distributions 
that are derived from the output of  N-body simulations.

\section{TESTS ON N-BODY SIMULATIONS}

A hybrid \rec of the observed \gal \distrbn by the method described
in \S2.2 incorporates a number of assumptions 
in addition to the core hypothesis that structure formed by the gravitational
instability of Gaussian  primordial fluctuations.
In decreasing order of importance, these assumptions are:
\begin{description}
\item[{(1)}] A value of the bias factor $b$.
\item[\ \ \ \ \ ] We need to assume a value of $b$ to determine the 
amplitude of mass fluctuations $\sigma_{8m}$ that corresponds to the 
observed galaxy number density fluctuations.
The value of $\sigma_{8m}$ is used (a) to determine the PDF of the 
final mass fluctuations used in the mapping step (H2B), 
(b) to choose the correction factors $C(k)$ used in the power
restoration step (H5), and, most importantly, 
(c) to fix the normalization of the initial conditions
when they are evolved forward in time.
\item[{(2)}] An explicit biasing scheme, i.e, a prescription for 
selecting galaxies from the underlying  mass distribution.
\item[\ \ \ \ \ ] In the final step (H6B), we evolve the reconstructed
{\it mass} density field forward in time using an N-body simulation.
Therefore, we have to adopt a specific biasing scheme to convert this 
evolved mass \distrbn to a galaxy \distrbn before we can compare it to the 
input galaxy data.
In principle, we can have many different biasing schemes all of which yield 
the same value of the bias factor, although the resulting galaxy 
distributions might be significantly different.
Matching the input data can yield constraints on the correct model
of biasing.
\item[{(3)}] A value of $\Omega$. 
\item[\ \ \ \ \ ] We have to assume a value for $\Omega$ when we evolve
the reconstructed initial conditions forward in time.
This assumed value has only a minimal effect on the resulting real space 
mass distribution (\cite{weinberg90}; \cite{nusser97}).  However, 
it directly affects the resulting peculiar velocity field 
($\vert v \vert \propto \Omega^{0.6}$ in the linear regime), and it therefore 
influences the \z space structure of the reconstructed \gal distribution.
The value of $\Omega$ also affects the normalization of the 
initial fluctuations because the 
clustering properties of the galaxies in redshift space 
(and hence $\sigma_{8g}$) 
are different from those in  real space (\cite{kaiser87}).
The value of $\Omega$ also affects the recovery of the initial
conditions in the first place because it is used in 
correcting input data from redshift space to real space, before defining
the smoothed final density field.  We describe this correction
procedure in \S4 below.
\item[{(4)}] A shape of the primordial power spectrum. 
\item[\ \ \ \ \ ] Although the amplitude of the initial power spectrum 
is constrained by 
the bias factor and the amplitude of galaxy number density fluctuations 
$\sigma_{8g}$, its shape is still an unknown quantity.
Information about this shape is required at two different steps in the 
\rec procedure: first, to compute the average PDF of the evolved mass \distrbn
and the correction factors $C(k)$, and later to add the random phase small 
scale waves for wavenumbers larger than $k_{\rm corr}$.
In practice, the \rec is insensitive to the assumed shape
of the power spectrum within a reasonable range because
the mass PDF and the correction factors are more sensitive to the 
amplitude of the power spectrum ($\sigma_{8m}$) than to its shape and
because the small scale waves that are added have only a modest 
influence on the evolved structure (\cite{lwp91}), so the
specific power spectrum used for them makes little difference.
\end{description}

In testing the \rec method, we are primarily interested in two
questions that deal with the effects of these assumptions :
\begin{description}
\item[{(1)}] If we make correct assumptions about the physics that 
produced the input \gal
\distrbn --- namely, the same bias factor, value of $\Omega$, and biasing 
prescription --- 
does the hybrid \rec method accurately reproduce the input data?
\item[{(2)}] If the method incorporates incorrect assumptions about the 
bias factor or $\Omega$, does it produce an identifiably erroneous
galaxy distribution?
\end{description}
The first question addresses the accuracy and robustness of the \rec
method, while the second addresses the sensitivity of the method as a
cosmological diagnostic test.
In the remainder of this paper, we will 
reconstruct input \gal distributions for which we
know the  correct set of assumptions.
For a fixed value of $\sigma_{8m}$, 
the value of $\Omega$ primarily affects the peculiar velocities of 
galaxies and has only a minimal effect on the evolved real space 
structure at zero redshift (\cite{weinberg90}; \cite{nusser97}).
The peculiar velocities influence the redshift space structure 
of the \gal \distrbn and will be important in the reconstructions of the mock
\z catalogs that we will consider in \S4.
However, in this section we work only with real space data,
so we simply adopt $\Omega=1$ and 
focus on the accuracy of the \rec method and on its ability
to detect incorrect assumptions about the bias factor.

In all our simulations, we use an initial power spectrum of the form
given by Efstathiou, Bond, \& White (1992),
\be
P(k) = \frac{Ak}{\left\{1+ \left[ ak+\left(bk\right)^{3/2}+\left(ck\right)^{2} 
\right\}^{\nu} \right]^{2/\nu}},
\label{eqn:pkdef}
\ee
where $a = (6.4/\Gamma)h^{-1}$Mpc, $b = (3.0/\Gamma)h^{-1}$Mpc, 
$c = (1.7/\Gamma)h^{-1}$Mpc, $\nu = 1.13$, and $A$ sets 
the normalization. 
This two parameter family of power spectra is characterized by the 
amplitude $A$ (or equivalent $\sigma_8$) and by the 
shape parameter $\Gamma$, which is equal to $\Omega h$
in cold dark matter models with a small baryon density 
and scale-invariant inflationary fluctuations.
We choose $\Gamma = 0.25$, a value that is consistent with 
the observed clustering properties of several galaxy catalogs 
(\cite{pd94}; \cite{maddox90}).
We choose random phases for the different Fourier components of the 
initial density field so that the resulting density field is Gaussian.

We define all the density fields in a periodic cube of side $200 h^{-1}$Mpc.
We follow the non-linear gravitational evolution of these density fields 
using a particle-mesh (PM) N-body code written by Changbom Park.
This code is described and tested in Park (1990).
We use $100^{3}$ particles and a $200^{3}$ force mesh in the PM simulations.
We start the gravitational evolution from a redshift $z = 23$ and follow
it to $z = 0$ in 46 equal incremental steps of the expansion scale factor 
$a(t)$.
For the unbiased case, we derive the galaxy distribution by randomly sampling
 the evolved mass \distrbn to the desired density.
In the case of biased distributions, we select galaxies from the
evolved mass \distrbn by assuming  a functional relationship between the local 
mass and galaxy densities.
We explain this biasing relation and our procedure for selecting 
galaxies in more detail in \S3.2.
We form the continuous galaxy density fields by cloud-in-cell 
(CIC) binning the discrete galaxy distributions onto a 
$100^{3}$ grid.
Since we would like to apply this \rec technique to real data sets
in the future, we test the \rec method on galaxy distributions whose number 
density and amplitude of fluctuations are typical of existing
data sets, ensuring
that the effects of sampling noise and non-linear gravitational evolution
are included  at a realistic level.
In all the tests shown below, we derive the galaxy density field
from a galaxy \distrbn whose average number density is
$n_{g} = 0.01 h^{3}$Mpc$^{-3}$ and whose rms fluctuation in 
spheres  of radius $8h^{-1}$Mpc is $\sigma_{8g} = 1.1$, which is
consistent with the value measured  from optical galaxy redshift survey 
catalogs (\cite{dp83}).

\subsection{\it Unbiased Reconstructions}

We choose the amplitude of the power spectrum of the true initial 
density field so that the $\sigma_{8g}$ of the non-linear
\gal \distrbn obtained by
randomly sampling the evolved mass \distrbn is $1.1$.
We recover the initial density field from this final,
unbiased galaxy \distrbn using all of the methods 
described in \S2.
Before the forward evolution steps in the \rec procedures, we multiply the 
reconstructed Fourier modes in the wavenumber range $0 < k/k_{f} \leq 20$ 
by correction factors $C(k)$ determined using an 
ensemble of reconstructions of N-body density fields with similar initial 
power spectra.
Here $k_{f} = 2\pi/L_{\rm box} = 0.0314\  h\;$Mpc$^{-1}$ is the 
fundamental wavenumber of the simulation box of side 
$L_{\rm box} = 200h^{-1}$Mpc.
For wavenumbers in the range $ 20k_{f} < k < k_{\rm Nyq}$, we
add random phase waves using the procedure  described in \S2, where 
$k_{\rm Nyq} = 50k_{f} = 1.57\ h$Mpc$^{-1}$ is the Nyquist frequency in the 
simulation box.

Figure 1 shows isodensity contours in a slice through the initial density 
fields convolved with a Gaussian filter $e^{-r^{2}/2R_{s}^{2}}$, with the  
smoothing radius $R_{s}=3h^{-1}$Mpc.
The slices correspond to the density field in the region
$(x1,y1) = (50,50) h^{-1}$Mpc to $(x2,y2) = (150,150) h^{-1}$Mpc
at a height of $z = 50h^{-1}$Mpc from the bottom of the 
periodic cube.
The contour levels range from $-2\sigma$ to $+2\sigma$ in intervals
of $0.4\sigma$, where $\sigma$ is the rms fluctuation in the density field.
The true initial density field is shown in panel (a).
The final unbiased galaxy \distrbn that is obtained by 
evolving this density field forward in time using the PM code 
is shown in Figure 4a below.
We show the density field reconstructed using the 
Zel'dovich-continuity dynamical scheme alone in panel (b). 
Clearly, the reconstruction in highly overdense regions is not satisfactory,
with the presence of ridge-like features surrounding the density peaks.
The perturbation theory approach underlying the dynamical scheme
breaks down in these highly overdense regions, resulting in the poor recovery.
Panel (c) shows the density field reconstructed by Gaussianizing the 
smoothed final density field.
The non-linear structures are recovered reasonably well, as they are
mapped onto the tails of the Gaussian distribution.
The results of the hybrid reconstruction are shown in panel (d).
Evidently, the reconstruction in non-linear regions is much better 
than that of the dynamical scheme alone.
The hybrid scheme also recovers more accurate positions for the density 
structures compared to Gaussianization alone.
This improvement is not obvious in Figure 1, but it will become  evident when 
we compare the locations of corresponding structures in the final galaxy 
distributions that are obtained by evolving these recovered initial 
density fields forward in time.

\begin{figure}
\centerline{
\epsfxsize=\hsize
\epsfbox[18 144 592 738]{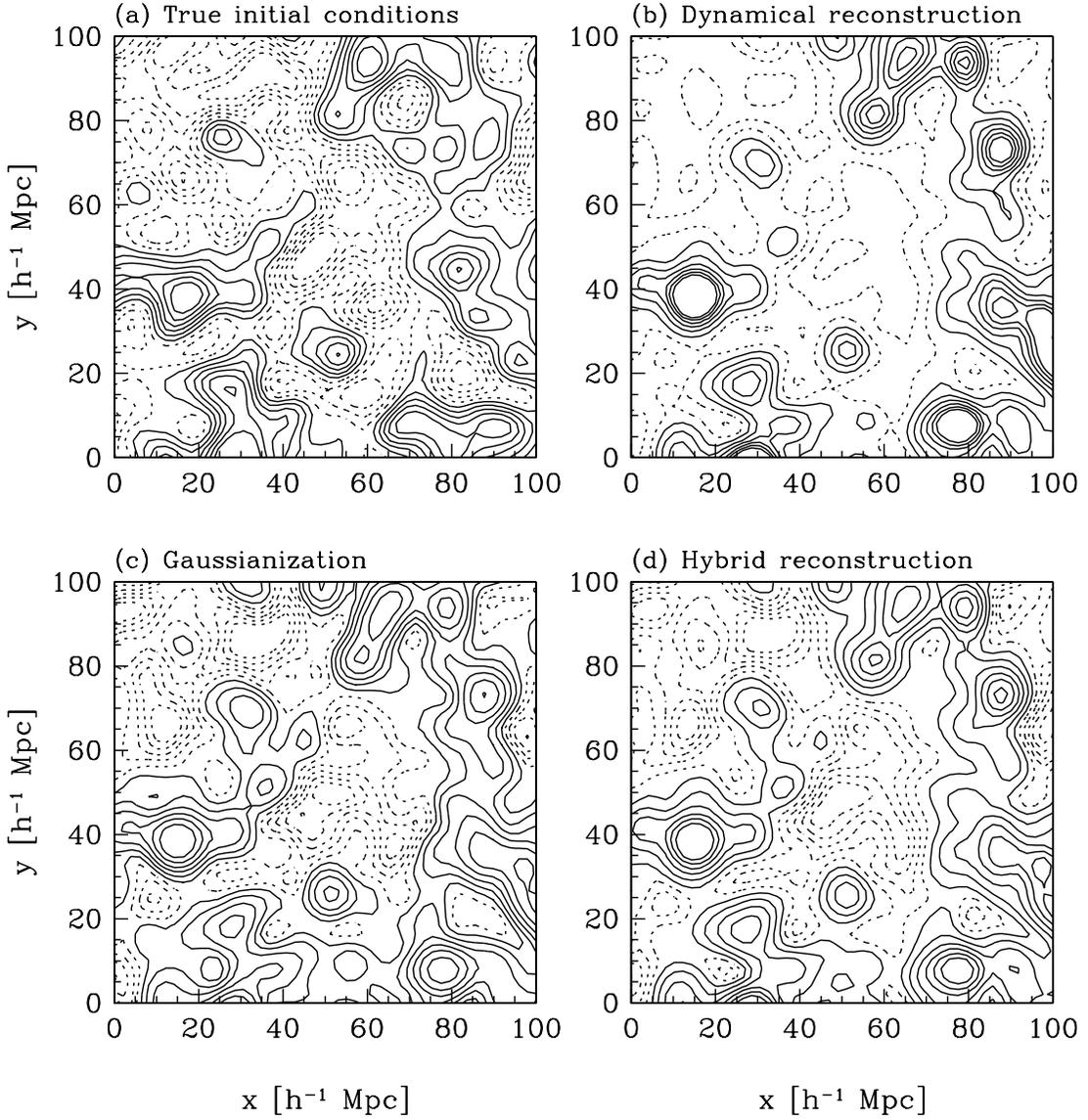}
}
\caption{Contours in a slice of the initial density field of a
test N-body simulation and its reconstruction.
The contour levels range from $-2\sigma$ to $+2\sigma$ in steps of $0.4\sigma$.
Solid contours correspond to overdensities, while dashed contours
correspond to underdensities.
({\it a}) True initial conditions, Gaussian with a $\Gamma=0.25 $ power 
spectrum.  A slice through the galaxy distribution evolved from this field
appears in Fig.\ 4a.
Remaining panels show the initial density field reconstructed from this
evolved distribution by 
({\it b}) the dynamical scheme alone,
({\it c}) Gaussianization alone, and
({\it d}) the hybrid scheme.
}
\end{figure}

Figure 2 shows scatter plots of the initial density fields.
The density contrast $\delta_{r}$ at any cell in the reconstructed field is
plotted against the true initial density contrast $\delta_{i}$ at the same cell.
We scale each \distrbn by its rms value because we determine the 
amplitude of the initial fluctuations only later by evolving this density
field forward and comparing it to the input data.
The scatter plot of the dynamically reconstructed field (panel a) 
clearly demonstrates the failure of this reconstruction method at the 
extremal regions ($|\delta| > 1$).
Gaussianization of the final density field (panel b) leads to  a better 
reconstruction in these extremal regions, but the scatter about the
perfect reconstruction line $(\delta/\sigma)_{r} = (\delta/\sigma)_{i}$ 
is quite large.  This scatter can be quantified by the correlation 
coefficient between the reconstructed and  
the true initial density fields, defined as 
\be
r = \frac{\left< \delta_{r}\delta_{i}\right>}
{\left<\delta_{r}^{2}\right>^{\frac{1}{2}}
\left<\delta_{i}^{2}\right>^{\frac{1}{2}}}.
\label{eqn:rdef}
\ee
This correlation is much smaller for the Gaussianization reconstruction than 
for the dynamical reconstruction.
The hybrid scheme shown in  panel (c) offers the best 
reconstruction of the three methods.
There is good recovery even in the extremal regions, and a
smaller scatter about the ridge line, leading to a much stronger 
correlation between the reconstructed and true initial density fields.

\begin{figure}
\epsfxsize=\hsize
\epsfbox[18 144 592 738]{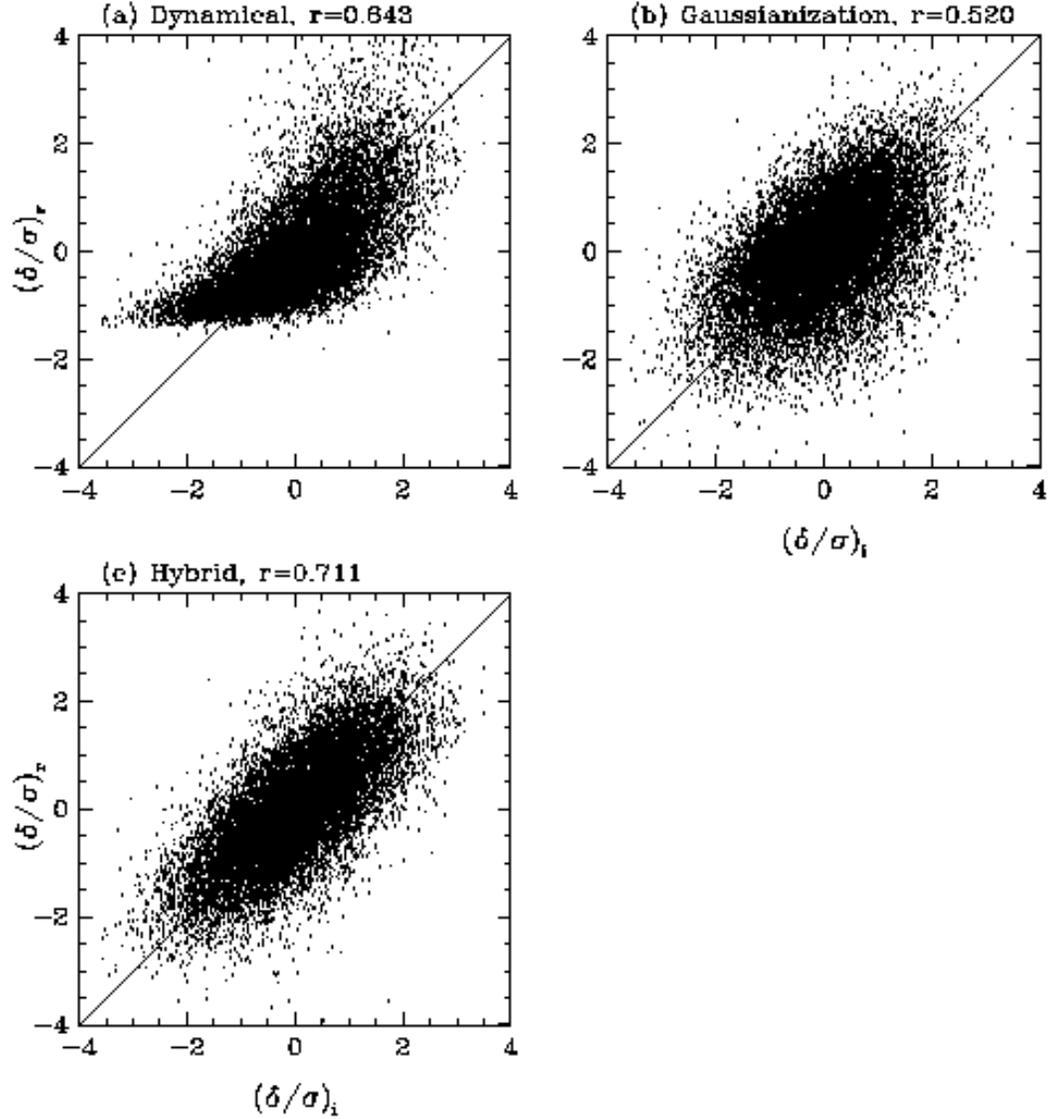}
\caption{ Cell by cell comparison of the reconstructed initial density contrast
$({\delta}/{\sigma})_{r}$ to the true initial density contrast 
$({\delta}/{\sigma})_{i}$ for 
({\it a}) dynamical reconstruction,
({\it b}) Gaussianization, and
({\it c}) the hybrid method.
All the density fields are smoothed with a Gaussian filter of radius 
$R_{s} = 3h^{-1}$Mpc  and scaled
by the rms fluctuation $\sigma$. The correlation coefficient $r$ is
indicated above each panel.
}
\end{figure}

Figure 3 shows the power spectrum of the true initial density 
field (dotted line), the reconstructed initial conditions (solid line),
and the reconstructed initial conditions prior to power restoration
(dashed line, with arbitrary normalization).
The dashed line displays the suppression of small scale power due to
non-linear evolution, but this is corrected adequately by the
power restoration step, as the good agreement in shape of the solid and
dotted lines demonstrates.
The power spectrum of the full hybrid reconstruction
has a slightly lower amplitude than the 
true initial power spectrum (about $10\%$ lower amplitude in the power 
spectrum corresponding to about a $5\%$ lower amplitude for $\sigma_{8m}$).
This may reflect the presence of residual non-Gaussianity in the
reconstructed field, which we detected as a slight ``meatball'' shift
in the genus curve (\cite{mwg88}).
Thus, although the 1-point probability \distrbn is Gaussian by 
construction, the N-point distributions of the recovered initial 
density field may be non-Gaussian.
However, any impact of residual non-Gaussianity on the derived 
$P(k)$ normalization is quite weak, as shown by the good agreement
between the true and the reconstructed power spectra in Figure 3.

\begin{figure}
\centerline{
\epsfxsize=\hsize
\epsfbox[18 144 592 738]{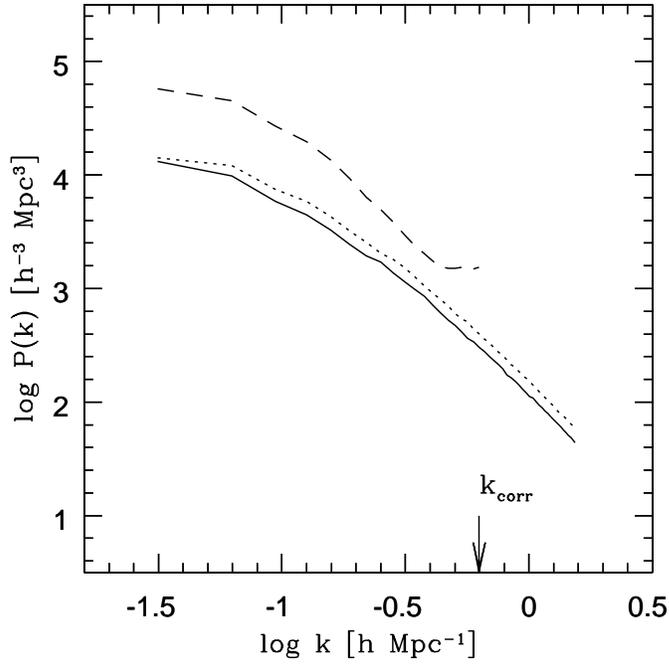}
}
\caption{ Power spectrum of the true initial density field (dotted line),
the density field reconstructed by steps (H1)-(H4) of the hybrid method 
(dashed line),
and the hybrid reconstructed density field after the power
restoration and amplitude matching procedures (solid line).
The dashed line has been multiplied by the factor $e^{k^{2}R_{s}^{2}}$ in 
the range 
$0 < k \leq k_{\rm corr}$ to restore the power lost in the Gaussian 
smoothing, and its amplitude has been fixed arbitrarily. 
}

\end{figure}

Figure 4 shows the true and the reconstructed final galaxy distributions.
We plot the locations of the ``galaxy'' particles, a random subset 
of all the N-body particles, that lie in a  region $40 h^{-1} $Mpc thick 
about the 
center of the cube and extend in the $x$-$y$ 
plane from $(50,50) h^{-1} $Mpc to
 $(150,150) h^{-1} $Mpc.
Comparing the locations of clusters in the three galaxy distributions, we see
that the hybrid scheme (panel b) in general, recovers more accurate 
positions for the clusters than does
Gaussianization alone (panel c).
This improvement is clear, for example, in the corresponding locations of 
the clusters
located near $(x,y) = (115,145) h^{-1} $Mpc and $(110,135) h^{-1} $Mpc in the 
true final galaxy distribution (panel a).
There is also a cluster at $(x,y) = (80,50) h^{-1}$Mpc in the \gau
reconstruction.
This cluster is located in an adjacent slice in the true and hybrid 
reconstructed galaxy distributions.
We will quantify the agreement in the cluster locations below.
Panel (d) shows the final galaxy distribution reconstructed by the hybrid scheme
assuming (incorrectly) that the galaxy distribution is biased with $b = 2$.
We explain the biasing scheme that we used to get this \gal \distrbn in \S3.2.
This biased  galaxy distribution clearly appears more diffuse 
compared to the true galaxy distribution.
We will quantify this diffuse appearance using the nearest neighbor statistic 
described below.

\begin{figure}
\epsfxsize=\hsize
\epsfbox[18 144 592 738]{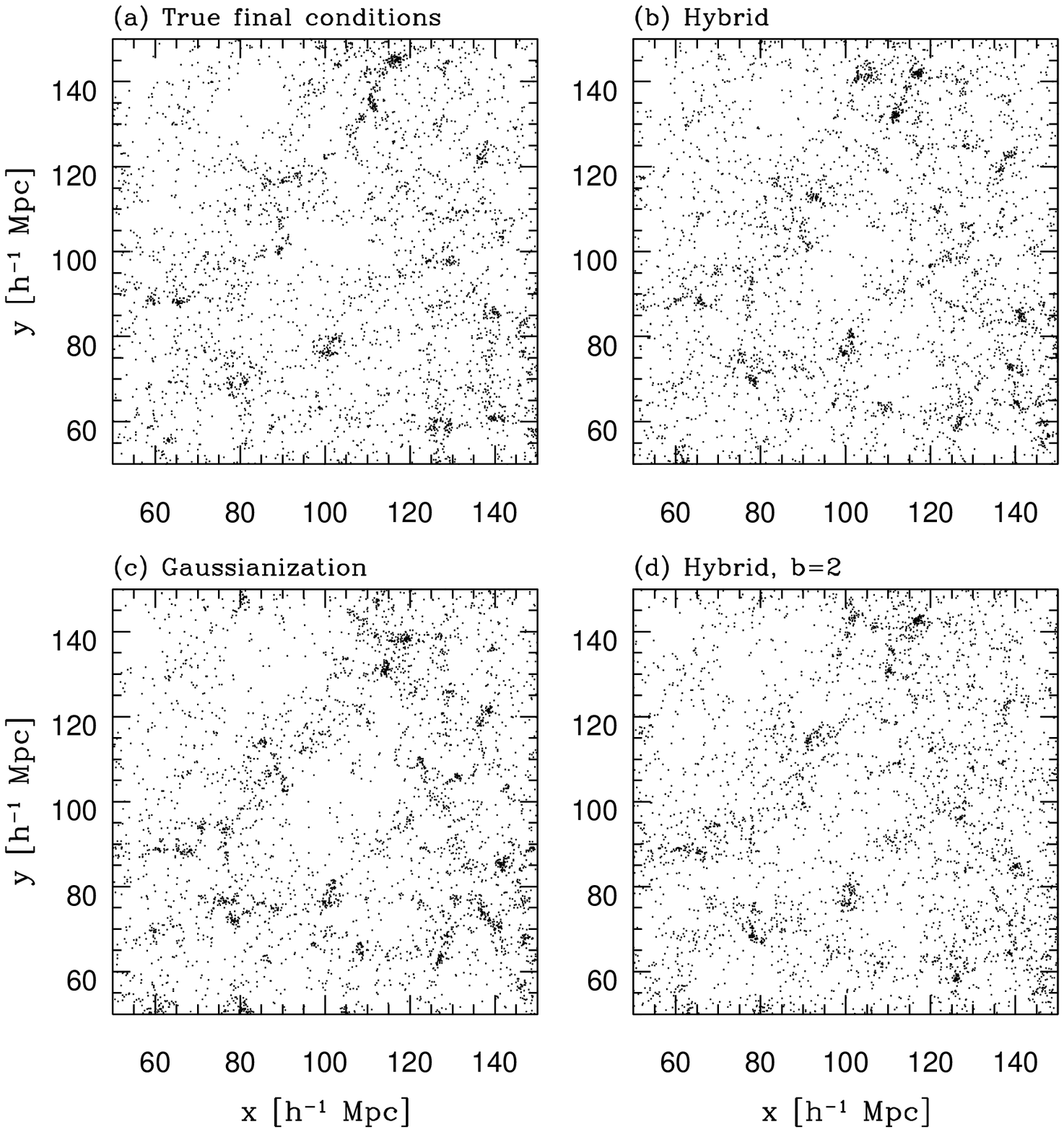}
\caption{ Final galaxy distributions, with $\sigma_{8g}=1.1$.
The panels show galaxy distributions in a slice $40 h^{-1}$Mpc thick
and spanning $100 h^{-1}$Mpc in the other two dimensions.
({\it a}) True final \gal distribution (unbiased).
({\it b}) Hybrid reconstruction assuming unbiased galaxy formation.
({\it c}) Gaussianization assuming unbiased galaxy formation.
({\it d}) Hybrid reconstruction assuming biased galaxy formation with $b=2$.
}

\end{figure}

Figure 5 shows a scatter plot of the final density fields after smoothing with 
a Gaussian filter of radius $R_{s} = 3 h^{-1} $Mpc and scaling by the 
rms fluctuation.
The correlation is much stronger for the hybrid \rec(panel a)
compared to Gaussianization alone (panel b), as would be expected from the 
greater dynamical accuracy of the hybrid method.

\begin{figure}
\epsfxsize=\hsize
\epsfbox[18 300 592 738]{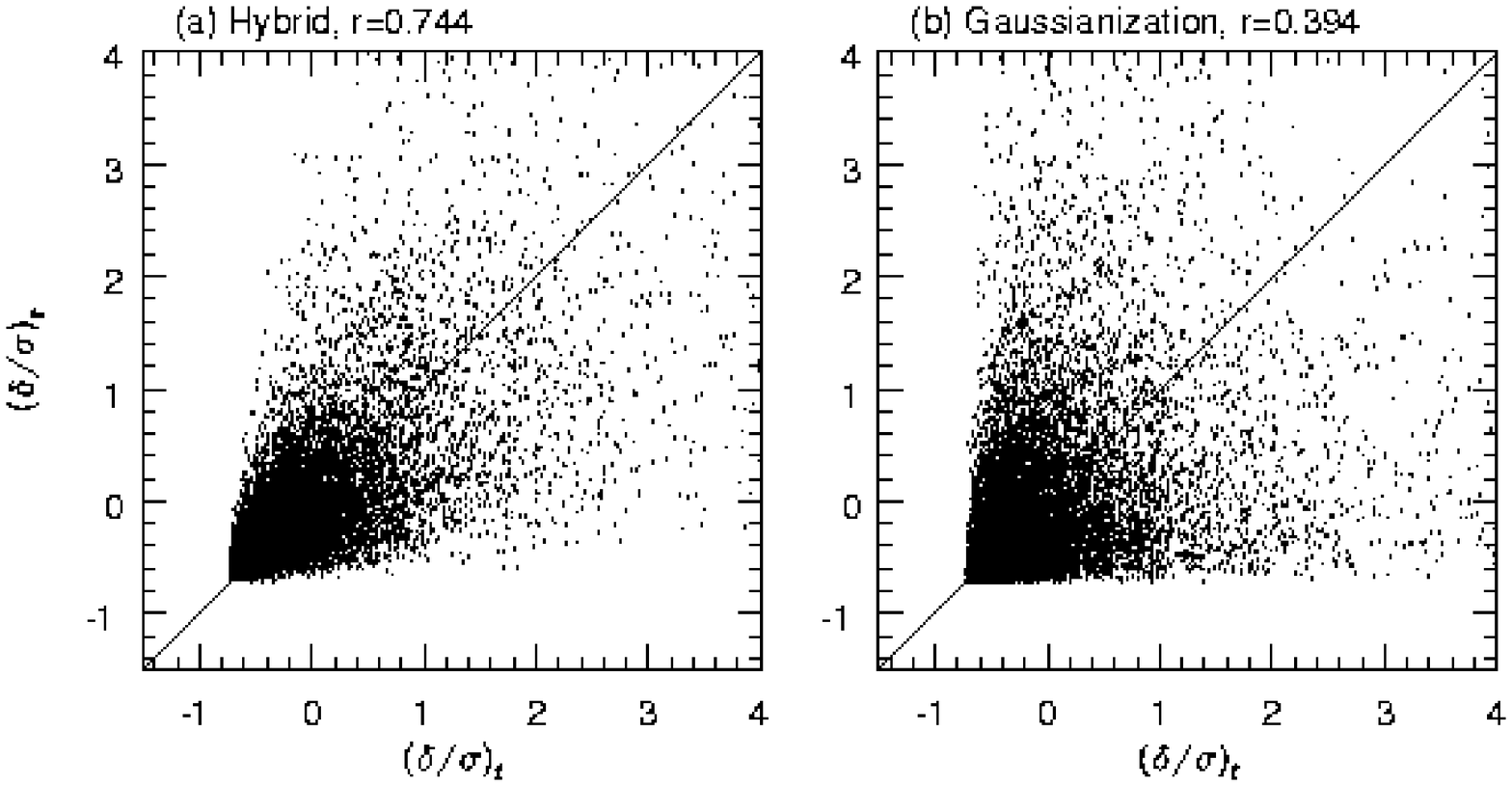}
\caption{ Cell by cell comparison of the reconstructed final density contrast
$(\delta/\sigma)_{r}$ to the true final density contrast 
$(\delta/\sigma)_{f}$ for 
({\it a}) Hybrid reconstruction and
({\it b}) Gaussianization.
The density fields are smoothed with a 3$h^{-1}$Mpc Gaussian filter and scaled
by the rms fluctuation $\sigma$. The linear correlation coefficient $r$ is
indicated above each panel.
}
\end{figure}

Clusters are the most massive collapsed structures in the final 
galaxy distributions.
The abundance and masses of clusters encode important 
information regarding the amplitude of mass fluctuations and 
the value of $\Omega$ (\cite{wef93}; \cite{ecf96}; \cite{cole97}; 
\cite{fan97}).
Therefore, we analyze the extent to which the locations and 
properties of clusters can be reproduced by the different 
reconstruction procedures.
We identify the clusters in the galaxy distributions using the standard 
friends of 
friends algorithm (\cite{davis85}), with a linking length parameter 
$ b = 0.2\bar d$, where $\bar d = 4.64 h^{-1}$Mpc is the mean 
inter-\gal separation.
The mean overdensity of clusters selected with this
linking parameter is approximately 250, corresponding roughly  
to the criterion for virial equilibrium.
We also require that a cluster contain at least 10 galaxies.

We match the clusters in the true and the reconstructed galaxy
distributions using the algorithm described by Weinberg, Hernquist \&
Katz (1997).
We first sort the cluster lists in descending order of cluster masses.
Then, for every cluster in the true final galaxy distribution, we find the 
most massive unmatched cluster in the reconstructed galaxy distribution
whose centroid lies within a distance $l = 12 h^{-1} $Mpc from the centroid
of the original cluster.
We repeat this procedure for progressively less massive clusters
until we complete the cluster list of the true \gal distribution.
The  results that we show below are not sensitive to reasonable 
variations in the values of $l$, although for a shorter matching 
length a larger fraction of clusters remains unmatched in the end.
The histograms in Figure 6 show the number of clusters that match between the 
true and the reconstructed final galaxy distributions as a function of the
distance between their centroids.
The solid and the dashed lines show this statistic for the hybrid and 
Gaussianization \rec schemes, respectively.
The dotted line shows the number of clusters that can  match randomly
between the true and the hybrid reconstructed \gal distributions.
We estimate this by interchanging the $x$ and $y$ coordinates of the clusters 
in the hybrid reconstruction and matching the clusters using the same 
algorithm.
Comparison of the solid and dashed histograms demonstrates the
clear superiority of the hybrid reconstruction method:
while the total number of matched clusters is similar for the two
reconstructions (about 400), the hybrid scheme puts clusters closer 
to their actual locations.
This is precisely the sort of improvement we expect from
the greater dynamical accuracy of the hybrid method.

\begin{figure}
\epsfxsize=\hsize
\epsfbox[18 144 592 738]{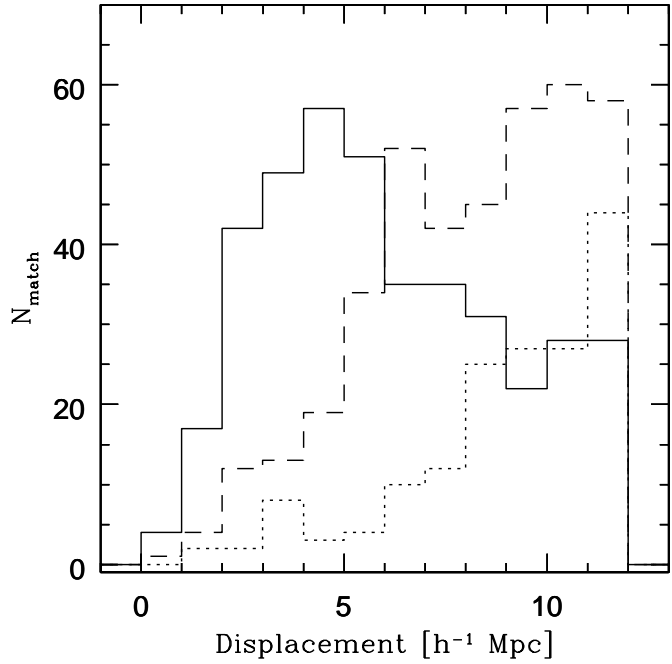}
\caption{ Cluster matching statistics for the final \gal
distributions of the  hybrid reconstruction scheme (solid line) and
Gaussianization (dashed line). Clusters in the true final \gal distribution 
are matched to the most massive unmatched cluster in the reconstructed 
\gal distribution within a radius of $12 h^{-1}$Mpc.
The dotted line shows the expected number of random matches; it is
obtained by interchanging the $x$ and $y$ coordinates of clusters
in the hybrid reconstruction and then matching.
}

\end{figure}

In Figure 7, we compare the multiplicities of the matched clusters in the 
true and the reconstructed \gal distributions.
Circles show the multiplicities of clusters that are matched between 
the true and the reconstructed \gal distributions.
Crosses parallel to either axis represent clusters that are present in one \gal
\distrbn (true/reconstructed), but not matched to a corresponding cluster
in the other (reconstructed/true) \gal distribution.
The scatter for the massive clusters (log $N_{c} > 1.5$) is much smaller
for the hybrid scheme (panel a) than for the \gau reconstruction
(panel b).
The hybrid scheme also matches a larger fraction of these clusters, as shown by 
the smaller number of crosses along either axis at $\log N_{c}  > 1.5$.

\begin{figure}
\epsfxsize=\hsize
\epsfbox[18 300 592 738]{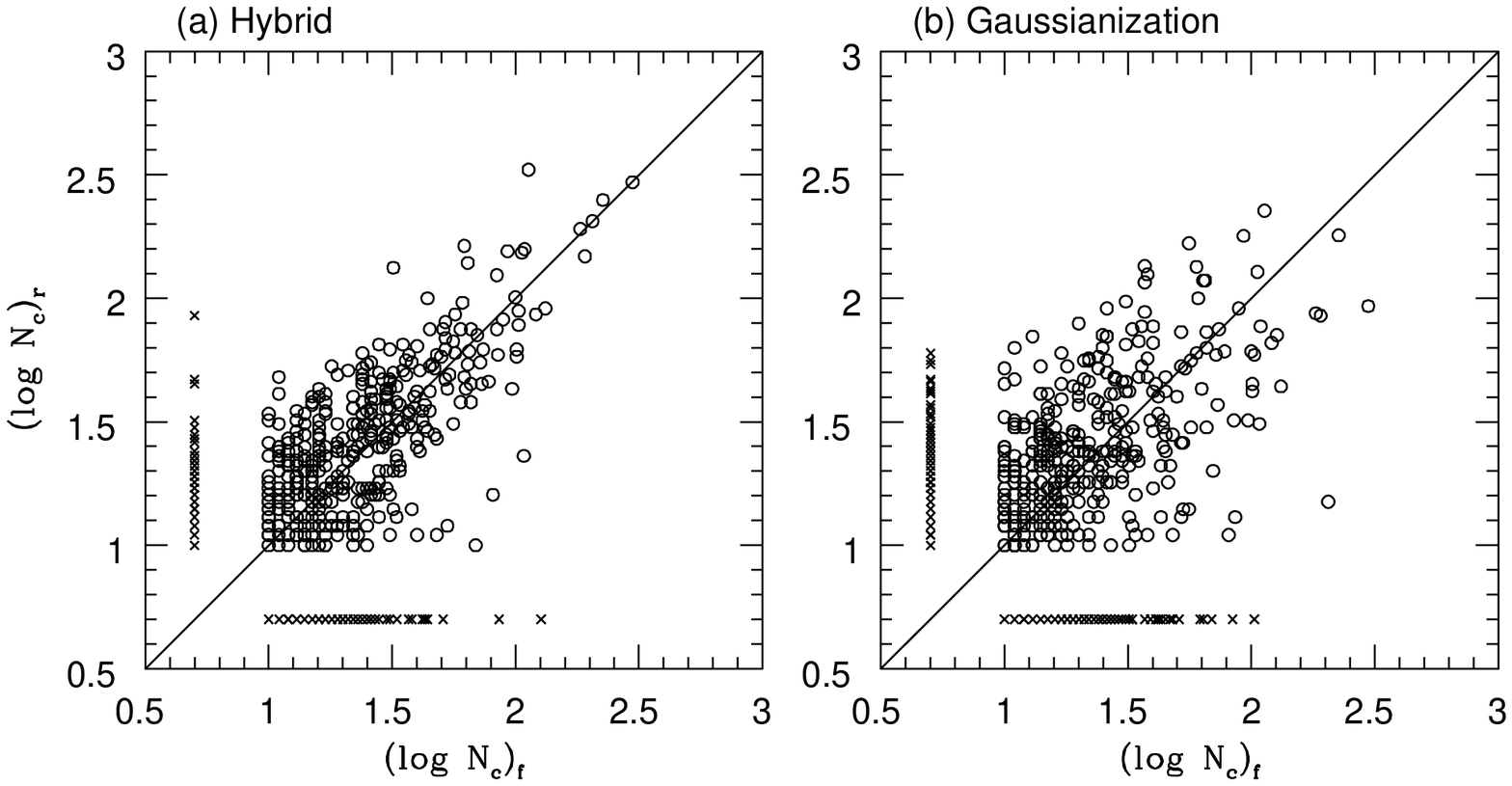}
\caption{ Comparison of the cluster multiplicities between the true final
\gal distribution and the reconstructed \gal distribution
for ({\it a}) the hybrid reconstruction scheme and ({\it b}) Gaussianization.
Crosses parallel to either axis represent clusters present 
in that \gal distribution alone.
}

\end{figure}

Thus far, we have directly compared the recovered initial density 
fields and the reconstructed final \gal distributions for the various
\rec methods with the true initial density field
and the true final \gal distribution.
These comparisons have helped us understand the accuracy of the \rec 
methods and have shown that the hybrid method is superior to the 
\gau and dynamical methods in its ability to reproduce the observed features.
We now compare the global statistical properties of the input and 
reconstructed \gal distributions.
Since the hybrid \rec has a significantly higher dynamical accuracy than 
the Gaussianization method, we show the results of our statistical comparisons
for the hybrid \rec only.

The main purpose of the global statistical
comparisons is to test the effects of the 
different assumptions that enter the \rec procedure.
In this section 
we focus on the bias factor, and we therefore
compare the results of an unbiased hybrid \rec of an unbiased 
true \gal \distrbn 
(the model) to the results of the hybrid reconstructions of the same 
model that assume 
(incorrectly) that the \gal \distrbn is  biased with bias factors of 
$b = 2$ or $b = 3$.
We perform the hybrid reconstructions of the input model by following the 
procedures described in \S2.
The specific biasing scheme that we use in step (H6B) is described in 
\S3.2 below.
We show the galaxy \distrbn that is reconstructed by the hybrid method
assuming $b = 2$ in Figure 4d.
We have applied numerous statistical measures to the input and the 
reconstructed \gal distributions, although we show only two of these here: the 
two-point correlation function $\xi(r)$ and the nearest neighbor
distribution $P(x_{n})$.
When we analyze the mock catalogs of redshift surveys in \S4, we will also
examine the angular anisotropy of the redshift space correlation 
function $\xi(s,\mu)$, which is induced by the peculiar velocities of galaxies.

Figure 8 shows the two-point correlation function $\xi(r)$ for the true
unbiased galaxy \distrbn (dotted line) and for the hybrid reconstructed 
\gal distributions with different  assumptions about bias.
We see that the $\xi(r)$ of the hybrid reconstruction
assuming (correctly) unbiased galaxy formation  matches the 
true $\xi(r)$ very well on all scales (solid line).
However, hybrid \rec assuming (incorrectly) $b = 3$  leads to a shallower 
$\xi(r)$, with a weak clustering strength on small scales (dot-dashed line).
For an observed value of $\sigma_{8g}$, the amplitude of mass 
fluctuations is a decreasing function of $b$.
Therefore, the mass distribution in an unbiased scenario
is more dynamically evolved and has a steeper $\xi(r)$ than in the 
corresponding biased case.
Biasing can amplify the mass clustering to match the input
$\xi(r)$ on large scales, but it cannot simultaneously achieve the 
strong small scale clustering that is produced by gravitational collapse.
The deficit of small scale clustering is not seen clearly
in $\xi(r)$ for the $b=2$ hybrid reconstruction (dashed line),
presumably because the effect of biasing is not strong enough compared 
to the effects of gravitational evolution at this level of bias.

\begin{figure}
\epsfxsize=\hsize
\epsfbox[18 144 592 738]{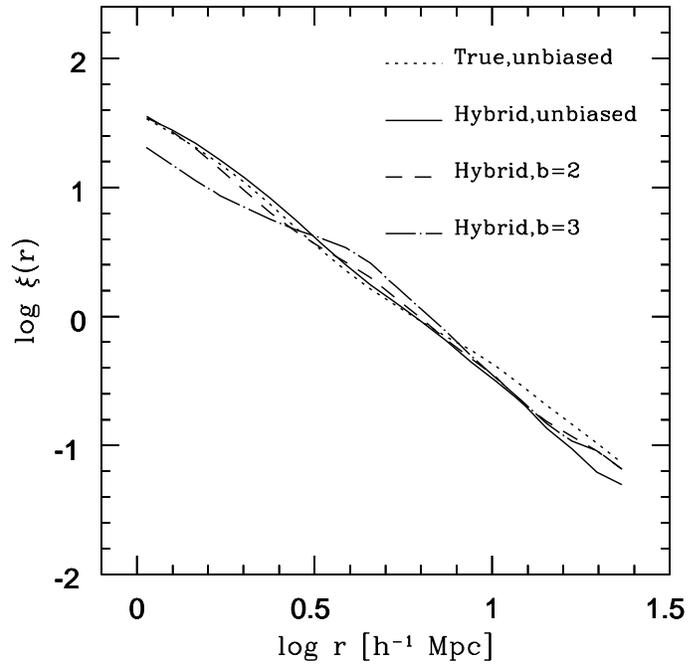}
\caption{ $\xi(r)$ for the final galaxy distributions:
true final galaxy distribution assuming unbiased galaxy formation 
(dotted line),
hybrid reconstruction assuming unbiased galaxy formation (solid line), 
and hybrid reconstructions assuming biased galaxy formation with $b=2$ 
(dashed line) and $b=3$ (dot-dashed line).
}

\end{figure}

In Figure 9, we plot the distribution of distances to the nearest neighbor
of each galaxy $P(x_{n})$ for the true unbiased galaxy \distrbn (dotted line) 
and for the reconstructions with different assumptions regarding bias.
To remove its dependence on the average density of galaxies, this \distrbn
is expressed in terms of $x_{n}$, which is equal to the separation $r_{n}$ 
divided by
the mean inter-galaxy separation $\bar d \equiv n_{g}^{-1/3}$ 
(i.e, $x_{n} = r_{n}/\bar d$).
At separations smaller than the force resolution of our PM code, the exact 
behavior of this distribution cannot be estimated reliably.
Therefore, we show only the mean level of the distribution for $x_{n} < 0.2$, 
corresponding roughly to distances $r_{n} < 1h^{-1}$ Mpc.
We normalize the distributions in Figure 9 so that
\be
\int_{0}^{\infty} P(x_{n})dx_{n} = 1.
\label{eqn:nnbrnorm}
\ee
We see that the \rec that correctly assumes an unbiased \gal \distrbn 
(solid line) reproduces the true $P(x_{n})$ (dotted line) very well.
The biased reconstructions have undergone 
weaker non-linear evolution, and they
therefore have fewer galaxy pairs at close
separations and correspondingly more pairs at $x_{n} \sim 0.4$.
This statistic clearly captures and quantifies the diffuse appearance 
of the $b=2$ reconstruction that is shown in Fig. 4d.
The effects of gravitational evolution are still weaker in the hybrid \rec with 
$b=3$, and the corresponding $P(x_{n})$ (dot-dashed line) is much
flatter than that of the true unbiased \gal distribution.

\begin{figure}
\epsfxsize=\hsize
\epsfbox[18 144 592 738]{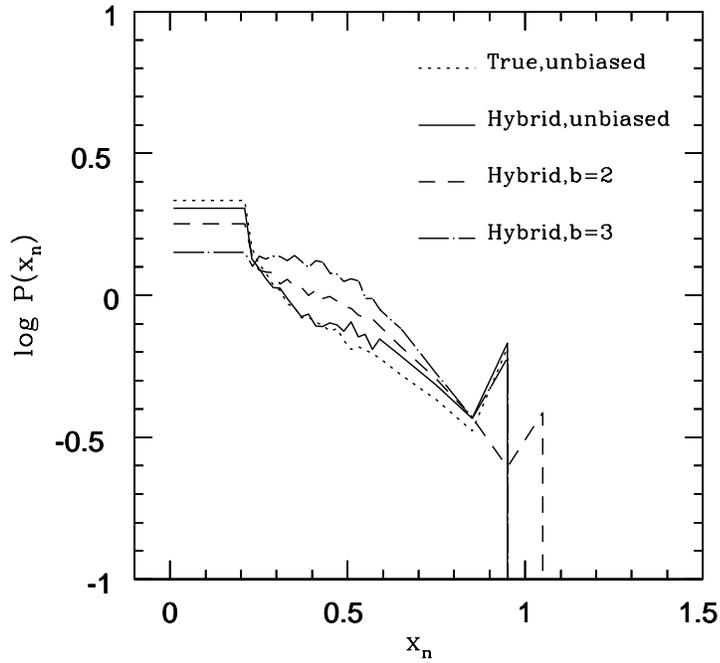}
\caption{Nearest neighbor distribution  for the final galaxy distributions
in terms of $x_{n}$, the separation $r_{n}$ divided by the mean
inter-\gal separation $\bar d $.
Dotted, solid, dashed, and dot-dashed lines show, respectively,
the true, unbiased final galaxy distribution,
the hybrid reconstruction assuming unbiased galaxy formation,
and hybrid reconstructions assuming biased galaxy formation with $b=2$ 
and $b=3$. 
}

\end{figure}

Based on the tests in this section, we can arrive at the 
following two conclusions regarding the performance and the use of the
hybrid \rec method.
\begin{description}
\item[{(1)}] The hybrid \rec method performs significantly better than either 
the dynamical scheme or the \gau method in reconstructing unbiased,
real space \gal distributions.
We also carried out the relevant comparisons while reconstructing 
biased \gal distributions and mock redshift catalogs and 
always found that the hybrid method yields the most accurate reconstruction.
Therefore, in subsequent sections, we will only show the results of the 
hybrid reconstruction.
\item[{(2)}] Biased reconstructions of unbiased models 
produce insufficient small scale clustering for a given level of 
fluctuations in the final \gal distribution ($\sigma_{8g}$).
We are able to detect this failure 
visually and by using statistical measures like the 
\nnbr distribution.
We conclude that \rec analysis can be used 
to test the hypothesis of biased \gal formation.
\end{description}

\subsection{\it Biased Reconstructions}

We now test the ability  of the hybrid scheme to reconstruct biased \gal 
distributions and further test its ability
to detect incorrect assumptions about bias.
We perform all the simulations using the same parameters as in the 
unbiased case except for the amplitude of the initial fluctuations.
We normalize the amplitude of the initial power spectrum so that
$\sigma_{8m} = \sigma_{8g}/b =0.55$ for a bias factor $b=2$.
We evolve this initial density field through a PM code and 
select ``galaxies'' from this evolved mass distribution
using a local power law biasing relation between the mass
density $\rho_{m}$ and the galaxy density $\rho_{g}$ (\cite{mph98}):
\be
\log \left( \frac{\rho_{g}}{\langle \rho_{g} \rangle} \right)
= A + B \log \left( \frac{\rho_{m}}{\langle \rho_{m} \rangle} \right).
\label{eqn:biasdef}
\ee
We choose the constants $A$ and $B$ so that the resulting \gal \distrbn
has the desired average number density $n_{g} = 0.01h^{3}$Mpc$^{-3}$ and 
rms fluctuation amplitude $\sigma_{8g} = 1.1$.
The probability that a mass particle in a region where the mass
density is $\rho_{m}$ is chosen as a \gal is proportional to 
$\rho_{m}^{B-1}$.
We compute the mass density $\rho_{m}$ in a 
sphere of radius $5h^{-1}$Mpc around the particle.
This biasing relation is similar to the one suggested by Cen \& Ostriker 
(1993) based on hydrodynamic simulations incorporating physical models
for galaxy formation (\cite{co92}), but it differs in that there
is no quadratic term that saturates the biasing relation at 
high values of the mass density.

In all the tests of reconstructions of biased \gal distributions, we adopt
as the true \gal distribution (the input data) a fiducial \gal distribution 
with $\sigma_{8g} = 1.1$, biased to $b=2$ using the prescription
defined by equation~(\ref{eqn:biasdef}).
We compare this true \distrbn to
a biased hybrid reconstruction that correctly assumes $b=2$.
We will also show some comparisons to reconstructions that incorrectly assume
unbiased galaxy formation or biased galaxy formation with $b=3$.
When biasing the
evolved mass distributions in step (H6B), we use the same
power-law biasing prescription that we adopted for the true model.

Figure 10 shows the contour plots of the true initial density field and the 
hybrid reconstructed density field assuming (correctly) $b=2$.
The contours are plotted in the same slice as in Figure 1.
Comparing Figure~10 to Figure~1, we see that the recovery of the 
initial conditions is more accurate in the biased model, because the effect
of non-linear gravitational evolution is smaller in the biased case.
Figure 11a shows a scatter plot of the true and reconstructed
initial density contrasts.
Comparison to Figure 2c again shows the
more accurate recovery of initial densities in the biased model,
quantified by the increase in the correlation coefficient from
$r=0.711$ to $r=0.813$.  The more accurate initial conditions yield
a more accurate final galaxy density field, as shown by comparing the
final density scatterplot (Fig.~11b) to the corresponding plot for
the unbiased model (Fig.~5a).

\begin{figure}
\epsfxsize=\hsize
\epsfbox[50 200 565 480]{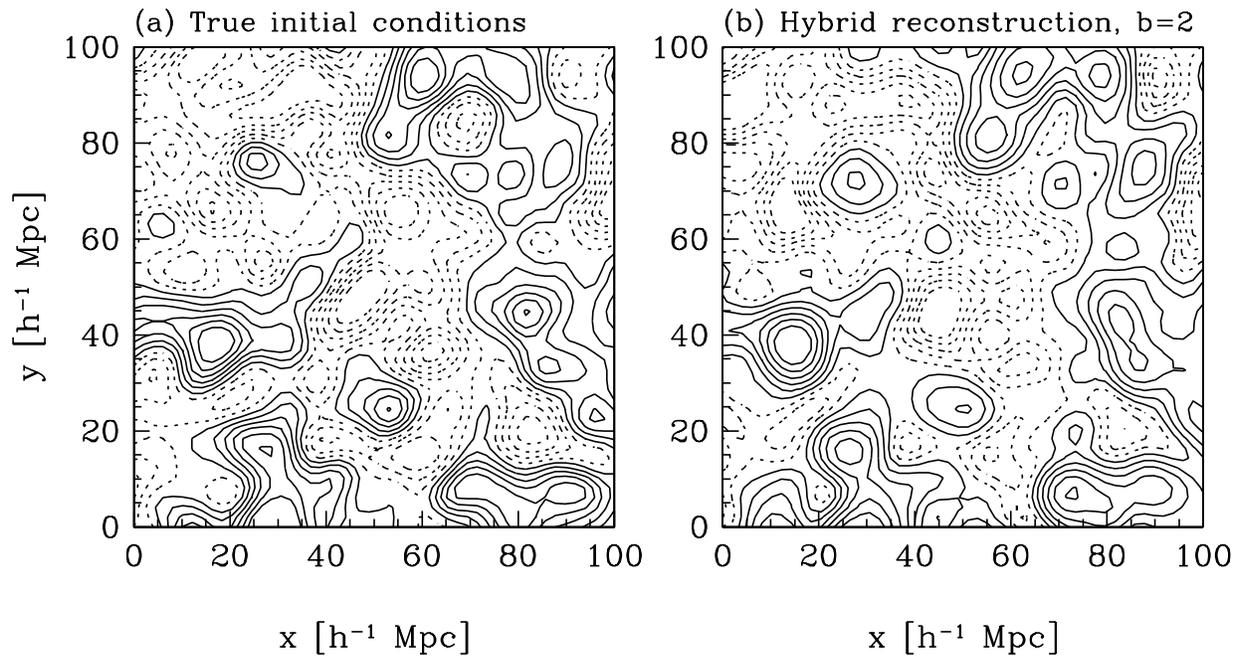}
\caption{Contours in a slice of the initial density field of a test N-body
simulation.
The contour levels range from $-2\sigma$ to $+2\sigma$ in steps of $0.4\sigma$.
Solid contours correspond to overdensities, while dashed contours
correspond to underdensities. 
({\it a}) True initial conditions, Gaussian with a $\Gamma=0.25 $ 
power spectrum.
A slice through the galaxy distribution obtained by gravitationally 
evolving this field and then selecting galaxies with $b=2$ is shown 
in Fig. 12a.
({\it b}) The initial density field reconstructed from this
biased galaxy distribution assuming $b=2$.
}
\end{figure}

\begin{figure}
\epsfxsize=\hsize
\epsfbox[18 300 592 738]{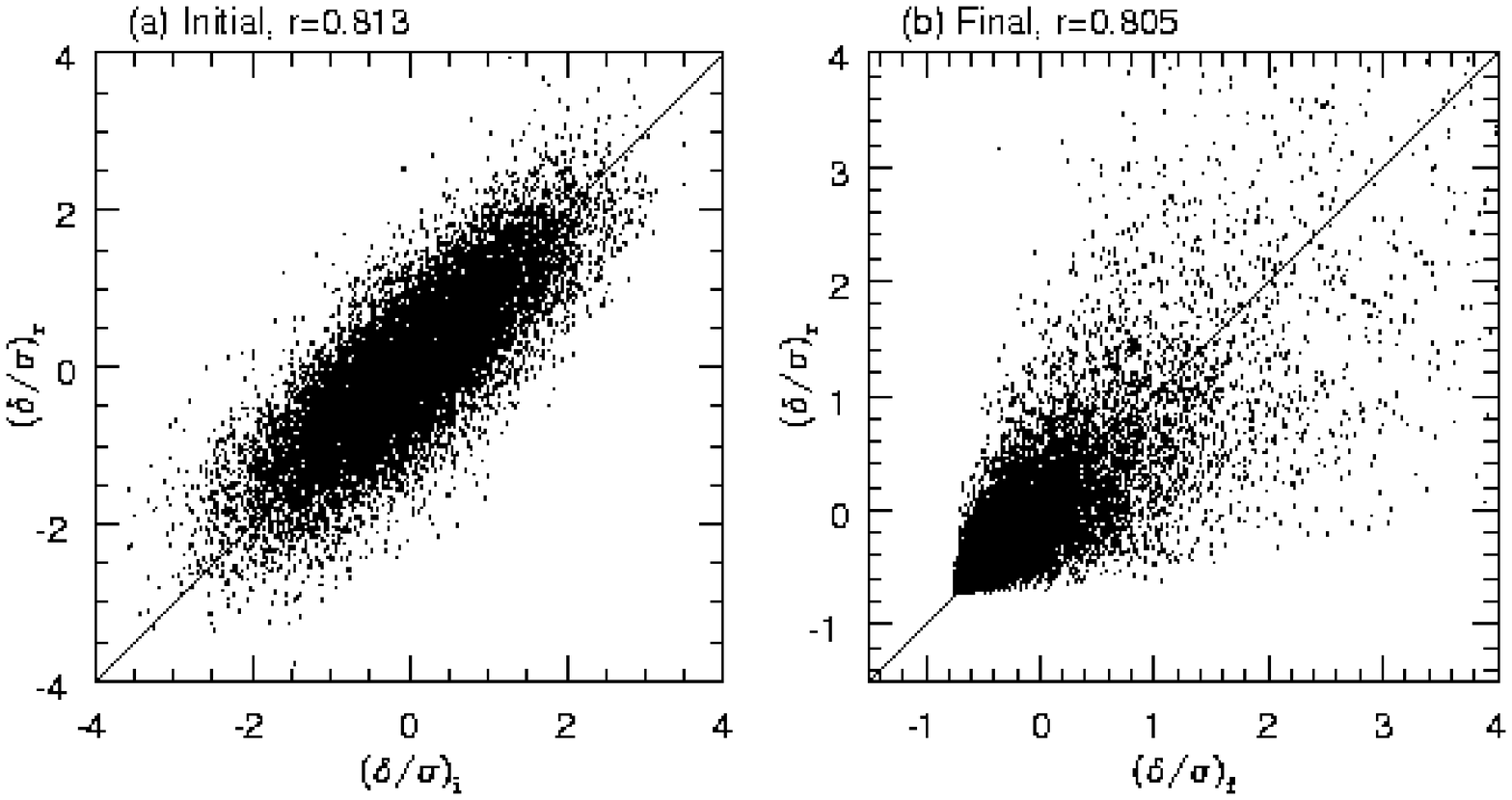}
\caption{ ({\it a}) Cell by cell comparison of the reconstructed 
initial density contrast $(\frac{\delta}{\sigma})_{r}$ and the true 
initial density contrast $(\frac{\delta}{\sigma})_{i}$ for the hybrid \rec
of the biased model.
({\it b}) Comparison of  the true and reconstructed final density contrasts.
All the density fields are smoothed with a Gaussian filter of radius 
$R_{s} = 3h^{-1}$Mpc and scaled by the rms fluctuation $\sigma$. The 
linear correlation coefficient $r$ is indicated above each panel.
}
\end{figure}

Figure 12 shows the power spectrum of the true initial density field 
by a dotted line.
The power spectrum of the density field reconstructed using the steps 
(H1), (H2B), (H3) and (H4) of the hybrid method (i.e., with
no power restoration) is shown by the dashed line.
The solid line shows the power spectrum after the power restoration
and the amplitude matching procedures.
By construction, the amplitude of the power spectrum is normalized so
that $\sigma_{8m} = \sigma_{8g}/b  = 0.55$ for the assumed value of
$b=2$.
The wavenumber beyond which random phase waves are 
added ($k_{\rm corr} = 20k_{f} = 0.628\ h $Mpc$^{-1}$) is  marked 
in the Figure.

\begin{figure}
\epsfxsize=\hsize
\epsfbox[18 144 592 738]{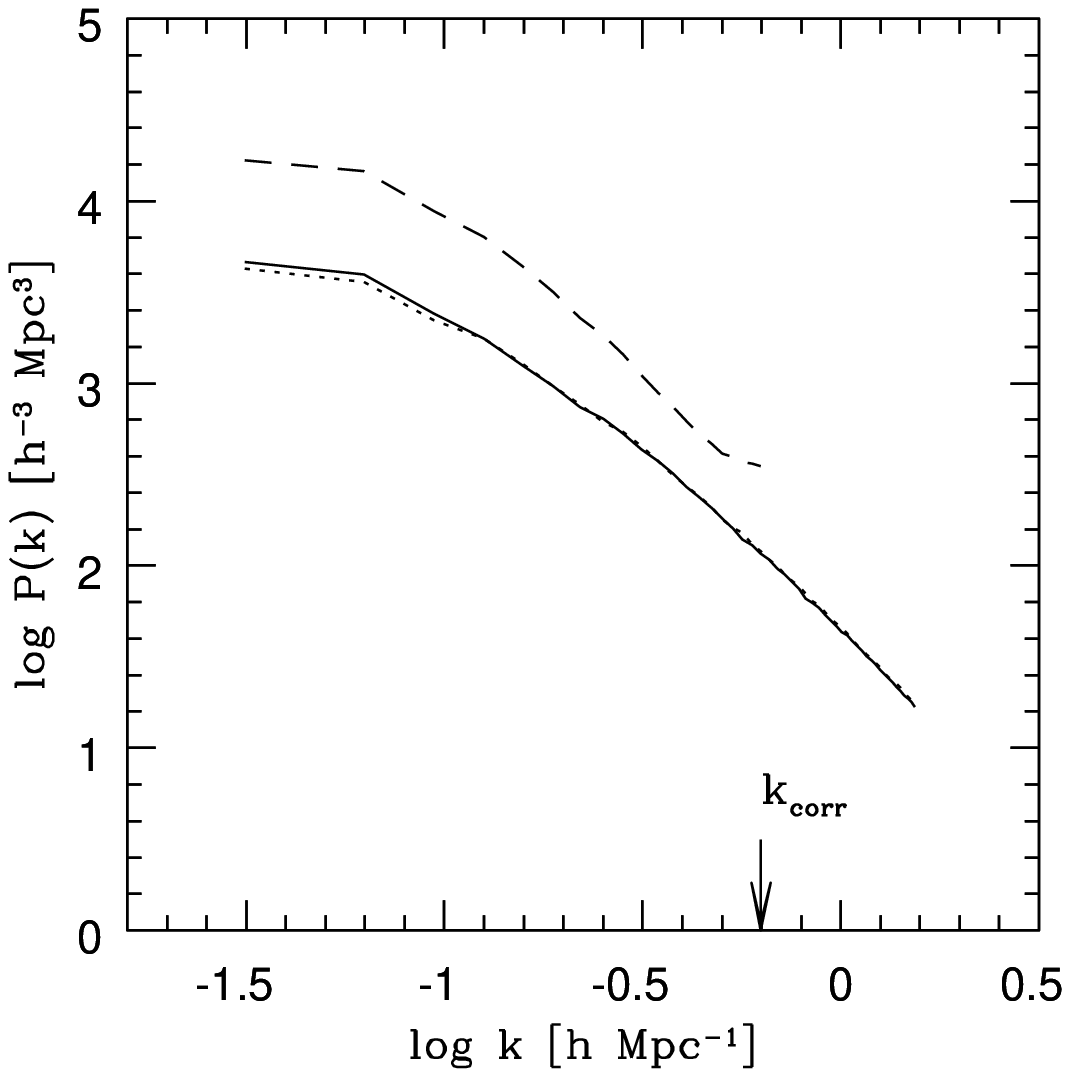}
\caption{ Power spectrum of the true initial density field (dotted line),
the density field reconstructed using steps (H1), (H2B), (H3) and (H4) of the 
hybrid method 
(dashed line), and the hybrid reconstructed density field after the power
restoration and amplitude matching procedures (solid line).
The final galaxy \distrbn is biased with $b=2$.
The dashed line  has been multiplied by the factor $e^{k^{2}R_{s}^{2}}$ 
in the range 
$0 < k \leq k_{\rm corr}$ to restore the power lost in the Gaussian 
smoothing, and its amplitude has been fixed arbitrarily. 
}
\end{figure}

Figure 13a shows the true final galaxy \distrbn when the galaxies are 
biased tracers of the mass \distrbn with a bias factor $b = 2$.
This \gal \distrbn is noticeably more diffuse than the unbiased
galaxy \distrbn shown in Figure 4a, although the rms fluctuation 
amplitude $\sigma_{8g}$ is identical for both distributions.
The galaxy \distrbn reconstructed by the hybrid scheme, assuming 
biased \gal formation with a correct value of $b=2$, is shown in 
Figure 13b.
The individual structures and the overall texture of the galaxy \distrbn 
appear very similar to those of the true distribution.
The statistical properties of this galaxy \distrbn closely match 
those of the true distribution, as shown below.
The reconstructed galaxy \distrbn assuming unbiased \gal formation 
(Fig. 13c) shows clear evidence for excessive dynamical evolution.
Clusters are more prominent and larger structures more clumpy than in
the true \gal distribution.
The \rec assuming $b=3$ (Fig. 13d) does not have 
enough non-linear structure and appears very diffuse.
This diffuse appearance can be easily quantified by the \nnbr 
statistic, as we will show below.

\begin{figure}
\epsfxsize=\hsize
\epsfbox[18 144 592 738]{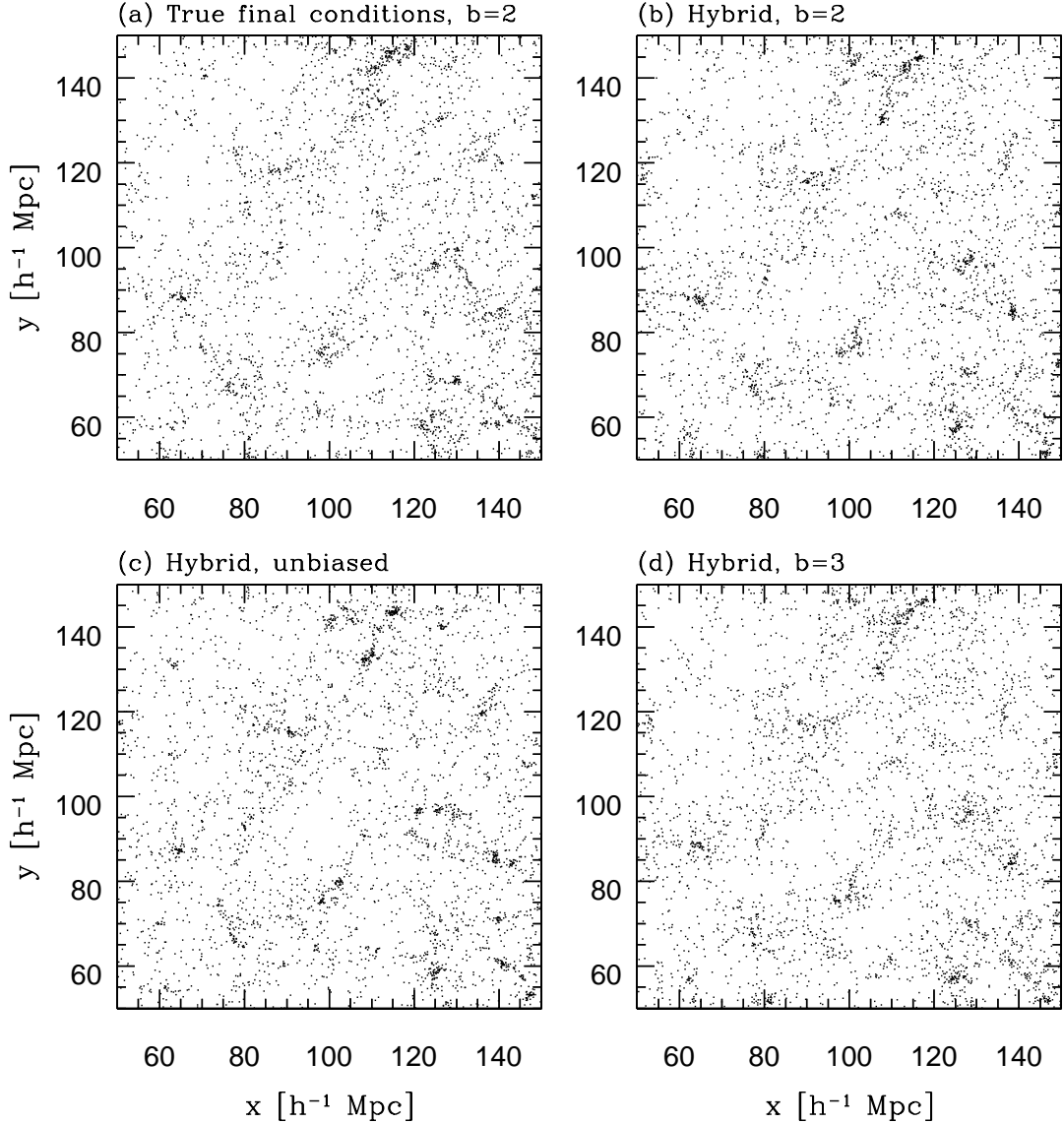}
\caption{ Final galaxy distributions, with $\sigma_{8g}=1.1$
({\it a}) True final galaxy distribution in the model with biased galaxy 
formation with $b=2$.
Remaining panels show hybrid reconstructions assuming 
({\it b}) biased galaxy formation with $b=2$,
({\it c}) unbiased galaxy formation and 
({\it d}) biased galaxy formation with $b=3$.
}

\end{figure}

Figure 14 shows the two-point correlation functions $\xi(r)$ of the 
true \gal distribution and the \gal distributions 
reconstructed with different assumptions about biasing.
The \rec with $b=2$  matches the true $\xi(r)$ closely on 
all scales.
Unbiased \rec leads to excessive clustering on small scales, resulting 
in a correlation function that is steeper than that of the input data.
The final \gal \distrbn in the $b=3$ \rec 
is less dynamically evolved and has a shallow $\xi(r)$ on small scales.

\begin{figure}
\epsfxsize=\hsize
\epsfbox[18 144 592 738]{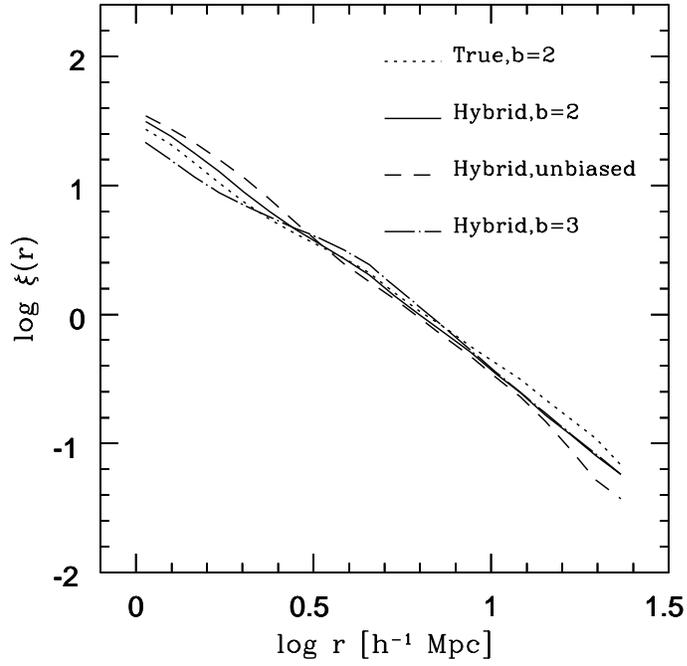}
\caption{ Correlation functions $\xi(r)$ for the true final galaxy distribution
of the biased model (dotted line) and for the hybrid reconstructions 
assuming biased galaxy formation with $b=2$ (solid line),
biased galaxy formation with $b=3$ (dot-dashed line), and unbiased galaxy 
formation (dashed line).
}

\end{figure}

The dotted line in Figure 15 shows the \nnbr \distrbn of the true \gal
distribution. 
The solid line that closely matches this dotted line corresponds to 
the hybrid \rec with the correct assumption for the bias factor $b=2$.
The excessive small scale clustering in the unbiased reconstruction
produces a steeper \distrbn (dashed line),
while the $b=3$ \rec has a flatter \nnbr \distrbn 
(dot-dashed line) that reflects its smaller degree of non-linear
evolution.
This statistic quantifies well the appearance of the \gal distributions 
in Figure 13, and it can therefore serve as a discriminatory statistic to 
distinguish between different assumptions about bias.

\begin{figure}
\epsfxsize=\hsize
\epsfbox[18 144 592 738]{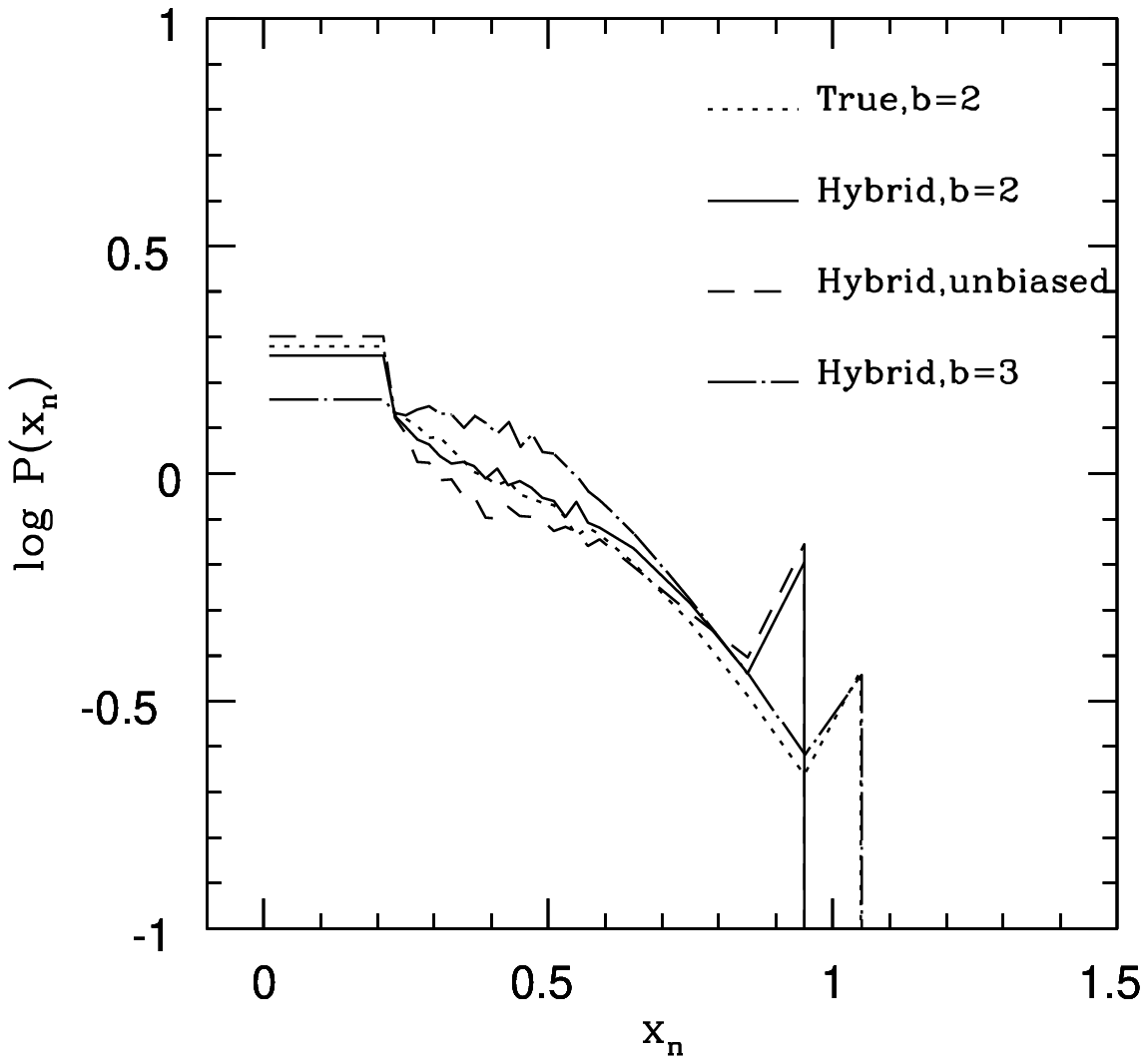}
\caption{ Nearest neighbor distribution for the final galaxy distributions,
with the same coding as in Fig. 14.
}

\end{figure}

The tests in this section show that the hybrid \rec 
scheme can  be applied successfully to biased \gal distributions.
Once again, we get the best recovery of the initial density fields and 
the final 
\gal distributions if we make the correct assumptions about the bias between
the final mass and \gal distributions.
Incorrect assumptions lead to \gal distributions that are incompatible with 
the input data, and this incompatibility can be quantified by the 
\nnbr \distrbn and the two-point correlation function,
though the latter is only marginally effective in distinguishing 
among reconstructions with modest differences in the bias factor.
We also find that, for a given level of $\sigma_{8g}$, the effects of bias are
more easily reversed than the effects of non-linear gravitational evolution.

\section{TESTS ON ARTIFICIAL REDSHIFT SURVEY CATALOGS}

The primary requirements for a redshift survey to be suitable
for \rec analysis are
good sky coverage and depth so that the gravitational 
influence of regions outside the survey boundaries is small,
dense sampling to reduce shot noise errors, and
a well understood selection function.
Of existing \z surveys, the IRAS-selected, Point Source Catalog Redshift Survey 
(PSCZ, see \cite{pscz1} and \cite{canavezes98}) best satisfies the above 
requirements.  However, IRAS and optically selected  galaxies are known to 
cluster differently (e.g., \cite{lahav90}; \cite{saunders92}; \cite{fisher94}),
so it is also desirable to analyze an optically selected galaxy \distrbn
using the \rec procedure, partly in order to understand the origin of 
this clustering difference.
Of course, the optical and the IRAS galaxies in a given region are both 
related to the {\it same} underlying {\it mass} distribution.
The Optical Redshift Survey (ORS, Santiago et al.\ 1995, 1996) is probably 
the best existing optical survey for \rec analysis because of its nearly full 
sky coverage, even though there are other surveys that contain more galaxies.
We hope to analyze both the PSCZ and the ORS using the hybrid
\rec procedure in the near future.
Here, we analyze artificial redshift catalogs that are designed to mimic 
these surveys, in order to test the ability of the \rec method to handle
redshift space input data with non-periodic survey boundaries and to
see what we can expect to learn from the \rec analysis of these catalogs.

We construct the mock redshift catalogs from the output of a PM simulation 
of an $\Omega = 0.4$ universe, assuming Gaussian initial 
fluctuations with a 
$\Gamma=0.25$ power spectrum.
This simulation evolves $100^{3}$ particles in a periodic cube of side 
$200h^{-1}$Mpc and uses a $200^{3}$ mesh to compute the gravitational forces.
We assume that the galaxies in the mock PSCZ catalog form in an unbiased manner
with $\sigma_{8g} = \sigma_{8m} = 0.75$, while the ORS galaxies are 
biased tracers of the same mass \distrbn with $\sigma_{8g} = 1.1$.
We reconstruct the \gal distributions  of these two mock catalogs using the 
hybrid \rec scheme.
In the power restoration step, we correct the power spectrum using 
empirical correction factors for 
$k_{f} \leq k \leq k_{\rm corr} = 15k_{f} = 0.471\ h$Mpc$^{-1}$,
and we add random phase waves for higher wavenumbers in the manner 
described in \S2.
We normalize the power spectrum by requiring that the final $\sigma_{8g}$ of
the reconstructed galaxy distribution match that of the mock catalog
in {\it redshift} space.
While in the previous section we showed how the degree of clustering on
small and large scales can be used to constrain the bias factor, here
we will focus mainly on our ability to constrain $\Omega$, given
the correct assumptions about the bias factor.
Therefore, we will reconstruct the two mock redshift catalogs assuming
both $\Omega = 0.4$ (the correct value) and $\Omega = 1$.
Any  systematic failure of the $\Omega = 1$ \rec to reproduce the 
input data will tell us about the discriminatory power of the \rec method.
We do, however, expect some tradeoff between $\Omega$ and $b$, 
if both parameters are allowed to vary simultaneously.

We select a Local Group observer from the final particle distribution
so that the velocity dispersion in a sphere of radius $5h^{-1}$Mpc around 
that observer is less than $250$ km s$^{-1}$, in accord with 
observations that imply
a cold velocity field near the Local Group (\cite{sandage86};
\cite{brown87}).
We assign each galaxy a redshift based  on its  real space distance and its
 radial peculiar velocity with respect to this Local Group particle.
We use the same Local Group observer for both the mock catalogs so that
the underlying mass distribution is identical in the regions where the 
two surveys overlap.
To create the mock redshift catalogs, we first select volume limited 
subsamples of the galaxy \distrbn extending to an inner radius $r_{\rm in}$. 
We supplement this volume limited sample with an extended magnitude 
limited sample out to a larger radius $r_{\rm out}$, so as to improve the 
\rec near the boundaries of the inner sample.
We reject all the galaxies in an angular mask about the observer 
to account for the  incompleteness of the surveys in the regions 
corresponding to the Galactic zone of avoidance (ZOA).

We form the final galaxy density fields by CIC binning the galaxies
in the mock redshift catalogs onto a 
$100^{3}$ cubical grid that represents a region $200h^{-1}$Mpc a side.
In the region $r < r_{\rm in}$, we assign equal weights to all the 
galaxies, as the catalog is volume limited up to that radius.
In the region $r_{\rm in} < r < r_{\rm out}$, we weight each galaxy by the 
inverse of the value of the selection function $\phi(r)$ at its location.
In the regions outside the survey boundaries, we set the density field
to be equal to its mean value inside the survey region.
We account for boundary effects in computing the smoothed density field
$\rho_{sm}({\bf r})$ by using the ratio method of Melott \& Dominik (1993), 
\be
\rho_{sm}({\bf r}) = \frac{\int M({\bf r'})\rho({\bf r'})W({\bf r-r'})
d^{3}{\bf r'}}{\int M({\bf r'})W({\bf r - r'})d^{3}{\bf r'}} ,
\label{eqn:smmask}
\ee
where $W({\bf r})$ is the smoothing filter and the mask array 
$M({\bf r})$ is set to 1 for pixels inside the survey region and
to 0 for pixels outside the survey region.

\subsection{\it Correction for Redshift Space Distortions}

Redshifts of galaxies reflect the combination of Hubble flow
at their real space locations and the radial component of the 
peculiar velocities acquired during gravitational evolution.
This peculiar velocity component distorts the mapping
of galaxy positions from real to redshift space, making the line of
sight a preferred direction in an otherwise isotropic universe.
However, we need the mass density field in real space in order to recover the 
initial mass density fields using the hybrid \rec method.
Therefore, we need to correct for these peculiar velocity induced 
distortions.
The effects of these distortions on the redshift space density field
are \diff  on different scales.

On small scales, the velocity dispersion associated with a cluster
stretches it along the line of sight into a ``Finger of God'' 
feature that points directly toward the observer.
This feature spreads a compact cluster in real space over a large radial 
distance in redshift space and thus reduces the amplitude of small scale 
clustering.
To correct for this effect, we first identify the clusters in redshift 
space using a friends-of-friends algorithm that employs different 
linking lengths in the radial and transverse directions 
(\cite{huchra82}; \cite{nw87}; \cite{mfw93}).
Here we use a transverse linking length of $0.6h^{-1}$Mpc and a radial
linking length of $500$ kms$^{-1}$ (\cite{grcg94}).
For each cluster, we shift the radial
locations of the member galaxies so that the 
resulting compressed cluster has a radial velocity dispersion of
$100$ kms$^{-1}$, roughly the value expected from Hubble flow across
its physical extent.

The distortions on large scales arise from coherent 
inflows into overdense regions and outflows from underdense regions
(\cite{sargent77}; \cite{kaiser87}).  
These bulk flows are generated by large scale density fluctuations
that can be reasonably assumed to be still in the quasi-linear
regime of gravitational evolution.
To remove these large scale distortions and estimate the
real space mass density field, we apply a modified version
of the iterative procedure suggested by Yahil et al. (1991) and 
Gramann, Cen \& Gott (1994) to the
cluster-compressed, redshift space galaxy distribution:
\begin{description}
\item[{(R1)}:] For biased galaxy density fields, we first apply a monotonic
local map to the redshift space {\it galaxy} density field that enforces a
 numerically  determined PDF of the real space 
{\it mass} density field  corresponding to the assumed value of the bias 
factor $b$.
This mapping provides our zero-th order estimate of the real space mass density
field, correcting for the effects of bias and peculiar velocity 
distortions on the PDF.
We could apply a similar mapping even for the unbiased case, in the hope
of having a more accurate starting point for peculiar velocity corrections.
In practice, however, we find that this mapping does not significantly 
improve the convergence of the iterative procedure, so we ignore it in the 
 unbiased reconstruction.
\item[{(R2)}:] We predict the velocity field from this {\it mass} density field
using Gramann's (1993b) second-order perturbation theory relation,
\be
{\bf v(r)} = f(\Omega)H\left[{\bf g(r)}+\frac{4}{7}\nabla C_{g}({\bf r})\right],
\label{eqn:vdelta2}
\ee
where ${\bf g(r)}$ is the gravitational acceleration field computed from 
the equation
$\nabla \cdot {\bf g(r)} = -\delta({\bf r})$ and $C_{g}$ is defined by 
equation~(\ref{eqn:cgdef}).
This step requires that we assume a value of $\Omega$ to 
compute the factor $f(\Omega)$.
\item[{(R3)}:] We use this velocity field information to correct the positions 
of galaxies so that their new positions are consistent with their
Hubble flow velocities and the peculiar velocities at their locations.
\end{description}
We iterate these three steps until the corrections to the
galaxy locations in step (R3) become negligible and the galaxy density field 
has converged.
In practice, we find that the positional corrections become very small in 
about three steps.
We use the mass density field derived from the inferred real space 
\gal \distrbn as the input to the hybrid \rec scheme.
In the last step of the reconstruction, after selecting galaxies from the 
evolved 
N-body mass \distrbn in an unbiased or biased manner, we project these galaxies 
into redshift space, so that we can compare the reconstructed and the 
true input \gal distributions directly in \z space.

\subsection {\it Reconstruction of a Mock PSCZ Catalog}
The PSCZ survey contains all galaxies in the IRAS Point Source 
Catalog whose $60 \mu m$ flux is greater than $0.6$Jy,
excluding the regions that are heavily contaminated by Galactic 
sources (mainly the low Galactic latitude zone $\vert b \vert < 5^{\circ}$).
The catalog contains about $15,500$ galaxies and covers about $83\%$ 
of the sky.
We create a mock catalog of this survey by selecting a volume limited sample 
from an unbiased galaxy distribution extending to 
$r_{\rm in}=55h^{-1}$Mpc 
at an average density of $0.01h^{3}$Mpc$^{-3}$.
We also include a magnitude limited sample 
to $r_{\rm out}=75h^{-1}$Mpc, 
with the selection function decreasing as $\phi(r) \propto r^{-4}$
in the region $r_{\rm in} < r \leq r_{\rm out}$.
We exclude all galaxies in a 10$^\circ$ wedge to mimic the survey's
Galactic plane cut.
We reconstruct the mock catalog assuming that the \gal \distrbn is 
unbiased with respect to the mass distribution.

Figure 16 shows isodensity  contours of the 
true and reconstructed initial density fields in a slice
through the center of the mock PSCZ survey.
The hybrid scheme recovers the true initial density field quite
accurately in the inner regions, although near the boundaries the density
field recovery is poor.
The clumping of contours at the edges is an artifact of the graphing
routine;
the true and reconstructed density fields are actually continuous 
across the boundaries.

\begin{figure}
\epsfxsize=\hsize
\epsfbox[18 300 592 738]{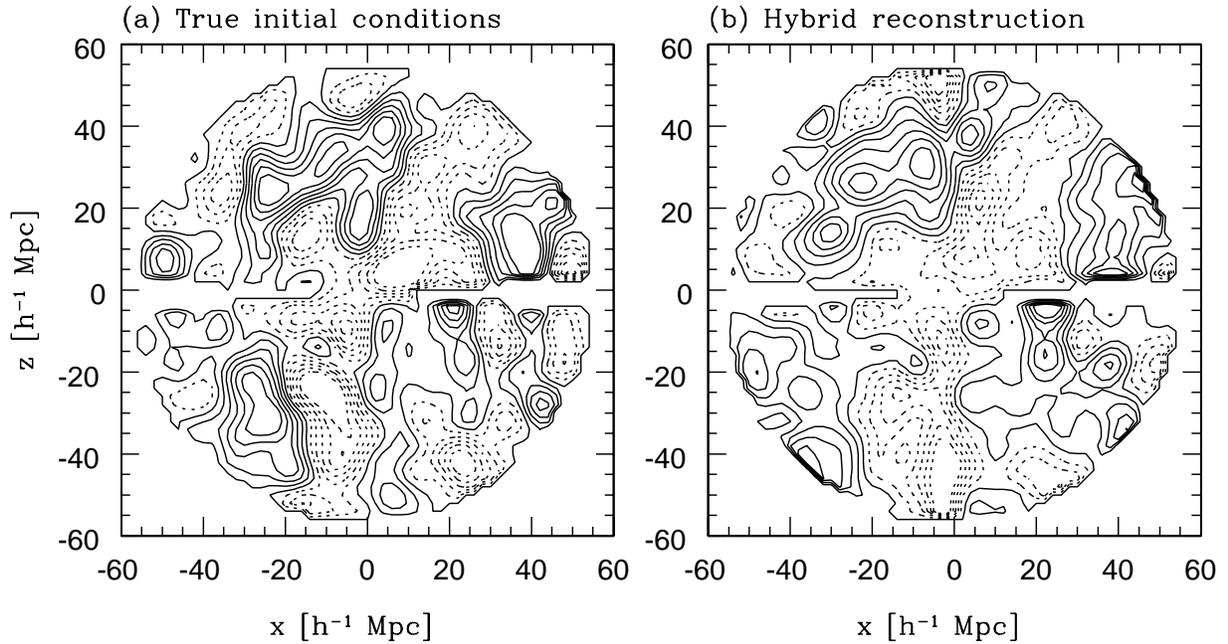}
\caption{ Contours in a slice of the initial density field for the 
mock PSCZ survey. 
The contour levels range from  $-2\sigma$ to $+2\sigma$ in steps of 
$0.4\sigma$. 
Solid contours correspond to overdensities, while dashed contours
correspond to underdensities.
({\it a}) True initial conditions from a $\Gamma=0.25 $ power spectrum.
A slice through the galaxy distribution evolved from this field
and selected using the PSCZ survey geometry appears in Fig. 18a.
({\it b}) The initial density field recovered by the hybrid \rec method.
The 10$^\circ$ Galactic plane cut can be seen near $z=0$.
}

\end{figure}

Figure 17a shows a scatter plot of the true and reconstructed 
initial density fields.
The  scatter is greater than that for the corresponding unbiased full 
cube \rec (Fig.~2c), even though the final \gal density field is 
less non-linear here ($\sigma_{8g} = 0.75$ for the mock PSCZ catalog, as 
opposed to 1.1 in the full cube simulations).
This larger scatter probably reflects the gravitational influence of
regions beyond the survey boundaries that cannot be accounted for due to
the finite volume of the survey.
Nevertheless, we see that it is possible to recover the initial density fields
quite accurately from a realistic galaxy catalog.
The comparison of the final density fields in Figure 17b shows that the hybrid 
scheme reproduces the true \gal density field without any major systematic
errors.

\begin{figure}
\epsfxsize=\hsize
\epsfbox[18 300 592 738]{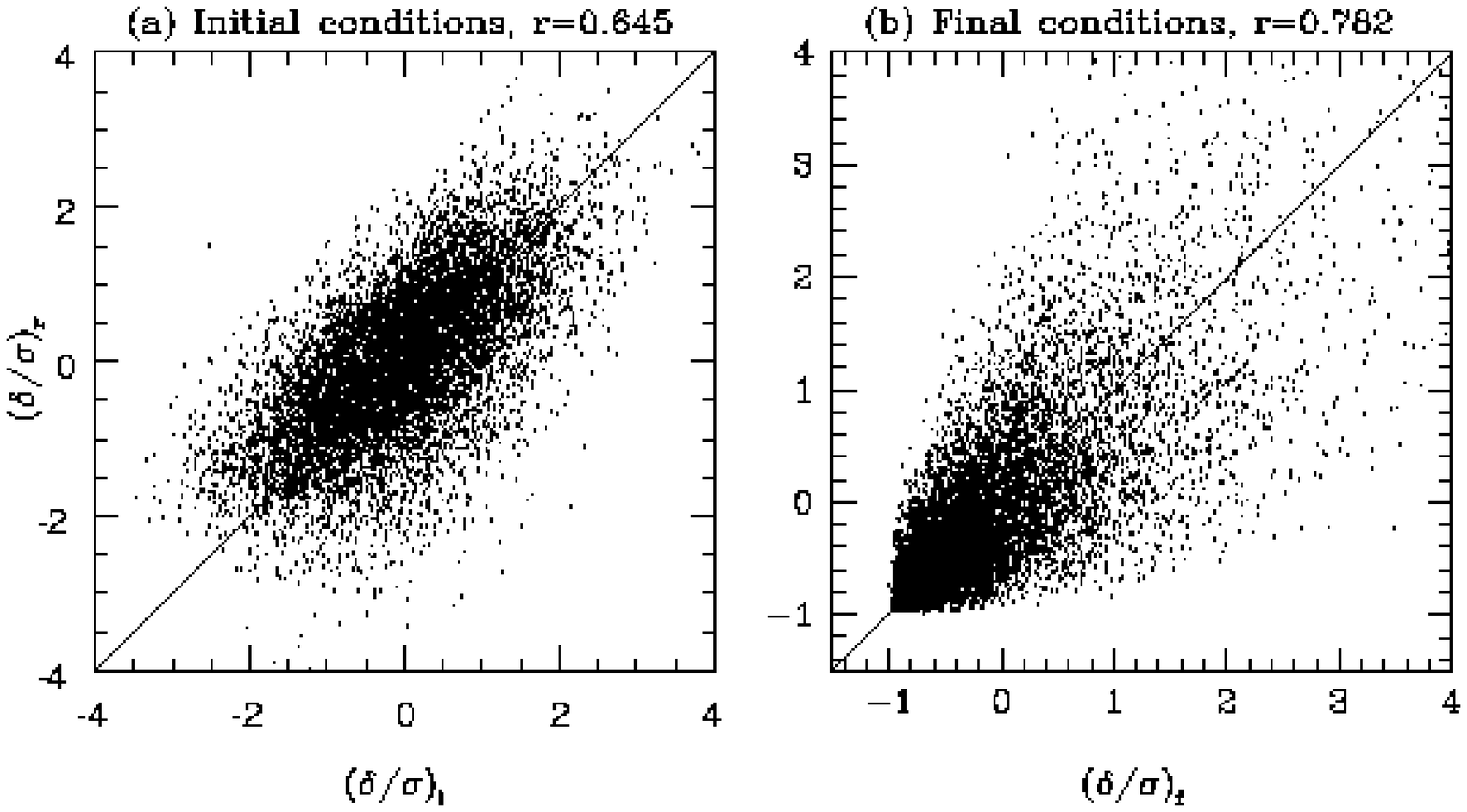}
\caption{({\it a}) Cell by cell comparison of the hybrid reconstructed 
initial density contrast $(\delta/\sigma)_{r}$ to the true initial density 
contrast $(\delta/\sigma)_{i}$ for the mock PSCZ catalog.
({\it b}) Comparison of the reconstructed final density contrast 
$(\delta/\sigma)_{r}$ 
to the true final density contrast $(\delta/\sigma)_{f}$, in redshift space.
All the density fields are smoothed with a Gaussian filter of radius 
3$h^{-1}$Mpc
and scaled by the rms fluctuation $\sigma$. The linear correlation 
coefficient $r$ is indicated above each panel.
}

\end{figure}

Figure 18 shows the power spectra of the true initial density field 
(dotted line) and the hybrid reconstructed density fields
 after the power restoration and amplitude normalization procedures.
The solid and the dashed lines show the reconstructed power spectrum  assuming 
$\Omega = 0.4$ (the correct value) and $\Omega = 1$ respectively.
The slight amplitude mismatch arises from the residual errors
present in the recovered initial density field.

\begin{figure}
\epsfxsize=\hsize
\epsfbox[18 144 592 738]{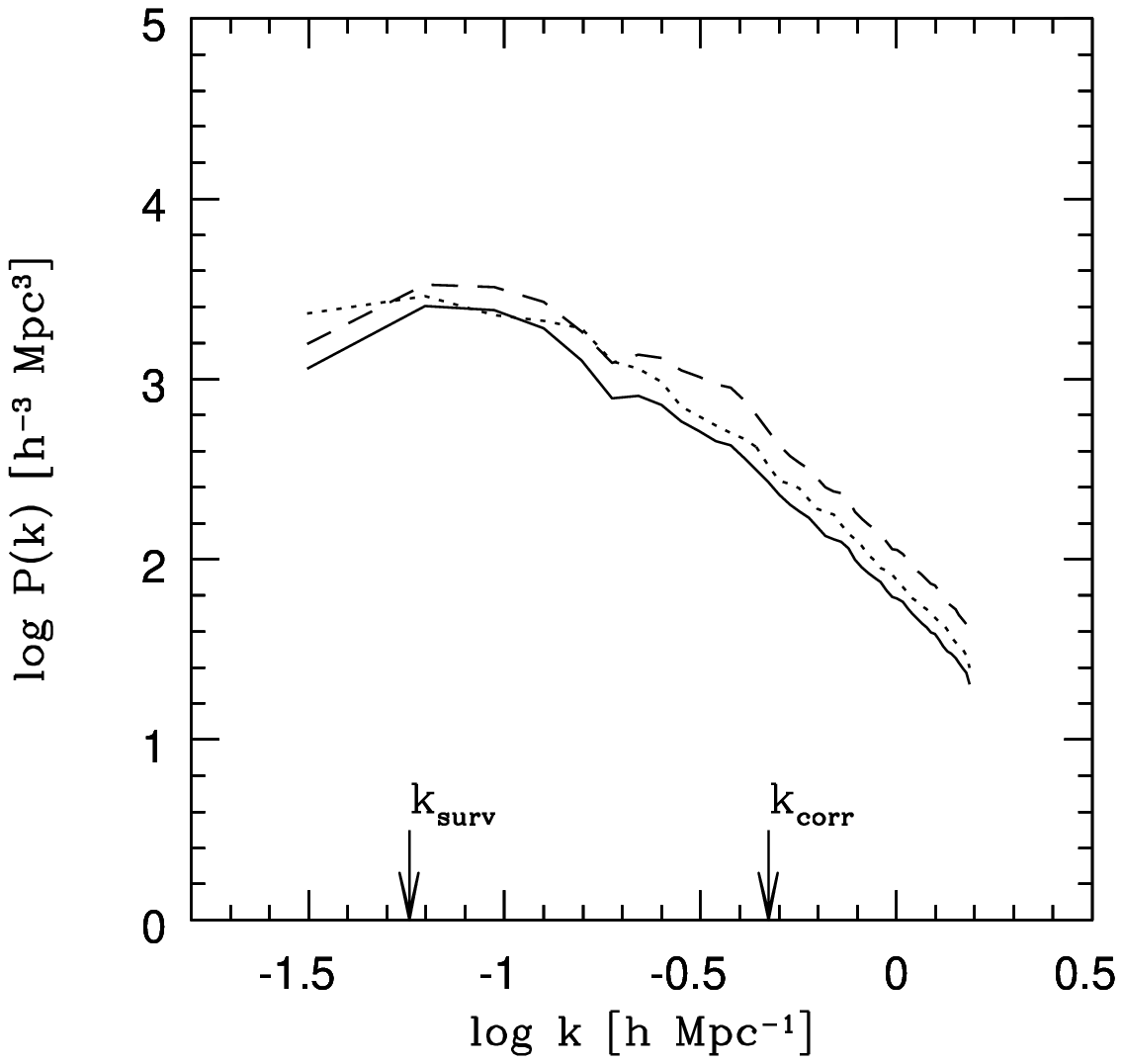}
\caption{ Power spectrum of the true initial density field of the
mock PSCZ survey (dotted line) and the normalized, hybrid reconstructed 
density 
field recovered for $\Omega = 0.4$ (the correct value, solid line) and for
$\Omega = 1$ (incorrect value, dashed line).
The arrows show the wavenumber that corresponds to the survey
size ($k_{\rm surv} = 2\pi/2r_{\rm in} = 0.0571\ h$Mpc$^{-1}$) and the 
wavenumber $k_{\rm corr} = 15k_{f} = 0.471\ h$Mpc$^{-1}$ beyond which 
random phase waves are added to the reconstructed field.
}

\end{figure}

Figure 19 shows the true and the reconstructed 
\gal distributions of the mock PSCZ survey in real space (top panels)
and redshift space (bottom panels).
All the galaxies in a $40h^{-1}$Mpc thick slice centered on the Local 
Group are shown.
Comparing panels (a) and (b), we see that the prominent clusters
are reproduced at the appropriate locations.
However, a notable failure is the absence in the reconstructed \gal \distrbn 
of the filamentary structure that runs from $(x,z) = (-5,20)h^{-1}$Mpc to 
$(5,40)h^{-1}$Mpc in the mock PSCZ survey.
This structure was not present in the adjacent slices either.
We did, however, find an extra cluster at that location in the slice that 
lies above the one shown in the Figure.
We found that this filamentary structure is actually comprised of clusters
that appear close together in projection.
One of the clusters that is closest to the top edge of the slice has moved 
to an adjacent slice during reconstruction, thereby destroying the 
apparent ``filament''.

\begin{figure}
\epsfxsize=\hsize
\epsfbox[18 144 592 738]{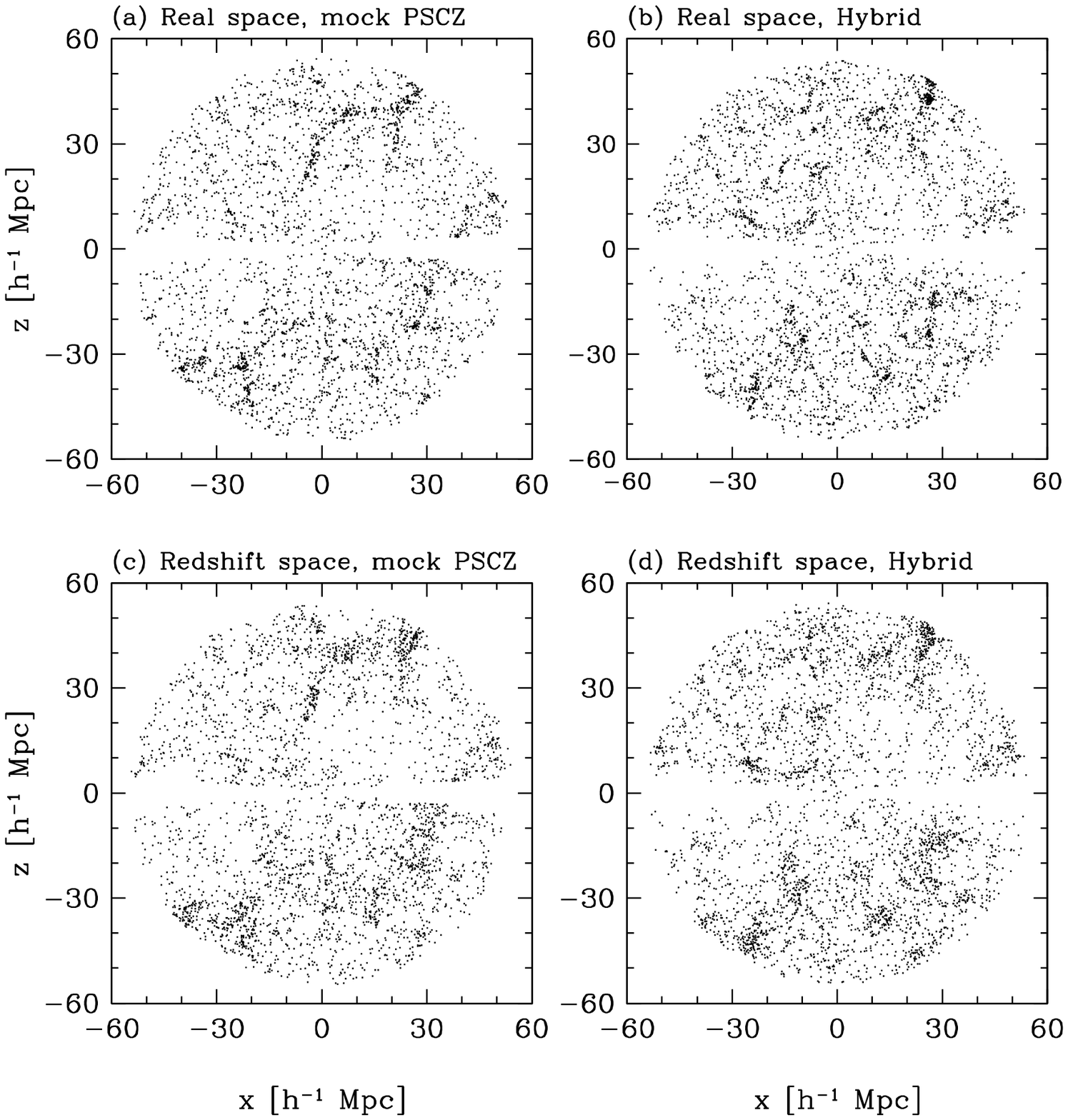}
\caption{ Final galaxy distributions for the mock PSCZ catalog \rec
in the region $-20 < y < 20 h^{-1} $ Mpc.
({\it a}) True final galaxy distribution in real space.
({\it b}) Final galaxy distribution of the hybrid reconstruction in real space.
({\it c}) True final galaxy distribution in redshift space.
({\it d}) Final galaxy distribution of the hybrid reconstruction in 
redshift space.
}

\end{figure}

Figures 20a and 20b compare the cluster multiplicities and 
cluster velocity dispersions between the true and the reconstructed mock PSCZ 
catalogs.
We  identify the clusters in the redshift space \gal distributions using the 
friends-of-friends algorithm described in \S4.1.
We match the clusters in the true and reconstructed redshift \gal
distributions using the algorithm explained in \S3.1.
The open symbols show the cluster comparison for a \rec  assuming
$\Omega = 0.4$, while the filled symbols show the comparison
for a \rec assuming $\Omega = 1$.
The squares parallel to either axis represent clusters present in that 
\gal \distrbn alone.

A larger number of clusters are matched in
the \rec that assumes the correct value of $\Omega = 0.4$.
The $\Omega=1$ reconstruction leaves several of the most massive
clusters unmatched.
Furthermore, the velocity dispersions of clusters in the $\Omega =1 $ \rec are 
systematically higher than those in the true input \gal distribution.
This behavior is expected because the amplitude of fluctuations $\sigma_{8}$ is
matched to that of the input \gal distribution in both cases, and
the average mass density is higher in an $\Omega =1 $ universe.
The clusters in an $\Omega=1$ reconstruction are therefore more massive and 
have a higher velocity dispersion.
This comparison
shows that the velocity dispersion of clusters in the reconstructed
\gal \distrbn can be used to constrain the value of $\Omega$, although the
constraint will be weakened if the bias factor is not known {\it a priori}.

\begin{figure}
\epsfxsize=\hsize
\epsfbox[18 300 592 738]{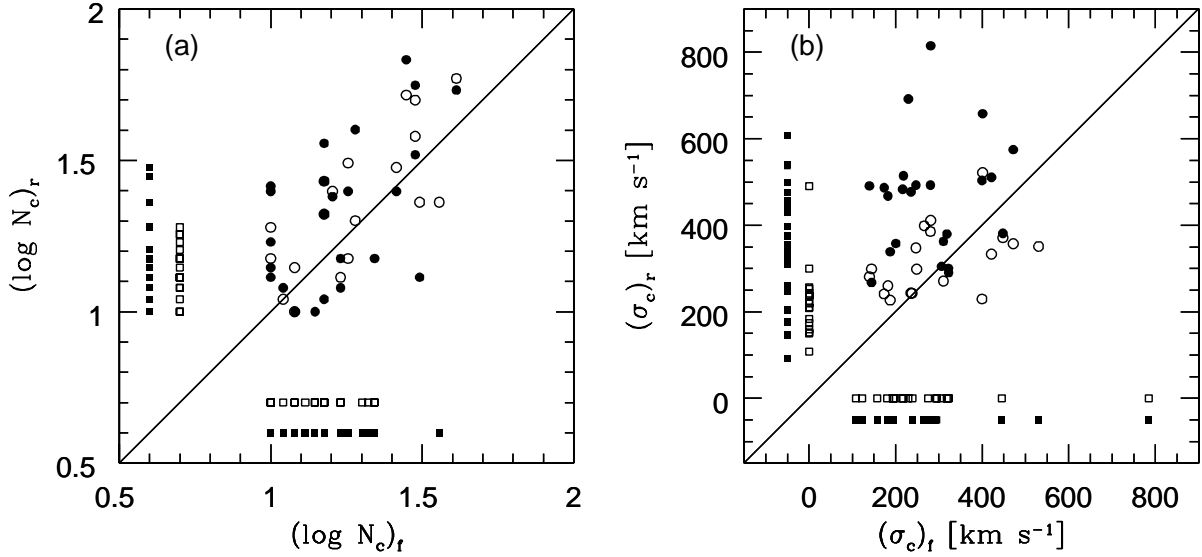}
\caption{ Comparison of ({\it a}) cluster multiplicities and ({\it b}) 
cluster velocity dispersions for the mock PSCZ catalog reconstructions.
Values from the \rec are plotted on the $y$-axis against those
from the true mock catalog \gal distribution.
Squares parallel to either axis represent clusters present in that 
galaxy distribution alone. 
Open symbols show the reconstruction with $\Omega = 0.4$ (correct value), 
and filled symbols show a \rec with $\Omega = 1$. 
}

\end{figure}

The peculiar velocities of galaxies affect the redshift space clustering 
on both small and large scales, as discussed in \S4.1.
The distortion caused by the velocity
dispersions of collapsed clusters can be used to constrain $\Omega$ and $b$ as
discussed above.
We can also use the large scale, coherent flow distortions 
to constrain $\Omega$ and $b$.
When the density fluctuations are small, the induced anisotropy
of the redshift space correlation function $\xi(s,\mu)$
can be derived from linear perturbation theory 
(\cite{kaiser87}; \cite{le89}; \cite{ajsh93a}).
This linear theory anisotropy depends solely on the 
parameter combination $\beta = \Omega^{0.6}/b$, which
can therefore be inferred from the measured $\xi(s,\mu)$.
A similar analysis can be performed using the redshift space power spectrum 
(\cite{cfw94}).
However, these results are valid only when the density fluctuations are 
strictly in the linear regime,
and this condition is generally violated on the scales
accessible to existing redshift surveys.
Attempts to estimate $\beta$ from redshift space distortions
often assume a simple model for a position-independent,
non-linear velocity component 
superposed on the linear flow (e.g., \cite{fisher94}; \cite{cfw95}).
The derived value of $\beta$ is only as good as the velocity model.
Reconstruction, on the other hand, predicts the fully non-linear velocity field
at the location of every galaxy.
Thus, we can constrain the values of $\Omega$ and $b$ more accurately by
demanding that a \rec reproduce the full angular anisotropy of the \z 
space correlation function.

Figure 21 shows the correlation functions $\xi(s,\mu)$ for the mock PSCZ
catalog and its $\Omega=1$ and $\Omega=0.4$ reconstructions,
in five different angular bins.
We compute the correlation functions using the estimator of Hamilton (1993b),
\be
\xi(s,\mu) = \frac{N_{DD}N_{RR}}{N_{DR}^{2}} - 1,
\label{eqn:xismu}
\ee
where $N_{DD}, N_{DR}$ and $N_{RR}$ are the number of galaxy-galaxy,
galaxy-random, and random-random pairs with a separation $s$ at an angle 
 $\theta = \cos^{-1}(\mu)$ to the line of sight in \z space.
We use a random catalog that has the same geometry and selection function
as the true galaxy distribution and contains about 50,000 points
distributed randomly within the survey volume.
We consider only those galaxy pairs that subtend an angle smaller than 
$\alpha_{max} = 60^{\circ}$ at the observer so that the lines of sight 
to both the galaxies in the pair are approximately parallel.
The filled circles show the real space correlation function $\xi(r)$.
Since the real space correlation function is isotropic, we  compute it using
all the galaxy pairs in the sample that are separated by a distance $r$.
We compress clusters before measuring
$\xi(s,\mu)$ so that the Finger-of-God suppression 
is minimized and it is easier to detect the large scale amplification,
which reaches its maximum value for separations along the 
line of sight ($\theta = 0, \mu=1$).
The enhancement is clearly seen in the panel corresponding to $\mu = 0.9$,
where the redshift space correlation functions lie above the 
real space correlation function in the range of
separations $ r < 10h^{-1}$Mpc. 
As expected, this large scale enhancement, which depends on
$\beta$, is larger for the $\Omega = 1$ \rec than for the $\Omega = 0.4$ 
reconstruction.

\begin{figure}
\epsfxsize=\hsize
\epsfbox[18 144 592 738]{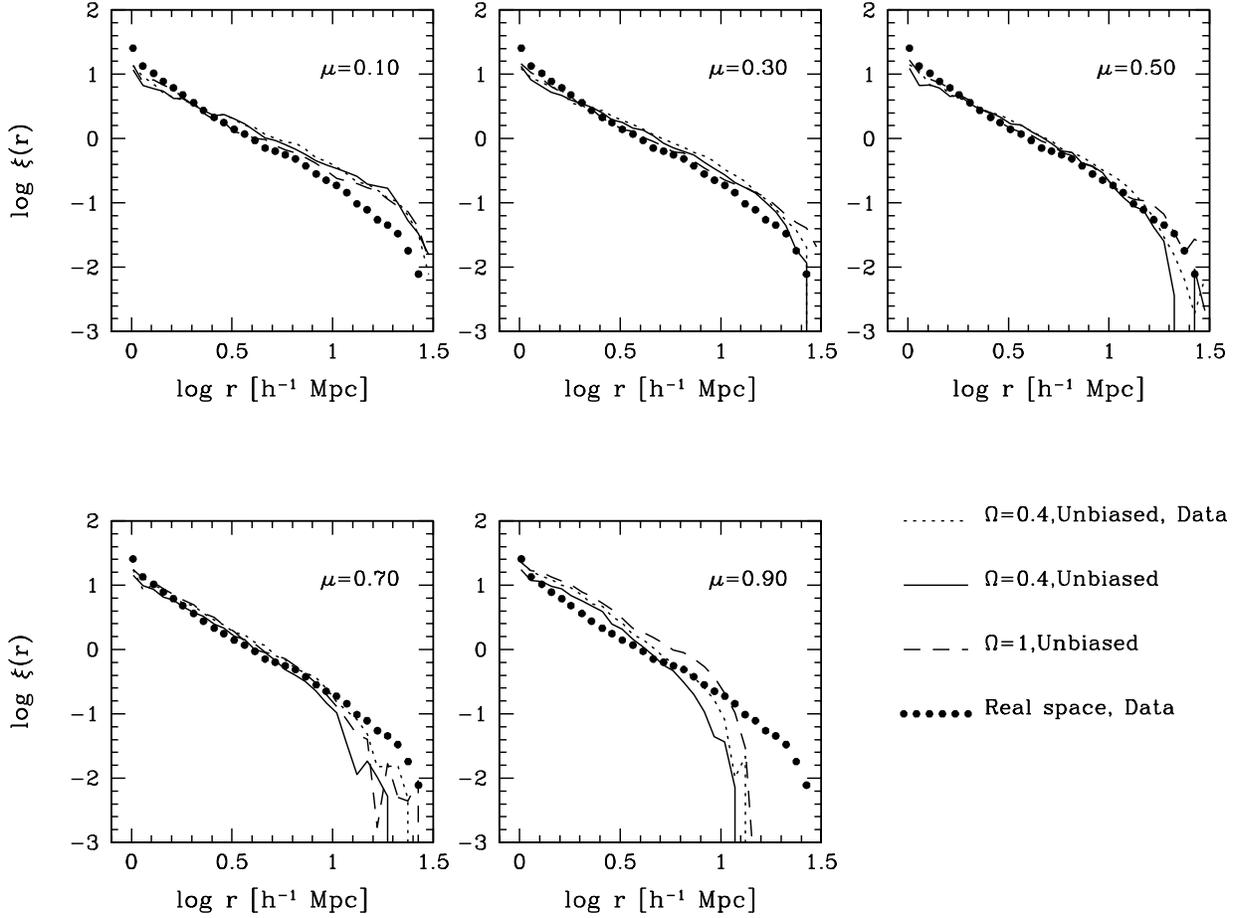}
\caption{ Correlation functions of the mock PSCZ catalog and its 
reconstructions after compressing the clusters.
The galaxy pairs contributing to the different panels have  different 
orientations relative to the line of sight, $\mu = \cos(\theta)$.
The filled circles show the real space correlation function of the 
mock catalog and are the same in all panels.
The dotted line shows the redshift space correlation function of the 
mock catalog.
The solid line shows the redshift space correlation function for a hybrid
reconstruction using $\Omega = 0.4$ (the correct value), while 
the dashed line is for a \rec with $\Omega = 1$. 
}
\end{figure}

Statistical uncertainties in measurements of redshift space distortions
arise mainly from the finite volume of the redshift surveys themselves.
``Cosmic variance'' noise has more impact on $\xi(s,\mu)$ measurements
than on $\xi(r)$ measurements because $\xi(s,\mu)$ is not averaged over
angles.  Each coherent sheet or filament in the redshift survey
causes an enhanced signal in the angular bin that corresponds to its
orientation.  The anisotropy signals from these randomly oriented structures 
would average to zero in an infinite survey, but in a finite volume they
mask the anisotropy caused by peculiar velocities and 
produce statistical uncertainty
in $\beta$ estimates.  Reconstruction overcomes the cosmic variance
error in redshift space distortion studies because a reconstruction
reproduces the physical structures in the survey volume with their
correct orientations.  The $\xi(s,\mu)$ for a reconstruction (or a 
real galaxy map) is not exactly isotropic even in the absence
of peculiar velocities, but the {\it differences} in
$\xi(s,\mu)$ for reconstructions with different values of $\Omega$ 
are due solely to the differences in peculiar velocities, not to 
changes in the physical orientations of coherent structures.

We demonstrate this point in Figure~22, which
shows the statistic $(\xi_{s}-\bar{\xi})/\bar{\xi}$,
 the fractional difference between the ensemble mean redshift space correlation 
function ${\overline{\xi(s,\mu)}}$ and the redshift space correlation function 
$\xi(s,\mu)$ for the true and the reconstructed mock PSCZ catalogs.
We compute the  mean correlation function $\overline{\xi(s,\mu)}$ and the 
cosmic variance band (shaded region) from an ensemble of 20 independent mock 
PSCZ catalogs, and we
plot this statistic only when $\overline {\xi(s,\mu)} > 0.1$.
The filled squares show the fractional difference for the true 
redshift space correlation function of the primary mock catalog, i.e.,
the departure from the mean $\overline{\xi(s,\mu)}$ in our single survey volume.
The solid line and the dashed line show the same statistic for the 
reconstructed \gal \distrbn assuming $\Omega = 0.4$ and 
$\Omega = 1$, respectively.
The angular anisotropy of the \gal \distrbn reconstructed with the correct
assumption for $\Omega$ matches the true angular anisotropy well 
within the cosmic variance band.
On the other hand, although the $\Omega = 1$ \rec clearly produces excessive 
anisotropy, especially so for $\mu = 0.9$, it could only be marginally rejected
in straight statistical comparisons because of the large cosmic variance band.
However, it is clearly inferior to the $\Omega = 0.4$ \rec and 
can be rejected at a large confidence level using the \rec analysis, mainly 
because a \rec with the correct $\beta$ can match the observed angular
anisotropies to much better than the cosmic variance limit.

\begin{figure}
\epsfxsize=\hsize
\epsfbox[18 144 592 738]{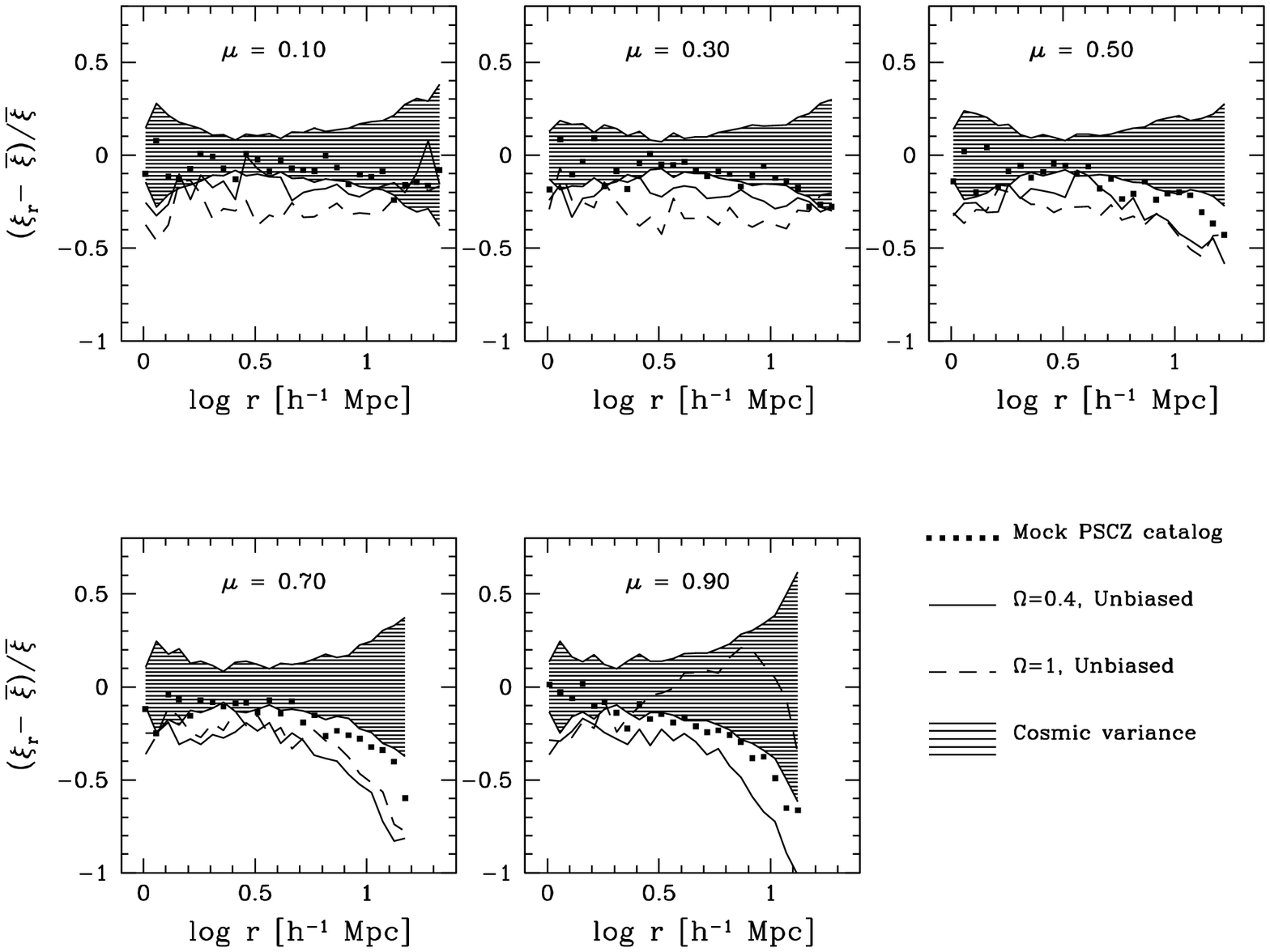}
\caption{ Fractional difference between the correlation function of the 
reconstruction and the mean correlation function in redshift space for
a mock PSCZ catalog.
The mean correlation function $\bar\xi$ and the $1 \sigma$ cosmic variance 
band (shaded region) are computed from an ensemble of 20 
independent mock PSCZ catalogs.
The filled squares show the redshift space correlation for the mock catalog.
The solid and dashed lines show the redshift space correlation function 
for hybrid reconstructions assuming $\Omega = 0.4$ and $\Omega = 1$
respectively.
}
\end{figure}

Figure 23 shows the \distrbn of nearest neighbors in the mock PSCZ 
catalog and its reconstructions.
If computed using the \z space \gal distribution, this statistic would
show a spurious peak at distances corresponding to the velocity dispersions 
of typical galaxy groups.
However, we would like to use this statistic to measure the degree of small 
scale clustering in the same manner as in the tests on full cube, real space 
\gal distributions, 
Therefore, we estimate the \nnbr \distrbn from the redshift space \gal
distributions  using the method suggested by Weinberg \& Cole (1992).
For every galaxy at a redshift $z$, we consider all the galaxies that lie
 within a redshift range $\Delta v < 1000\ {\rm km s^{-1}}$ to be its
potential nearest neighbor.
Of these candidate neighbors, we then choose the galaxy that lies closest
to this galaxy in the transverse direction, and we compute
the distribution of this transverse separation $R_{t}$ divided by 
the mean inter-particle separation $\bar d$ (i.e, $x_{n} = R_{t}/\bar d$).
The dotted line shows this \distrbn for the mock PSCZ catalog, while the
solid and dashed lines show this statistic for the \gal
distributions that are reconstructed assuming $\Omega = 0.4$ and 
$\Omega = 1$ respectively.
Both the reconstructions recover the \nnbr \distrbn of the input data quite 
accurately.

\begin{figure}
\epsfxsize=\hsize
\epsfbox[18 144 592 738]{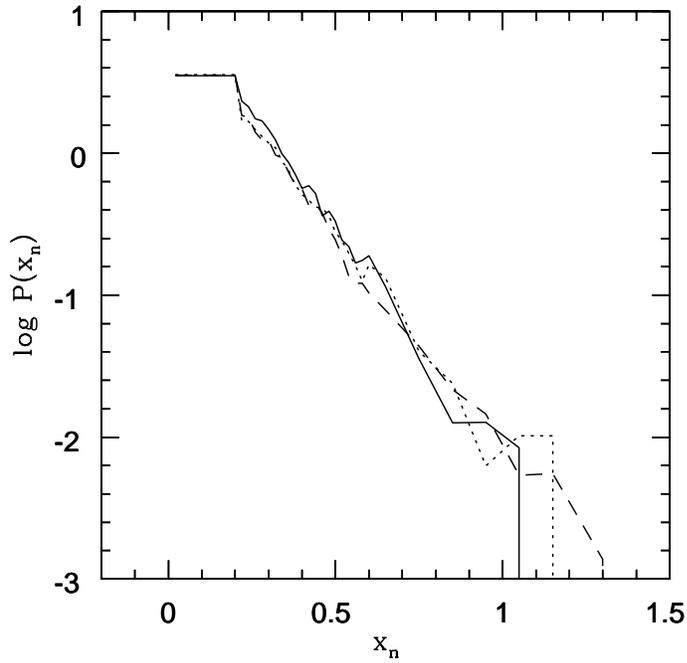}
\caption{ Nearest neighbor distribution for the final galaxy
 distributions of the mock PSCZ catalog and its reconstructions.
The nearest neighbor distribution is computed  in \z space using 
tangential separations with a $\Delta v = 1000\ {\rm km s^{-1}}$ line of
sight cut.
The dotted line shows the nearest neighbor distribution of the true final 
galaxy distribution in the mock PSCZ catalog.
Nearest neighbor distributions of the hybrid reconstructions 
are shown for $\Omega=0.4$ (solid line) and $\Omega=1$ (dashed line).
}
\end{figure}

The redshift space correlation function and the \nnbr \distrbn 
use the data from the redshift catalogs alone.
We can constrain the cosmological parameters more effectively if we also have
the data about the peculiar velocities of galaxies.
Comparison between predicted and observed peculiar velocity fields
is one of the main motivations for all-sky redshift surveys like
PSCZ and ORS.
The predictions often use the linear theory relation between the
density and velocity fields and thus break down in non-linear
regions characterized by multi-stream flows.
Attempts to correct for this breakdown either use quasi-linear approximations 
between the density and velocity fields or assume that the true peculiar 
velocity field is a combination of the linear theory predicted field
and a position-independent, random velocity dispersion.
The power of these comparisons is then limited by the validity of the
model for the non-linear components of the peculiar velocity field.
Reconstruction, on the other hand, predicts the fully non-linear peculiar
velocity field at each point in \z space, thereby giving a velocity field
that can be directly compared to peculiar velocity data without the need for
any additional modeling or approximations.
We now compare the velocity field reconstructed with different assumptions 
about $\Omega$ to see the accuracy to which we can reproduce the fully 
non-linear, true final velocity field.

Figure 24 shows the $x$ and $z$ components of the velocity field of the mock 
PSCZ survey and its reconstructions.
This plot shows the velocity field in the same slice whose density 
field is plotted in Figure 16.
We compute the velocity  and the velocity dispersion at any point
as the mean and the $1\sigma$ dispersion about this mean of the 
velocities of all the galaxies located within $5h^{-1}$Mpc of  this point.
Panel (a) shows the true velocity field of the mock PSCZ catalog.
Panel (b) shows the \vel field of the reconstructed \gal \distrbn
assuming (correctly) $\Omega = 0.4$.
Panel (d) shows the same field for the $\Omega = 1$ reconstruction.
We also plot, in panel ({\it c}), the \vel field predicted from the galaxy 
density field by the linear theory relation, for $\Omega = 0.4$.
Although the linear theory reproduces the true \vel field quite 
accurately in the low density regions, it systematically overestimates 
it in the high density regions as it does not account
for the deviation of the velocity vector from the gravitational vector
during the evolution of overdense regions (\cite{gr93b}).
The \rec assuming $\Omega = 1$, on the other hand,
systematically overestimates all the velocities, and the reconstructed 
\vel field is everywhere too hot compared to the true \vel field.
The \rec with the correct assumption of $\Omega = 0.4$ provides the
best recovery of the true \vel field.
The amplitude of the velocities is comparable to the true values, 
and the non-linear component in high density regions is recovered quite well.

\begin{figure}
\epsfxsize=\hsize
\epsfbox[18 144 592 738]{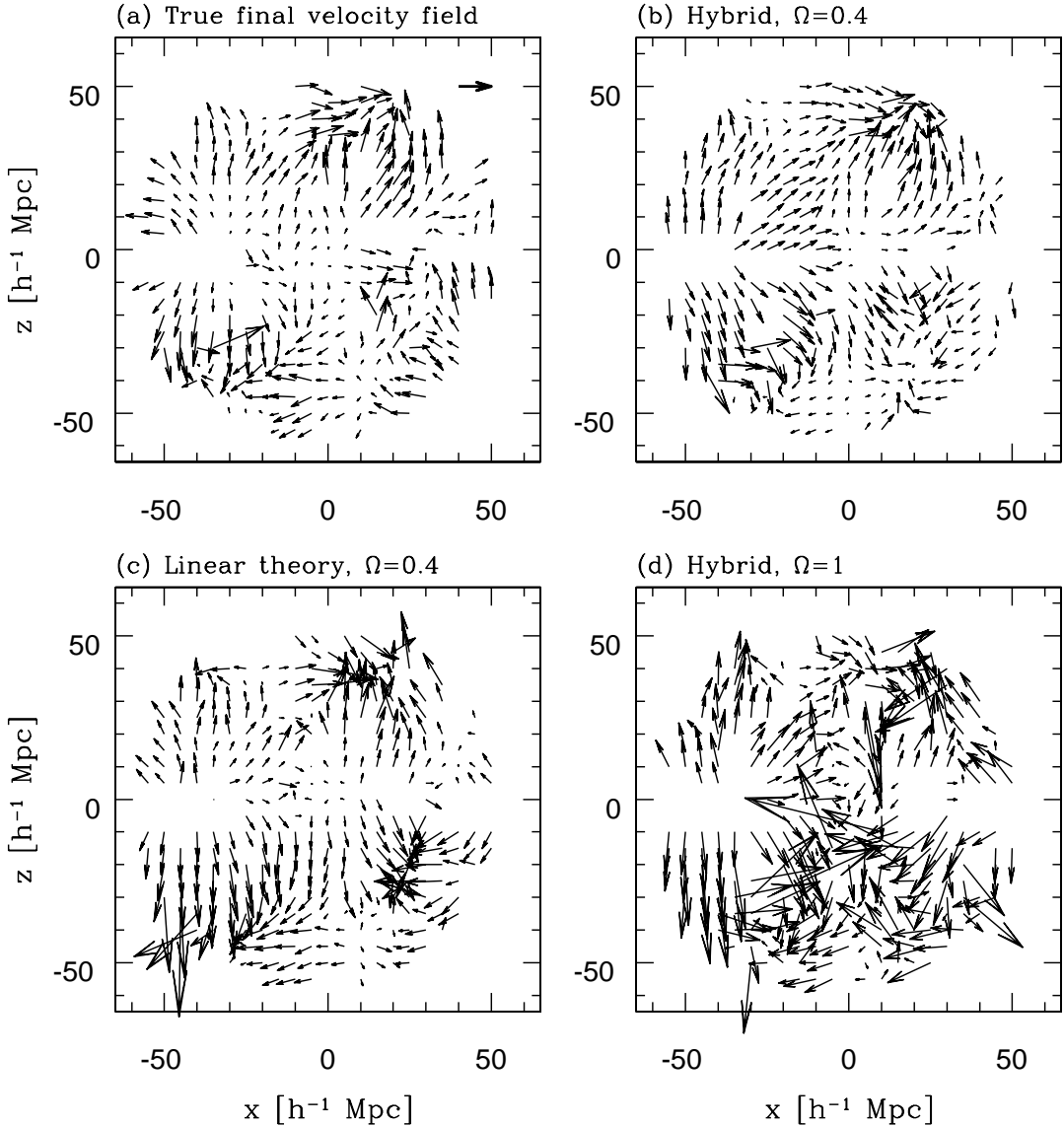}
\caption{ Velocity fields of the true and reconstructed galaxy distributions
for the mock PSCZ survey, averaged over a $5h^{-1}$Mpc top hat window.
Only the $x$ and $z$ components of the velocity field in a slice through 
the center of the survey are shown.
({\it a}) Velocity field of the true mock PSCZ catalog.
({\it b}) Reconstructed velocity field assuming $\Omega=0.4$ 
({\it c}) Linear theory prediction for $\Omega = 0.4$.
({\it d}) Reconstructed velocity field assuming $\Omega=1$.
The length of the dark arrow in the top right corner in panel ({\it a}) 
corresponds to 500 kms$^{-1}$.
}
\end{figure}

We show the velocity dispersion field in Figure 25,
where the radius of the circle at each field point is proportional to 
the value of the velocity dispersion at that point.
We compute the velocity dispersion at a field point only if there are
at least four galaxies within $5h^{-1}$Mpc of it.
The different panels correspond to the same galaxy distributions as in 
Figure 24.
Here again, the reconstructed \gal \distrbn with $\Omega = 0.4$
matches the true \vel dispersion better than either the $\Omega = 1$
\rec or the linear theory prediction.
The \vel dispersion of the $\Omega = 1$ \rec is systematically larger
than the true value, reinforcing the conclusions from the cluster 
velocity dispersions (Fig. 20).
In practice, it is very difficult to reliably map the velocity dispersion 
field from the noisy peculiar velocity field because of the large errors
in the redshift-independent distances to individual galaxies.
However, the velocity dispersion affects the redshift space structure of the 
\gal distribution, so we need to correctly account for it before we can 
reliably compare model predictions with the galaxy redshift data.
From this Figure, it is clear that the velocity dispersion is a 
highly variable function of position and that this positional variation is 
reproduced quite accurately by the reconstruction with the correct assumptions.
Therefore, a measurement of $\beta$ using the full velocity dispersion field 
predicted by the \rec should be more accurate than a $\beta$ measured 
assuming a position-independent velocity dispersion (\cite{wsdk97}).

\begin{figure}
\epsfxsize=\hsize
\epsfbox[18 144 592 738]{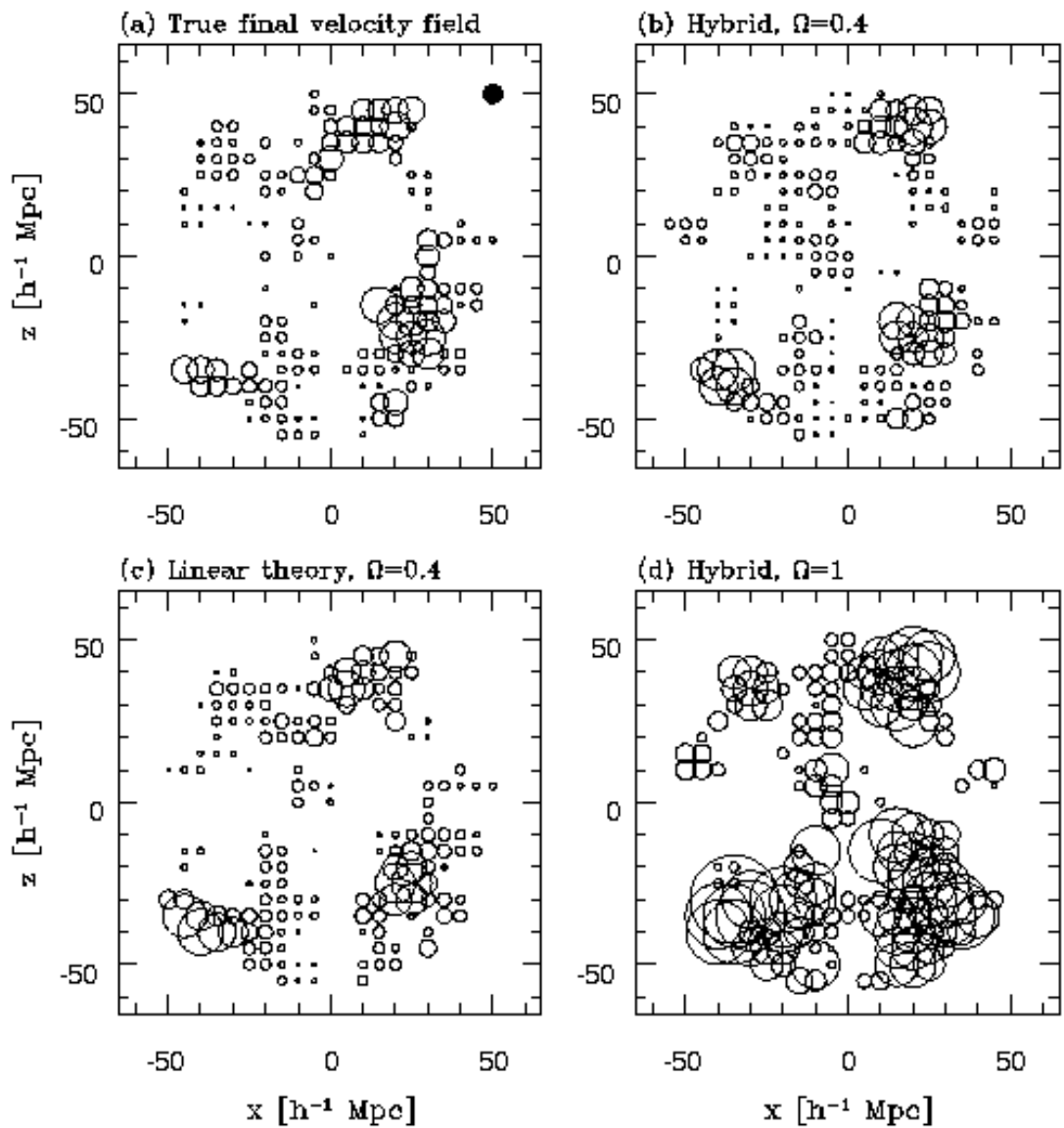}
\caption{ Velocity dispersion field of the true and reconstructed galaxy 
distributions for the mock PSCZ catalog.
({\it a}) True mock PSCZ catalog.
({\it b}) Reconstructed galaxy distribution assuming $\Omega=0.4$ 
({\it c}) Linear theory prediction for $\Omega=0.4$ 
({\it d}) Reconstructed galaxy distribution assuming $\Omega=1$ .
We compute the velocity dispersion only if there are at least four galaxies
within $5h^{-1}$Mpc of the field point.
The radius of the filled circle centered at (50,50) in panel ({\it a}) 
corresponds to a velocity dispersion of 400 kms$^{-1}$.
}

\end{figure}

\subsection {\it Reconstruction of a Mock ORS Catalog}

The ORS (\cite{ors1}) is a \z survey of optically selected galaxies covering
about $98\%$ of the sky with Galactic latitude $|b| > 20^{\circ}$.
It is drawn from three different catalogs,  the Uppsala Galaxy Catalog
(UGC), the European Southern Observatory Galaxy Catalog (ESO), and
the Extension to the Southern Galaxy Catalog (ESGC).
It has 2 subcatalogs, one magnitude-limited subsample complete to 
a $B$ magnitude of 14.5 and another subcatalog complete to a $B$ major
axis diameter of $1.9'$.
There are about $8500$ galaxies in the catalog distributed over a solid angle 
of 8.09sr, with the magnitude-limited subsample containing about $5700$ 
galaxies.
We make a mock ORS catalog using the same mass \distrbn and Local 
Group observer used to create the PSCZ mock catalog.
We first select ``galaxies'' from this mass \distrbn using the power law 
biasing scheme described by equation~(\ref{eqn:biasdef}), so that the rms
fluctuation amplitude of the resulting \gal \distrbn is
$\sigma_{8g} = 1.1 \simeq 1.5\sigma_{8m}$.
We then select a volume limited subsample out to a radius of $40h^{-1}$Mpc 
so that the average density of galaxies in this volume is 
$0.008h^{3}$Mpc$^{-3}$.
We include an outer magnitude limited sample up to a radius of
$60h^{-1}$Mpc, where the selection function decreases as $\phi(r) \propto
r^{-3}$, and we exclude all galaxies in 
a $40^{\circ}$ wedge about the Local Group to mimic the survey's
Galactic plane cut.
Finally, we ``observe'' this \gal \distrbn in \z space in the frame of the 
Local Group observer.

We reconstruct this mock ORS catalog using the hybrid method as applied to 
biased \gal distributions.
The details of this \rec are similar to those of the mock PSCZ catalog
reconstruction.  The differences are:
(1) After correcting for \z space distortions using the method described 
in \S4.2, we map the real space \gal density field to an empirically determined
mass PDF with the appropriate $\sigma_{8m}$.
(2) We fix the amplitude of fluctuations in the reconstructed 
initial density field so that $\sigma_{8m} = \sigma_{8g}/b$.
(3) We choose ``galaxies'' from the reconstructed mass \distrbn 
using the power law biasing relation defined by equation~(\ref{eqn:biasdef}).
We set the parameters $A$ and $B$ of this relation so that the galaxy 
density is $n_g=0.008 h^3\; {\rm Mpc}^{-3}$ and the rms galaxy fluctuation
in redshift space $\sigma_{8g}$ matches that of the input mock catalog.

We present the results of the \rec analysis of the mock ORS catalog
in Figures~26 to~33, in the same manner as for the mock PSCZ catalog
reconstruction.
We show the density fields and the \gal distributions in the same slice as
for the mock PSCZ catalog.
Figure 26 shows the isodensity contours of the initial density field.
Although the gross features are recovered, the recovery is generally
poor, especially near the survey boundaries.
This poor recovery could in principle reflect either the 
small outer radius limit ($40\hmpc$ vs.\ $55\hmpc$ for PSCZ)
or the large angular mask of the mock ORS catalog.
To check which of the two effects is dominant, we reconstructed 
the mock ORS catalog assuming a smaller angular mask.
We found that the initial density field was recovered very well and the 
correlation
between the true and reconstructed fields was comparable to that for the mock
PSCZ reconstruction.
This suggests that we could significantly improve the \rec of the ORS 
catalog by 
filling in the large Galactic plane mask with the density field mapped by the 
PSCZ catalog.
We should of course, normalize the PSCZ density  field to have the 
same fluctuation amplitude as the ORS density fluctuations before filling 
in this region.
We have not followed this filling-in procedure here, but we may do so when
analyzing the real ORS data.

We show the scatter plot of the true and reconstructed initial and final 
density fields in Figures 27a and 27b respectively. 
The weak correlation between the true and recovered initial density fields 
quantifies 
the poor recovery seen in Figure 26.
Figure 28 shows the power spectrum of the true initial density field 
(dotted line)
and of the reconstructed initial fields after the power restoration 
and amplitude matching procedures.
The amplitudes of the reconstructed initial density fields are 
normalized so that $\sigma_{8m} = \sigma_{8g}/b = 0.75 $.
This normalization is accurate essentially by construction.
However, although the overall slopes of the recovered power spectra are correct,
there are  substantial oscillations in the recovered power spectrum 
that are not present in the true initial density field.

\begin{figure}
\epsfxsize=\hsize
\epsfbox[18 300 592 738]{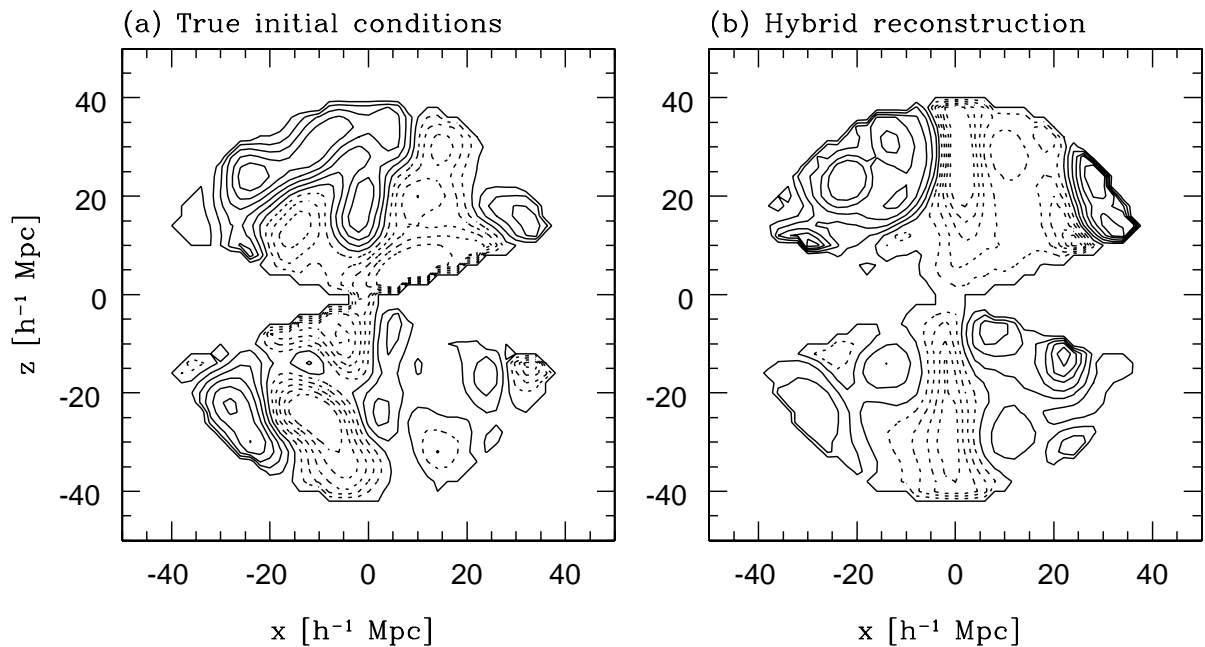}
\caption{ Contours in a slice of the ({\it a}) true and ({\it b}) reconstructed 
initial density fields for the mock ORS catalog in the same format as Fig. 16.
A slice through the galaxy distribution obtained by evolving the field 
in ({\it a}), and selecting galaxies in a biased manner is shown in Fig. 29a.
}
\end{figure}

\begin{figure}
\epsfxsize=\hsize
\epsfbox[18 300 592 738]{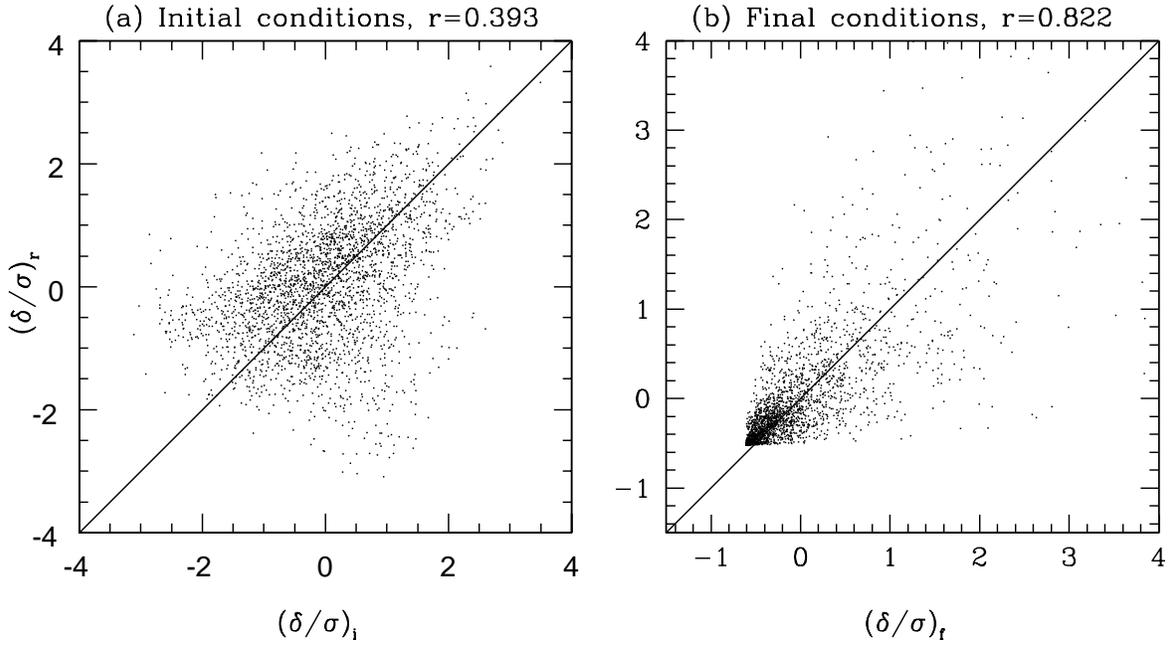}
\caption{({\it a}) Cell by cell comparison of the hybrid reconstructed 
initial density contrast $(\frac{\delta}{\sigma})_{r}$ to the true initial 
density 
contrast $(\frac{\delta}{\sigma})_{i}$ for the mock ORS catalog.
({\it b}) Comparison of the reconstructed final density contrast 
$(\frac{\delta}{\sigma})_{r}$ to the true final density contrast 
$(\frac{\delta}{\sigma})_{f}$, in redshift space.
All the density fields are smoothed using a Gaussian filter of radius 
3$h^{-1}$Mpc
and scaled by the rms fluctuation $\sigma$. The linear correlation 
coefficient $r$ is indicated above each panel.
}
\end{figure}

\begin{figure}
\epsfxsize=\hsize
\epsfbox[18 144 592 738]{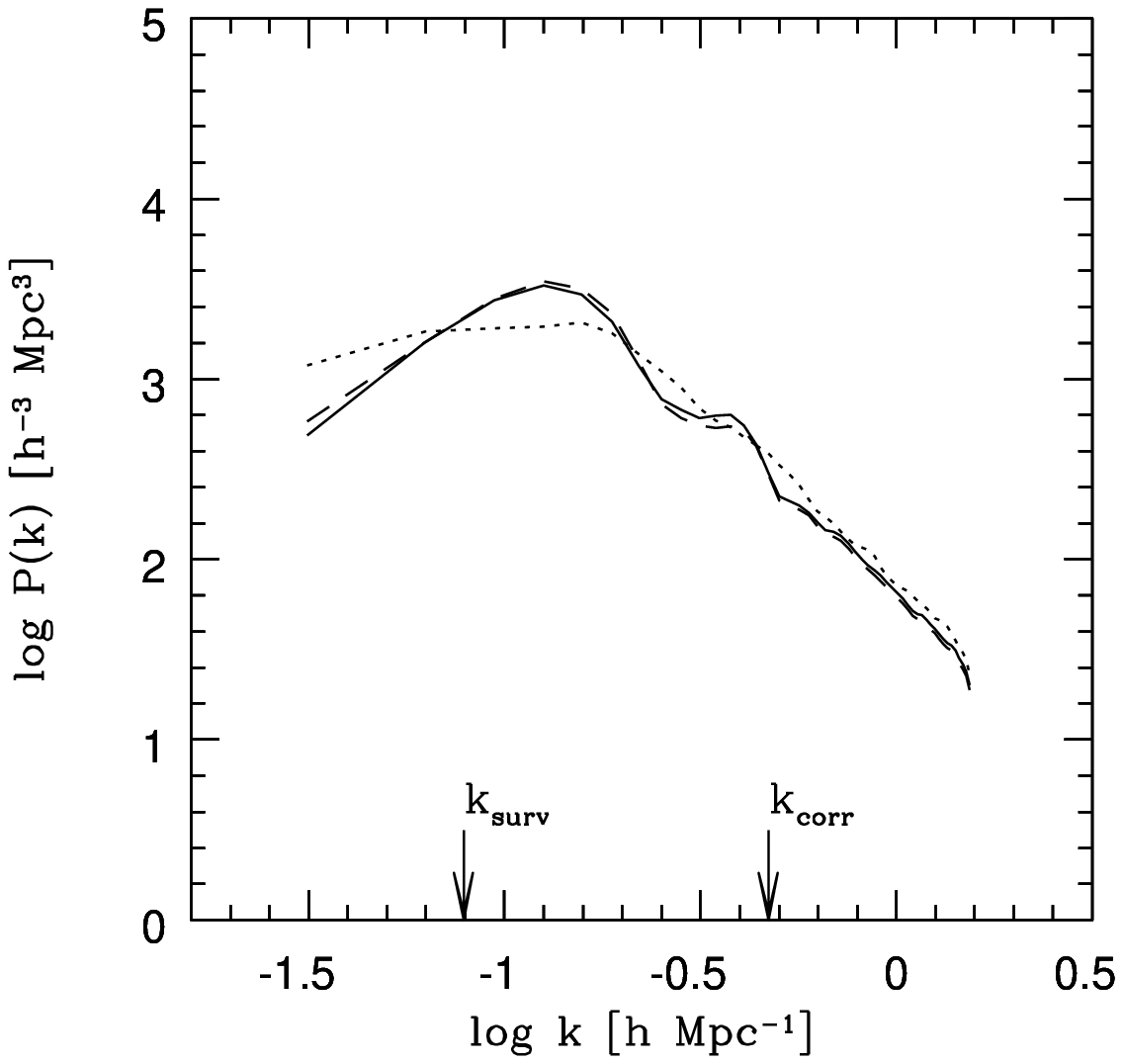}
\caption{ Power spectrum of the true initial density field of the
 mock ORS survey(dotted line) and the normalized hybrid reconstructed density 
field recovered for $\Omega = 0.4$ (the correct value, solid line) and for
 $\Omega = 1$ (incorrect value, dashed line).
The arrows show the smallest wavenumber that corresponds to the survey
size ($k_{\rm surv} = 2\pi/2r_{\rm in} = 0.078\ h$Mpc$^{-1}$) and 
the maximum wavenumber 
$k_{\rm corr} = 15k_{f} = 0.471\ h$Mpc$^{-1}$ beyond which random phase 
waves are added to the reconstructed field.
}
\end{figure}

Figure 29 shows the \gal distributions of the mock ORS catalog and its
reconstruction assuming  $\Omega = 0.4$, in real space (panels a and b) 
and redshift space (panels c and d).
Here again, as in the \rec of the mock PSCZ catalog, the filamentary 
structure running from $(x,z) = (-10,20)h^{-1}$Mpc to $(10,35)h^{-1}$Mpc 
is absent in the reconstruction.
There are also a few spurious features that are present in the reconstruction
alone, such as the clusters seen at $(x,z) = (20,35)h^{-1}$Mpc and 
$(-15,-15)h^{-1}$Mpc.
The cluster at $(x,z) = (25,-20)h^{-1}$Mpc, although recovered at the 
proper location, appears very rich in the reconstruction.
These features are also seen in \z space, where the clusters have 
prominent ``Fingers of God'' in the \rec but not in the true \gal 
distribution.
We also see that the reconstructed \gal \distrbn appears more dynamically
evolved than the true \gal distribution, although the final $\sigma_{8g}$
of the two \z space \gal distributions are identical and the power spectra  
are similar.
We show the multiplicities and velocity dispersions of clusters in panels (a) 
and (b) of Figure 30.
Although the cluster multiplicities are similar for the $\Omega = 0.4$
and $\Omega = 1$ reconstructions, the cluster \vel dispersions 
are significantly higher than the true values for the  $\Omega = 1$ 
reconstruction. 

\begin{figure}
\epsfxsize=\hsize
\epsfbox[18 144 592 738]{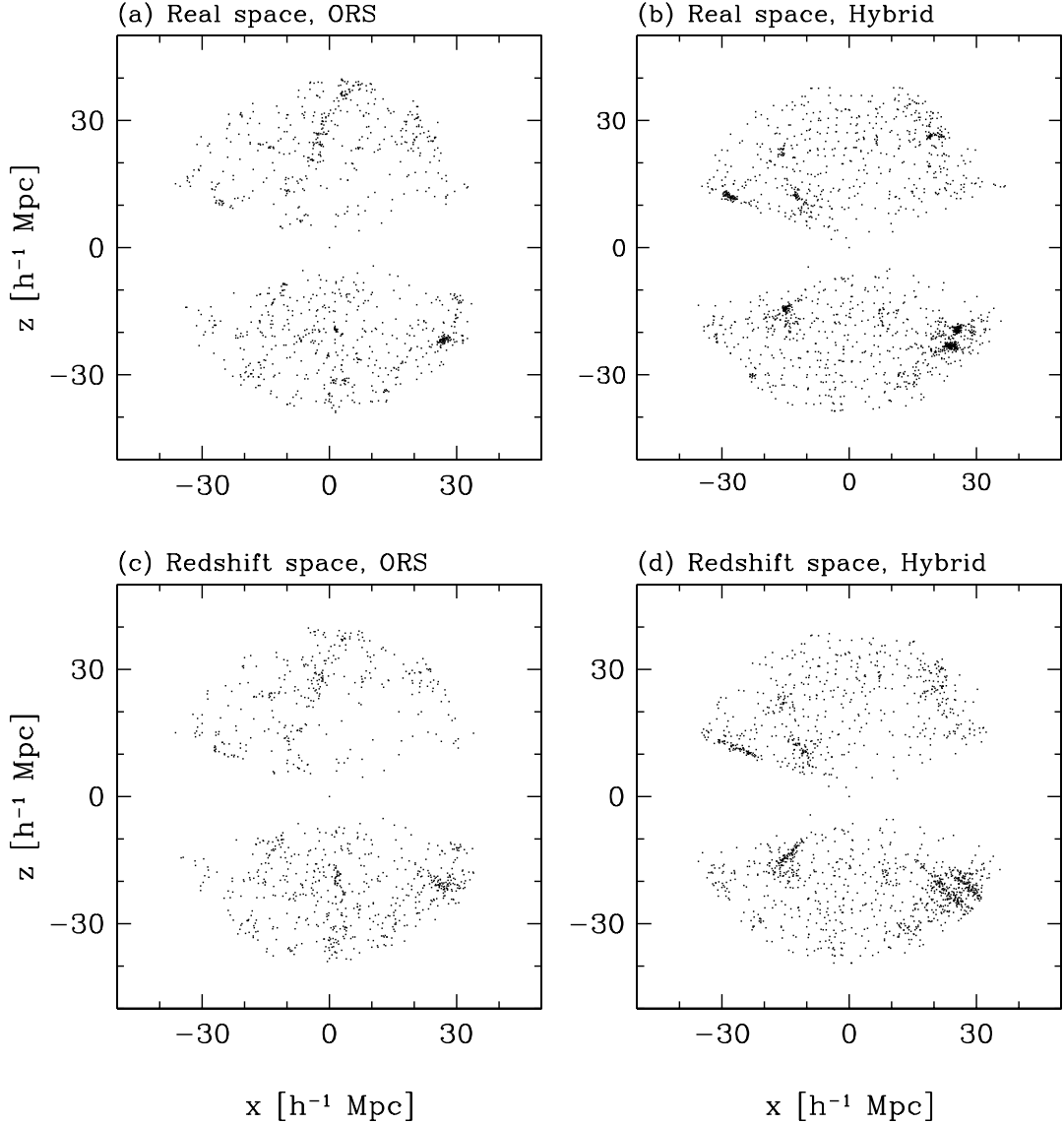}
\caption{ Final galaxy distributions for mock ORS catalog
in the region $-20 < y < 20 h^{-1} $Mpc.
({\it a}) True final galaxy distribution in real space.
({\it b}) Final galaxy distribution of hybrid reconstruction in real space.
({\it c}) True final galaxy distribution in redshift space.
({\it d}) Final galaxy distribution of hybrid reconstruction in redshift space.
}

\end{figure}

\begin{figure}
\epsfxsize=\hsize
\epsfbox[18 300 592 738]{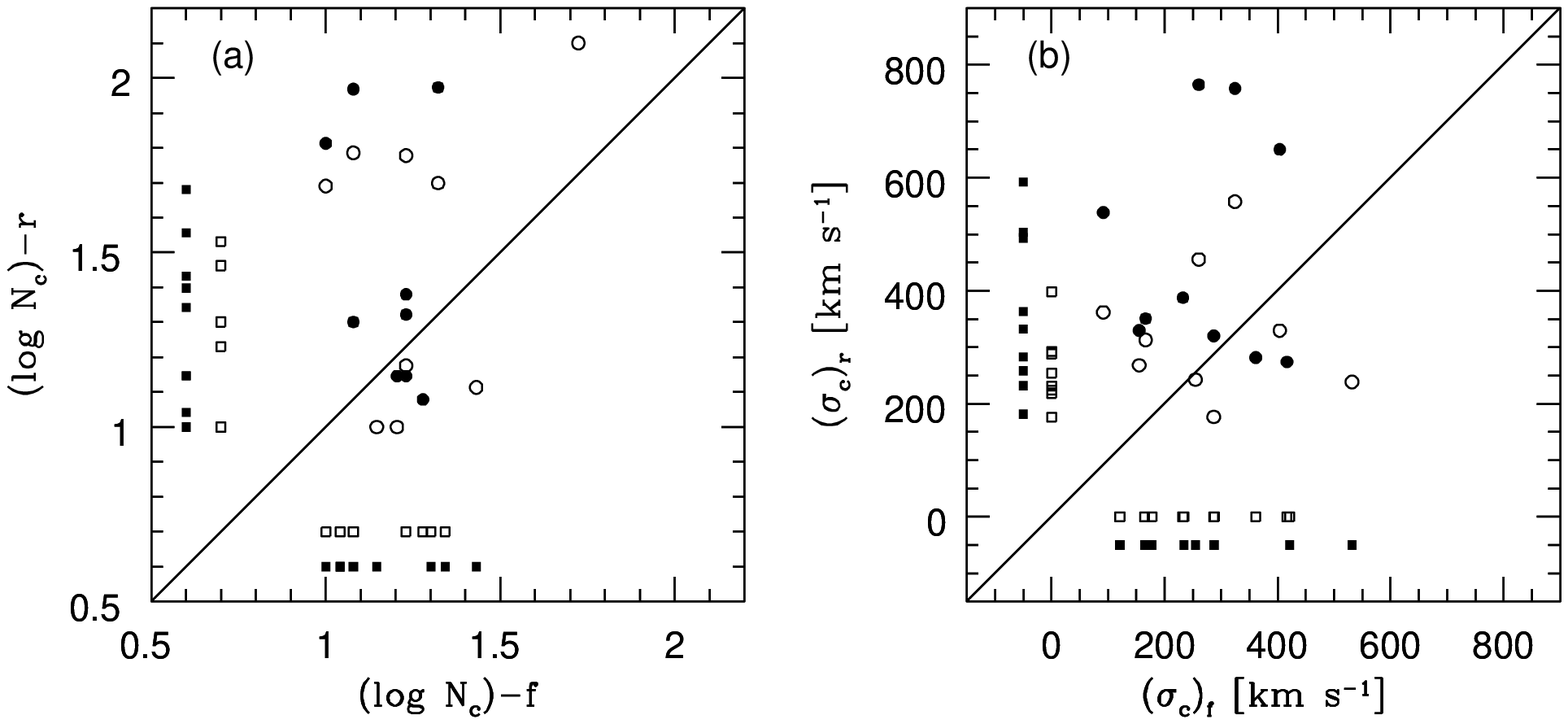}
\caption{ Comparison of ({\it a}) cluster multiplicities and ({\it b}) cluster 
velocity dispersions for the mock ORS catalog reconstruction.
Squares parallel to either axis represent clusters present in that 
galaxy distribution alone. 
Open symbols show the reconstruction with $\Omega = 0.4$ (correct value), 
and  the filled symbols show a \rec with $\Omega = 1$. 
}

\end{figure}

Figure 31 shows the two-point correlation functions $\xi(s,\mu)$ of the mock 
ORS catalog and its reconstructions after compressing the clusters.
The various symbols have the same meaning as in Figure 21.
The large scale clustering amplification along the line of sight is much 
 stronger for the $\Omega = 1$ \rec compared to that of the true \distrbn and 
the $\Omega=0.4$ reconstruction.
Figure 32 shows the fractional difference between the mean and the observed 
correlation functions for the ORS \rec in a manner similar to 
Figure 22 for the PSCZ catalog.
The mean correlation function $\overline{\xi(s,\mu)}$ and the cosmic variance 
band are computed from an ensemble of 20 independent mock ORS catalogs.
This cosmic variance band is broader than that for the PSCZ catalog 
in Figure 22 because the ORS catalog surveys a smaller volume and employs a
sparser sampling (at the chosen limiting radius).
We see that the $\xi(s,\mu)$ for the \rec with the correct assumption 
of $\Omega = 0.4$
always matches the true $\xi(s,\mu)$ to much better than the cosmic 
variance limit.
On the other hand, the $\Omega = 1$ \rec has a significantly higher 
degree of angular anisotropy and is a poor match to the true anisotropies.
This shows that we can effectively use the large scale amplification in the 
correlation function for $\mu \geq 0.7$ as a good diagnostic of 
$\Omega$ (or at least $\beta$), despite the errors caused by the angular
mask of the ORS.

\begin{figure}
\epsfxsize=\hsize
\epsfbox[18 144 592 738]{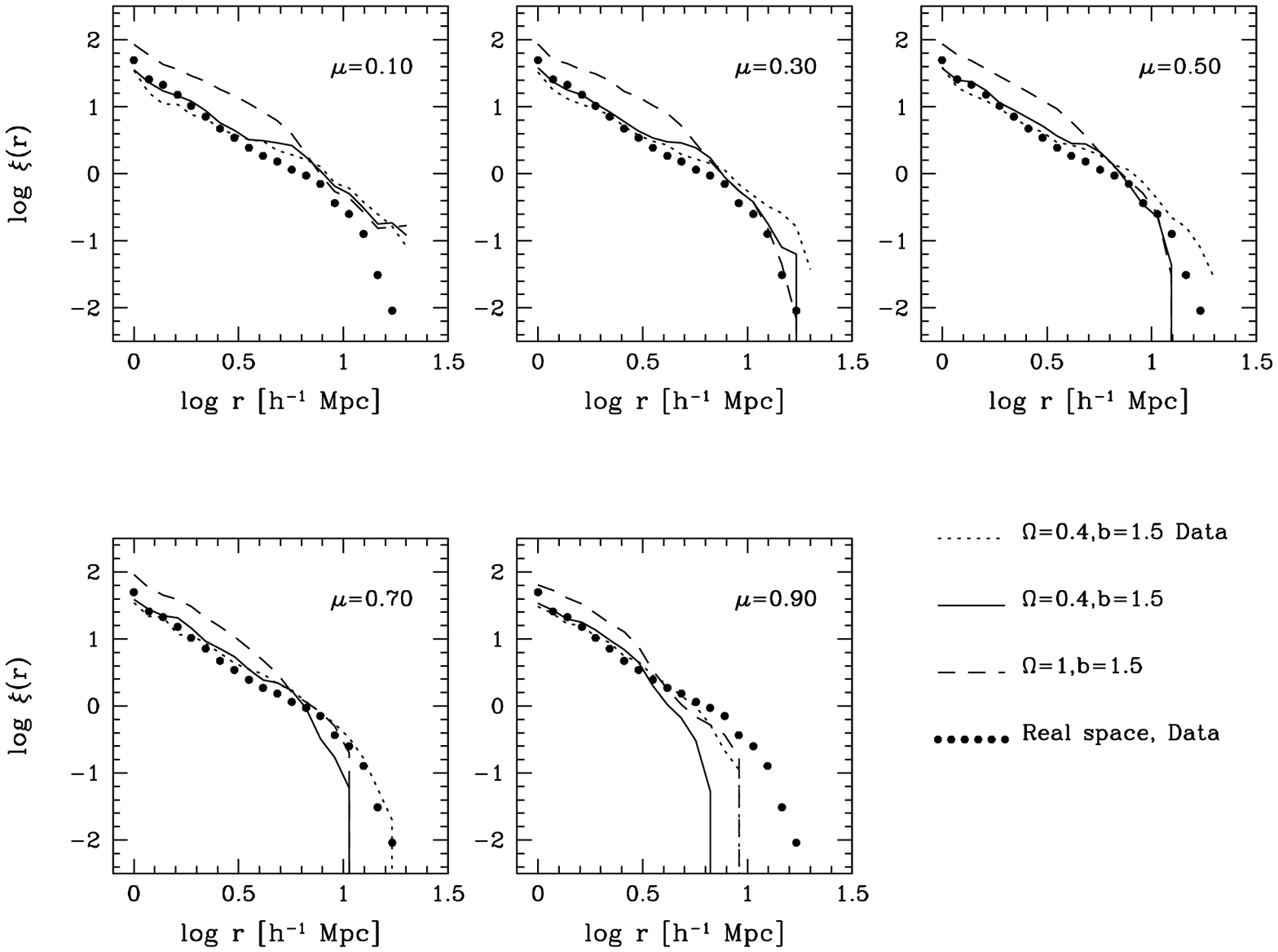}
\caption{ Correlation functions of the mock ORS catalog and its reconstructions
after compressing the clusters.
The \gal pairs contributing to different panels have different 
orientations relative to the line of sight, $\mu = \cos(\theta)$.
The filled circles show the real space correlation function of the 
mock catalog and are the same in all panels.
The dotted line shows the redshift space correlation function of the 
mock catalog.  The solid line shows the redshift space correlation 
function for a hybrid reconstruction using $\Omega = 0.4$ 
(the correct value), while the dashed line is for a \rec with $\Omega = 1$. 
}
\end{figure}

\begin{figure}
\epsfxsize=\hsize
\epsfbox[18 144 592 738]{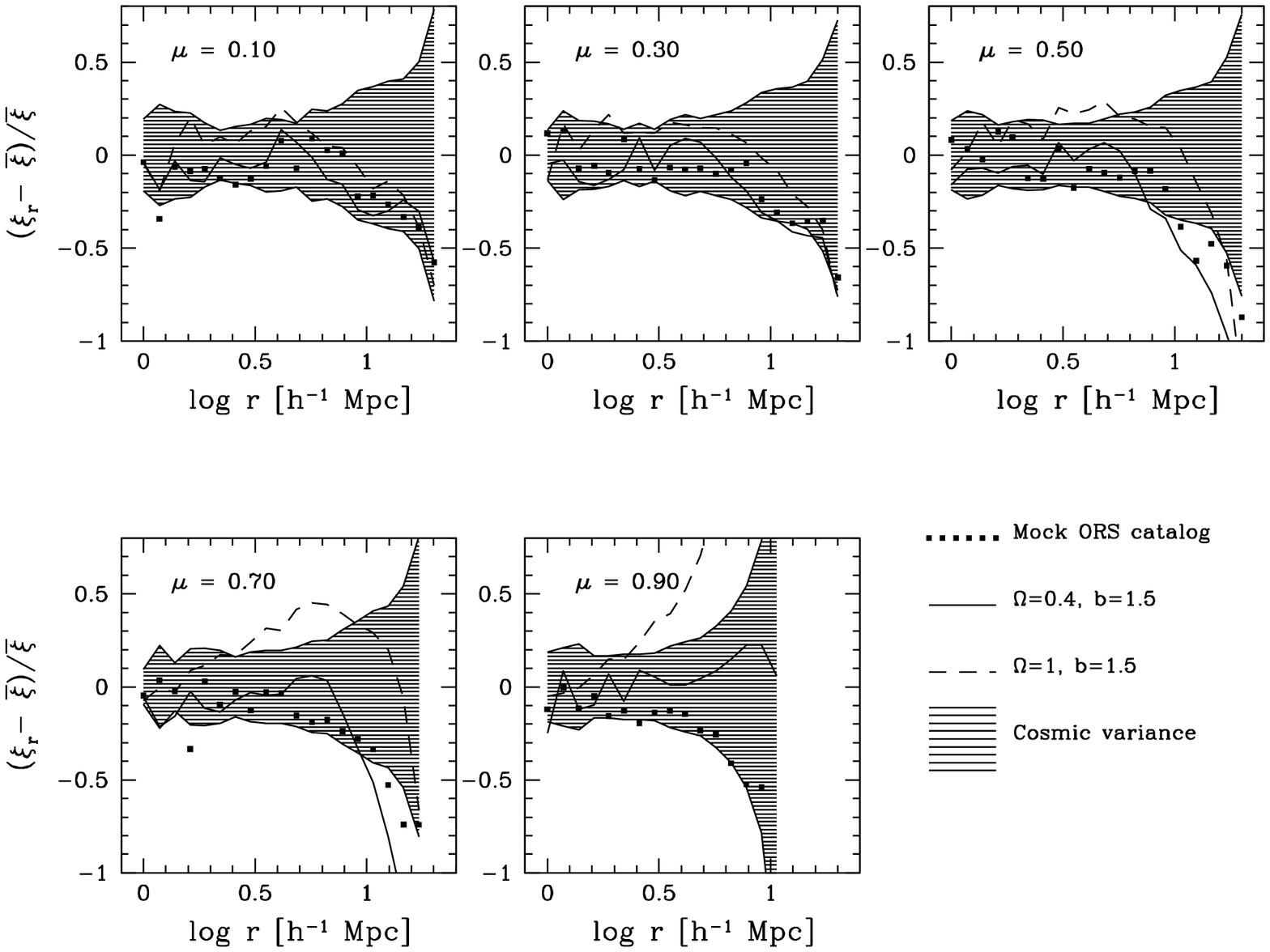}
\caption{ Fractional difference between the correlation function of the 
reconstruction and the true correlation function in redshift space for
the mock ORS catalog. 
The mean correlation function $\bar \xi$ and the cosmic variance band
(shaded region) are computed from an ensemble of 20 independent
mock ORS catalogs.
The filled squares show the redshift space correlation for the mock catalog.
The solid and dashed lines show the redshift space correlation function 
for hybrid reconstructions assuming $\Omega = 0.4$ and $\Omega = 1$,
respectively.
}
\end{figure}

We computed the \nnbr \distrbn for the mock ORS catalog and its 
reconstructions assuming $\Omega = 0.4$ and $\Omega = 1$,
in the same manner as described for the mock PSCZ catalog.
We found that both the reconstructions recovered the \nnbr \distrbn of 
the input data quite accurately.

Figure 33 shows the $x$ and $z$ components of the \vel fields
for the mock ORS catalog and its reconstructions.
These fields are computed in the same manner as described for the mock PSCZ 
catalog, and the fields are plotted in that slice whose density contours are
shown in Figure 26.
The linear theory predicted \vel field is derived from the density contrast 
field $\delta_{m}$ that is obtained by dividing the real space galaxy density 
contrast field by the bias factor, i.e, from $\delta_{m} = \delta_{g}/b$.
We find that the correct assumption \rec ($\Omega = 0.4$) provides the best 
match
to the true field in both the amplitude and the non-linear components.
The linear theory velocity field does not reproduce well the 
small scale incoherent velocities, and
the velocity field of the $\Omega = 1$ \rec 
has a much higher amplitude and is very hot compared to the true field.
We also computed the velocity dispersion field of the true and the reconstructed
ORS catalogs in the same manner as for the PSCZ catalog.
Here also, we found that both the amplitude and the spatial variation of this 
 velocity dispersion is best recovered by the \rec that  correctly assumes
$\Omega = 0.4$ and  $b=1.5$.

\begin{figure}
\epsfxsize=\hsize
\epsfbox[18 144 592 738]{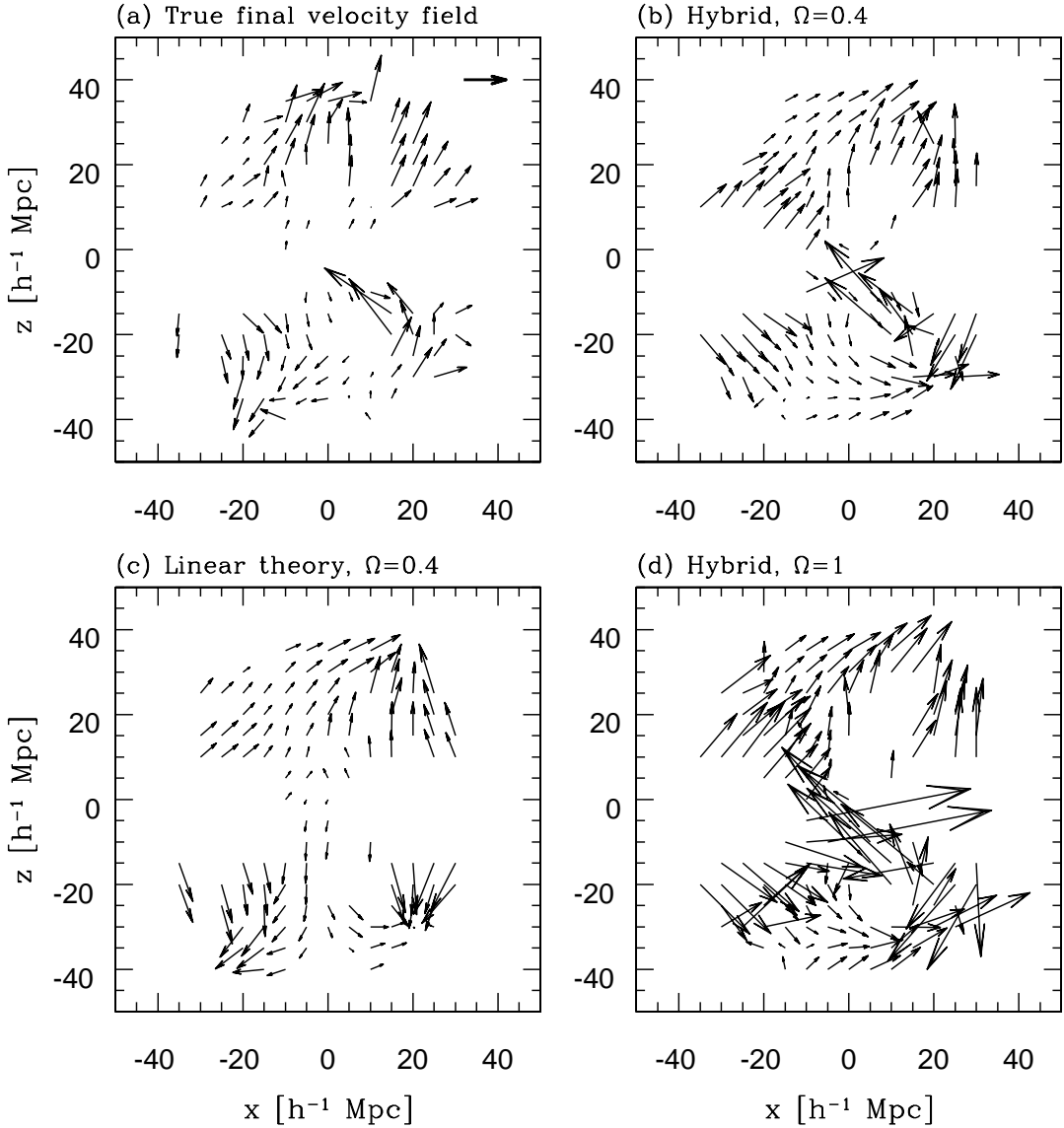}
\caption{ Velocity fields of the true and reconstructed galaxy distributions
for the mock ORS survey, averaged over a $5h^{-1}$Mpc top hat window.
Only the $x$ and $z$ components of the velocity field in a slice through 
the center of the survey are shown.
({\it a}) Velocity field of the true mock ORS catalog.
({\it b}) Reconstructed velocity field using $\Omega=0.4$.
({\it c}) Linear theory prediction for $\Omega = 0.4$.
({\it d}) Reconstructed velocity field using $\Omega=1$.
The length of the dark arrow in the top right corner in panel ({\it a}) 
corresponds to 500 kms$^{-1}$.
}
\end{figure}

\subsection {\it Reconstruction with Incorrect Bias Factors}

In all of the reconstructions of the mock catalogs shown so far, we have 
assumed that we know the correct value of the bias factor.
This is not the situation when we reconstruct the observed redshift \gal 
distributions.
The anisotropy of $\xi(s,\mu)$ and the cluster velocity dispersions
constrain most directly a parameter combination that is similar to $\beta$
(it is exactly equal to $\beta$ in the linear regime).
Translating this to a constraint on $\Omega$ requires 
independent information about the bias factor.
To check if we can constrain $\Omega$ and $b$ individually, we 
reconstructed the mock PSCZ and ORS catalogs with different assumptions 
about $\Omega$ {\it and} $b$.
Apart from the cases considered above, we used other combinations of
$\Omega$ and $b$ that have roughly the same $\beta$ as the respective mock 
catalogs.
We do not show the results of these additional reconstructions in detail.
Our main conclusions from these are:
\begin{description}
\item[{(1)}] The \rec with the correct assumption for $\Omega$ and $b$ always 
provides the best match to the input \gal distribution.
\item[{(2)}] Although the large scale clustering amplification depends only on 
$\beta$ in linear theory, we find that the fractional difference between the 
true and the reconstructed $\xi(s,\mu)$ is the smallest for the \rec with the 
correct $\Omega$ and $b$, when we compare in all the angular bins.
\item[{(3)}] The nearest neighbor distributions computed using the transverse 
separations do not discriminate between the different assumptions about 
the bias factor for the mock PSCZ catalog reconstructions.
The evolved structure is very similar for the range of $b$ considered.
We do see a marginal difference of the expected direction in the \nnbr 
distributions of the mock ORS catalog reconstructions, although this
difference may be too small to be detected at a reasonable confidence level.
\item[{(4)}] The cluster velocity dispersions and the peculiar velocity field 
are primarily determined by the value of $\beta$.
\end{description}

\section{DISCUSSION}

In this paper, we have described and tested a hybrid \rec scheme that 
can be used to reconstruct the observed distribution of galaxies.
When the galaxy \distrbn is an unbiased tracer of the mass distribution,
this scheme consists of the steps (H1)-(H7) listed in \S2.2.
If the \gal \distrbn is biased with respect to the underlying mass 
distribution, we replace the steps (H2) and (H6) by the steps 
(H2B) and (H6B).

A hybrid reconstruction using this method incorporates a number of assumptions.
The need for these assumptions and their effects on the \rec are 
discussed at the beginning of \S3.
The most fundamental of these assumptions is the hypothesis that the
primordial density fluctuations form a Gaussian random field,
as predicted by simple inflation models for the origin of the fluctuations
(\cite{guth82}; \cite{hawking82}; \cite{starobinsky82};
\cite{bardeen83}).  Other assumptions include
the values of the density parameter $\Omega$ and the bias factor $b$
and the biasing model used to select galaxies from the evolved 
mass distribution.
Given a \z space \gal distribution, we can reconstruct it
using different combinations of the latter assumptions, within the general 
framework of Gaussian initial fluctuations.
We can then use both local comparisons of structure
and global statistical comparisons to check what
combinations of the assumptions, if any, can best reproduce the input data.
In application to observational data, these comparisons
will enable us to test the validity of the different assumptions and
to constrain the allowed ranges of cosmological parameters.
If there is no reasonable combination of assumptions for 
which the reconstructed \gal \distrbn accurately reproduces the input \gal 
distribution, we will be forced to question the Gaussian assumption 
itself and explore alternative scenarios for the origin of structure.

We tested the hybrid \rec method on idealized galaxy distributions 
derived from the mass distributions of N-body simulations.
We tested it on both unbiased and biased galaxy distributions.
We also tested this \rec scheme on mock galaxy redshift catalogs
that are designed to mimic the geometry and depth of the PSCZ and ORS surveys.
In all these tests, we were primarily interested in checking whether
the hybrid \rec
method can accurately reproduce the input galaxy distribution for the correct
set of assumptions and discriminate against incorrect assumptions.
Our conclusions from these tests are as follows:
\begin{description}
\item[{(1)}] The hybrid method recovers the initial density fluctuations 
much better than either the Gaussianization or the dynamical scheme alone.
The hybrid reconstructed galaxy \distrbn matches the local and global
properties of the true input \gal \distrbn more accurately than the \gal 
\distrbn reconstructed by Gaussianization.
\item[{(2)}] A \rec that incorporates correct assumptions about $\Omega$
and $b$ always yields the best match to the input data.
Reconstructions with wrong assumptions about these parameters produce a
\gal \distrbn that is identifiably different from the input \gal distribution.
\item[{(3)}] The morphology of the true and reconstructed \gal 
distributions can be used to constrain the bias factor $b$ independent of 
$\Omega$.
For a fixed value of $\sigma_{8g}$, a biased galaxy \distrbn is less
dynamically evolved than an unbiased one.
This difference in the degree of dynamical evolution can be easily detected
when comparing galaxy distributions with $b=1,b=2$, and $b=3$, using the \nnbr
distribution.
However, it may be difficult to distinguish between more moderate 
values of $b$, say between $b=1$ and $b=1.5$, using the PSCZ and the ORS 
catalogs.
An uncertain {\it form} of the biasing relation between galaxies and mass
will add another degree of freedom, extending the range of values of 
$b$ that provides acceptable reconstructions, and it will thus reduce the 
discriminatory power of the \rec method.
However, we can expect improvements on this front, as we hope to get 
reasonable bias prescriptions through a better understanding of the galaxy 
formation process using hydrodynamical simulations 
(see, e.g., \cite{co92}; \cite{khw92}).
\item[{(4)}] Reconstruction allows the parameter $\beta = \Omega^{0.6}/b$  
to be constrained more accurately than in conventional analyses of the
anisotropy of the redshift space correlation because:
\begin{description}
\item[{(a)}] The reconstructed peculiar velocity field is fully non-linear and
automatically includes the spatial variations of the non-linear component.
Hence, while estimating $\beta$ from the angular anisotropies in $\xi(s,\mu)$, 
we are not restricted to the large scales  where linear theory is
a good approximation, but we instead use the correlation 
function information in the entire range of pair separations.
\item[{(b)}] Using the correct assumptions for $\Omega$ and $b$ (or at least 
for the combination $\beta = \Omega^{0.6}/b$), we can reproduce the 
$\xi(s,\mu)$ of the input \gal \distrbn more accurately than the cosmic 
variance band, because reconstructions automatically reproduce
the orientations of large scale features that are the main source of
noise in the purely statistical approach to redshift space distortions.
This result is demonstrated in Figures~22 and~32 for the mock PSCZ 
and ORS surveys respectively. 
\end{description}
\item[{(5)}] Reconstruction predicts both the density and the fully non-linear
velocity field starting from the redshift data alone.
Thus, at any location in \z space, we can construct a predicted
distribution of the peculiar velocities of galaxies in its vicinity 
that is more accurate than that provided by linear theory.
This prediction can be used to correct for the inhomogeneous Malmquist bias 
which plagues the estimates of $\beta$ from the comparison between  
observed density and velocity fields.
It also has the potential to improve the performance of velocity-velocity 
comparisons (\cite{wsdk97}; \cite{willick98}), 
as the velocity and velocity dispersion fields 
can be predicted more accurately.
\item[{(6)}] The mock PSCZ catalog can be reconstructed more accurately 
than the mock ORS catalog.
The difference primarily reflects the larger sky coverage of the PSCZ survey
and not its greater depth.
This result
suggests that the \rec of the ORS catalog can be improved by filling 
in the large angular mask region with an appropriately normalized 
PSCZ density field.
Alternatively, the initial density field recovered by the PSCZ catalog
can be used to reconstruct the ORS catalog by using  the forward evolution
 steps that are appropriate for a biased reconstruction.
However, even without these modifications, the $\xi(s,\mu)$ of the 
reconstructed ORS catalog can match its input values to better than the 
cosmic variance limit, and the reconstructed velocity field is a better match
to the true velocity field compared to  the linear theory prediction.
Thus, \rec will improve the ability of these surveys to constrain $\beta$ 
both from  the analysis of clustering anisotropies and from peculiar velocity
comparisons.
\end{description}

Reconstruction analysis is a complement to the statistical approach
to large scale structure, not a replacement for it.
Its strength lies in its ability to break the cosmic variance
barrier and its ability to constrain $\Omega$ and $b$ by simultaneously
using information from linear and non-linear scales.
The penalty is that a reconstruction is not perfectly accurate even
if it is based on correct assumptions, but the systematic
errors of reconstruction on a particular data set can be calibrated
using N-body mock catalogs.  Reconstruction analysis can enhance
the power of galaxy redshift surveys to constrain the density parameter
and the relation between galaxies and mass and to test the hypothesis
that large scale structure originated in the gravitational instability
of Gaussian primordial density fluctuations.

\acknowledgments

We acknowledge valuable conversations over the course of several
years with Avishai Dekel, Carlos Frenk, Mirt Gramann, and Adi Nusser.
We thank Changbom Park for allowing us to use his PM N-body code.
This research was supported by NASA Astrophysical Theory Grant
NAG5-3111 and NSF Grant AST-9616822.

\end{document}